\documentclass[aps, pra,reprint,superscriptaddress,
               amsmath,amssymb,longbibliography]{revtex4-2}
\usepackage{physics}
\usepackage{braket}
\usepackage{graphicx}
\usepackage{verbatim}
\usepackage{microtype}
\usepackage{bbm}
\usepackage[colorlinks=true, linkcolor=blue, citecolor=blue, urlcolor=blue]{hyperref}
\usepackage{titletoc}
\titlecontents{section}[1.6em]
  {\vskip2pt}
  {\contentslabel{1.6em}}
  {\hspace*{-1.6em}}
  {\titlerule*[0.5pc]{.}\contentspage}
\titlecontents{subsection}[3.9em]
  {}
  {\contentslabel{2.3em}}
  {\hspace*{-2.3em}}
  {\titlerule*[0.5pc]{.}\contentspage}

\renewcommand{\eqref}[1]{\ref{#1}}

\begin{document}

%\title{Sequential Measurements as a Resource for Quantum Metrology}
%\title{Sequential Measurements in Ancilla-Assisted Metrology for Displacement Sensing}
\title{Sequential Measurements as a Resource in Ancilla-Assisted Quantum Metrology}
%\title{Sequential Ancilla-Assisted Metrology for Displacement Sensing}

\author{Koray Mentesoglu}
\email{kmente@uw.edu}
\affiliation{Department of Electrical and Computer Engineering, University of Washington, Seattle, Washington 98195, USA}

\author{Rahul Trivedi}
\affiliation{Max Planck Institute of Quantum Optics, Hans-Kopfermann-Stra\ss e 1, 85748 Garching, Germany}
\affiliation{Munich Center for Quantum Science and Technology, Schellingstra\ss e 4, 80799 M\"unchen, Germany}

\author{Sara Mouradian}
\email{smouradi@uw.edu}
\affiliation{Department of Electrical and Computer Engineering, University of Washington, Seattle, Washington 98195, USA}

\date{\today}

\begin{abstract}
We present a protocol for simultaneous estimation of both quadratures of a displacement channel on a quantum harmonic oscillator using a single ancilla qubit, without complex state preparation or adaptive control. The qubit is coupled to the quantum harmonic oscillator via controlled displacements of strength $A$ and undergoes $N$ rounds of measurement and reset, while the oscillator coherently accumulates the displacement signal. The non-commuting displacements imprint an order-dependent geometric phase at each round, giving a quantum Fisher information that scales as $\mathcal{O}(A^{2} N^{3} )$. Although the measurement record space grows as $2^N$, we show that exact likelihoods for any individual measurement record can be computed in $\mathcal{O}(N^{2})$ time. With this, we show that measurement in the computational basis approximately saturates the quantum bound and maintains the $\mathcal{O}(A^2 N^3)$ scaling. Furthermore, the periodic information extraction makes the protocol robust to decoherence. Our results provide a framework for the design and analysis of sequential measurement in ancilla-assisted protocols.
 
\end{abstract}

\maketitle

%\tableofcontents

\section{Introduction}
\label{sec:intro}

Quantum metrology aims to estimate a physical parameter of a signal via the response of a controllable quantum system. Generally, quantum sensing protocols follow a simple pattern~\cite{Degen17}; the system is prepared in an engineered initial state, then the system interacts with the signal of interest, and finally the system's state is measured. Non-classical resources such as entanglement~\cite{Conlon23, Marciniak22, Zhang15, Hempel13} or squeezing~\cite{Eckner23, Burd19, Tse19, Bond26} paired with optimal measurements can provide quantum-enhanced precision~\cite{Grochowski25, Dey25, Valahu25, Labarca26}, though these generally must be optimized with respect to prior information about the signal~\cite{Kaubruegger21}. Quantum control throughout the integration time can be used to improve sensitivity~\cite{Kotler11, Mouradian21, Allen25, Naghiloo17}, though this also often relies on learned information about the signal. 

While most quantum metrology protocols use measurements solely at the end of signal integration, work in quantum error correction and non-classical state preparation has shown that mid-circuit measurements can be embraced as a resource~\cite{Fluh19, Ofek16, Sivak23, Sayrin11, Blok14, Camp20}. To that end, recent works have explored using measurements as a resource in metrology as well~\cite{Direkci26, Chen25, Yang25, Yang24, OConnor24, Monte22, Ritboon22, Ilias22, Schmitt17}. However, the exponentially growing trajectory space generated by repeated measurements has limited existing work to studying the pre-asymptotic regime, leaving open how signal information is distributed across measurement outcomes and how it can be most efficiently extracted.

In this manuscript, we present a metrology protocol that uses measurements as a resource for quantum-enhanced single- and multi-parameter displacement sensing in a quantum harmonic oscillator (QHO). Displacement sensing in a QHO underlies a wide array of applications, including force sensing~\cite{Wang19}, gravitational wave detection~\cite{Aasi13}, magnetometry~\cite{Eisenach21}, electric-field sensing~\cite{Gilmore21}, and exotic particle detection~\cite{Braggio25}. The presented protocol borrows the ancilla-assisted structure of Kitaev's phase estimation algorithm and its descendants~\cite{Kitaev02, Kimmel15, Higg07, Terhal16}, but critically does not require the ancilla to be prepared in an eigenstate of the channel of interest. The measured sequence of outcomes forms a correlated stochastic process~\cite{OConnor24, Pasquale17, Asadian14} that distributes signal information across time; a central contribution of this work is a framework for identifying where in this record the information resides. The presented protocol is similar to other approaches that rely on a system's native evolution for in-situ computation, including reservoir computing~\cite{Senanian24}, pulse characterization~\cite{Dong25}, and analog simulations~\cite{Geim26} and is compatible with any system supporting oscillator-ancilla coupling such as trapped ions, superconducting qubits, or neutral atoms~\cite{Leibfried03, Aspelmeyer14, Liu25,Li26}. In contrast to previous demonstrations of simultaneous estimation of non-commuting variables~\cite{Bond26, Valahu25}, no complex non-classical state preparation or adaptive experimental control is required. 

\begin{figure*}[t]\centering  \includegraphics[width=\textwidth]{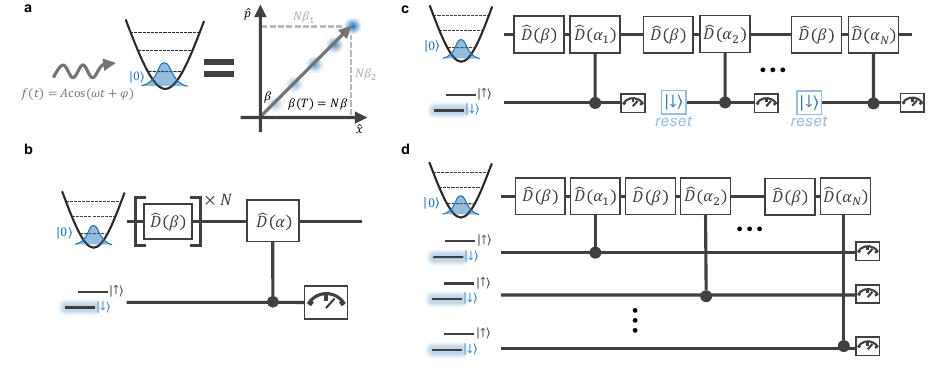}\caption{\textbf{a} A QHO feels a force $f(t)$, which manifests as a displacement in phase space. A single interrogation time consists of $N$ signal interactions. \textbf{b} \textit{Single Measurement Protocol:} Phase is accumulated on the qubit to gain a factor of $N$ advantage in precision. The signal displacement $\hat{D}(\beta)$ is followed by an ancilla-assisted measurement, which includes a state-dependent coupling displacement and projective measurement on the qubit in $ \{  \ket{\uparrow},  \ket{\downarrow} \}$. \textbf{c} \textit{Sequential Measurement Protocol:} The coupling and measurement operations are repeated at every round. The qubit is reset into the ground state before the next round, and the QHO evolution continues. \textbf{d} An equivalent interpretation to \textbf{c}, where $N$ ancillae are initialized and never reset. Instead, each of the $N$ ancillae are measured once, after the last controlled displacement. }\label{fig}\end{figure*}

This manuscript is organized as follows. In Sec.~\ref{sec:weak_meas}, we outline the building blocks for ancilla-assisted measurement of a QHO and introduce a single-measurement protocol as a baseline. In Sec.~\ref{sec:sequential}, we present the sequential-measurement protocol central to this work and develop the analytical framework used throughout. By mapping the repeatedly measured ancilla onto a register of ancillae and deferring measurements, we are able to compress the exponentially large record space onto a permutation-invariant reference state of polynomial dimension, with the signal entering only through single-qubit rotations. In Sec.~\ref{sec:qfi_single}, we use this structure to separate the quantum Fisher information (QFI) into a coherence contribution and a closed-form contribution due to the ancilla-assisted measurements, showing that information grows with measurement strength. In Sec.~\ref{sec:qfi_two}, we extend the framework to simultaneous estimation of both quadratures within a single interrogation time, and find that the QHO--ancilla coupling schedule can be used as a design parameter to maximize the QFI. Notably, the two-parameter protocol matches the total information of its single-parameter counterpart at an equal number of signal interactions. In Sec.~\ref{sec:cfi}, we restrict our analysis to measurements in the computational basis and use the structure developed for the QFI to compute exact measurement-record likelihoods at polynomial cost, showing that the classical Fisher information (CFI) approximately saturates the quantum bound for the chosen basis. Furthermore, we are able to identify how information about the signal is distributed across per-round outcomes and their correlations. We also show that the sequential protocol is naturally more resilient to phase wrapping, extending the sensor's dynamic range. Finally, in Sec.~\ref{sec:decoherence}, we consider the effect of noise showing increased resilience to coupling to a thermal bath, but more sensitivity to dephasing.

\section{Ancilla-Assisted Displacement Sensing}

We consider the joint estimation of the quadratures $(\beta_1,\beta_2)\in\mathbb{R}^2$ of a displacement operator $\hat{D}(\beta) = \exp(\beta\hat{a}^\dagger - \beta^*\hat{a})$ where $\beta = \beta_1 + i\beta_2$ (Fig.~\ref{fig}(a)). We treat $\hat{D}(\beta)$ as an unknown unitary channel applied $N$ times to a known probe state. Concretely, this could arise from a resonant force with a constant but unknown amplitude and phase, e.g. an electric field on a trapped ion, for a total time $T=N\tau$ where $\tau$ is the duration of a single application.

If a local oscillator is available, homodyne or heterodyne detection with a classical input state saturates the one- or two-parameter standard quantum limit (SQL), respectively~\cite{Genoni13, Duv17} (see Table~\ref{tab:protocol_comparison}). We consider the case where no local oscillator is available, and information is instead obtained by coupling the QHO to an ancilla qubit via a state-dependent displacement $\hat{D}(\hat{\sigma}_{\hat{n}} \alpha) = e^{(\alpha \hat{a}^{\dagger} - \alpha^{*} \hat{a})\hat{\sigma}_{\hat{n}}}$, where $\alpha \in \mathbb{C}$ sets the displacement of the QHO state in phase space and $\hat{\sigma}_{\hat{n}}$ sets the qubit basis on which it is conditioned. Measuring the ancilla provides information about the QHO without fully projecting its state~\cite{Aharonov88}. Such controlled displacement operations are available across superconducting, trapped-ion, neutral-atom, and photonic platforms~\cite{Liu25}.

In the following, we present and compare protocols. While a single terminal ancilla-assisted measurement matches the $\mathcal{O}(N^{2})$ information scaling of local-oscillator detection, we find that interleaving $N$ measurements and resets improves the scaling to $\mathcal{O}(A^{2}N^{3})$, as quantified by the Fisher information $\mathcal{F}$. These results are presented concisely in Table~\ref{tab:protocol_comparison} and formally outlined in Secs.~\ref{sec:qfi} and~\ref{sec:cfi}.

\begingroup
\renewcommand{\arraystretch}{1.7}
\setlength{\tabcolsep}{4pt}
\begin{table}[t]
\caption{\label{tab:protocol_comparison}
Scaling of single- and two-parameter estimation, reflected in the scalar Fisher information $\mathcal{F}$ and the trace $\operatorname{Tr}\mathcal{F}$ of the Fisher information matrix, per interrogation of $N$ signal displacements $\hat{D}(\beta)$, without decoherence. $A = |\alpha|$ is the controlled-displacement amplitude; LO = local oscillator; qb = ancilla qubit; MCMR\ = mid-circuit measurement and reset. The single- and sequential-measurement protocols implement $\hat{D}(\hat{\sigma}_{x}\alpha)\,\hat{D}(\beta)^{N}$ and $\big[\hat{D}(\hat{\sigma}_{x}\alpha)\,\hat{D}(\beta)\big]^{N}$, respectively.}
\begin{tabular}{cccc}
\hline
Protocol & Resources &$\mathcal{F}$ & $\operatorname{Tr}\mathcal{F}$  \\
\hline\hline
Homodyne & LO & $\mathcal{O}(N^{2})$ & --  \\
\hline
Heterodyne & LO & $\mathcal{O}(N^{2})$ & $\mathcal{O}(N^{2})$  \\
\hline
Single meas. & 1 qb & $\mathcal{O}(N^{2})$ & --  \\
\hline
Sequential & 1 qb, $N$ MCMR &
$\mathcal{O}(A^{2}N^{3})$ &
$\mathcal{O}(A^{2}N^{3})$  \\
\hline
\end{tabular}
\end{table}
\endgroup

\subsection{ Single-measurement protocol }
\label{sec:weak_meas}

The simplest ancilla-assisted scheme for single-quadrature displacement sensing is shown in Fig.~\ref{fig}(b). The ancilla is prepared in $\ket{\downarrow}$ and the QHO in the vacuum state $\ket{0}$. The accumulated signal displacement produces a coherent state, $\hat{D}(N\beta) \ket{\downarrow} \ket{0} = \ket{\downarrow} \ket{N\beta}$, after which the controlled displacement entangles the ancilla and QHO:
\begin{align}\label{eq:entangling}
\ket{\Psi_{\beta} } =  \hat{D}\left( \hat{\sigma}_{x} \alpha \right) \ket{\downarrow} \ket{N\beta} = \frac{1}{\sqrt{2}} \Big( & e^{i \theta} \ket{+} \ket{N\beta + \alpha}   \notag \\ & + e^{-i \theta} \ket{-} \ket{N\beta - \alpha} \Big),
\end{align}
where $\theta \equiv N \ \operatorname{Im}(\alpha \beta^{*})$, $\ket{\pm} = (\ket{\downarrow} \pm \ket{\uparrow})/\sqrt{2}$, and we have set $\hat{\sigma}_{ \hat n } = \hat{\sigma}_{x}$. Measurement of the ancilla in the computational basis yields outcome $b\in\{\downarrow,\uparrow\}$ with probability $p_{b}(\theta) = \bra{\Psi_{\beta}} \left(\ket{b}\bra{b}\otimes\mathbbm{1}\right) \ket{\Psi_{\beta}} = \frac{1}{2}\left[ 1 \pm e^{-2|\alpha|^{2}} \cos(4\theta) \right]$ (App.~\ref{app:single_meas}), the positive (negative) sign corresponds to $b = \,\downarrow$ ($b = \,\uparrow$). 
After the qubit measurement, the normalized QHO state, $\ket{\psi_b(\beta)} = (\bra{b}\otimes\mathbbm{1})\ket{\Psi_\beta}/\sqrt{p_b(\theta)}$, depends on both the signal and measurement outcome:
\begin{equation}\label{eq:conditional} \ket{\psi_b(\beta)} = \frac{1}{2\sqrt{p_b (\theta) }}  \big( e^{i \theta} \ket{ N \beta + \alpha} \pm e^{-i\theta}\ket{ N \beta - \alpha}\big). 
\end{equation}
The phase of $\alpha$ sets the quadrature to be estimated: setting $\alpha = i A$ makes $p_{b}(\theta)$ maximally sensitive to $\operatorname{Re}(\beta)$ and insensitive to $\operatorname{Im}(\beta)$~\cite{Sidhu20}. However, as the magnitude $|\alpha| = A$ is increased, $p_{b}(\theta) \rightarrow 0.5$, removing signal dependence from the outcome statistics~\cite{Fluh18}. Setting $N = 1$ and performing $M$ independent shots provides an estimate of one quadrature of $\beta$ with a variance that scales as $1/M$, the SQL. With an accumulated signal of $N\beta$ (Fig.~\ref{fig}(b))  estimation variance scales as $1/(MN^{2})$ in the absence of decoherence. As seen in Table~\ref{tab:protocol_comparison}, this method matches both the scaling of a homodyne measurement~\cite{Luis02} and the Heisenberg limit scaling achieved with multi-probe entanglement by trading spatial resources for time~\cite{Higg07}. In the following, we refer to this protocol as the single-measurement protocol.

\subsection{Sequential-measurement protocol}
\label{sec:sequential}

In this manuscript, we introduce a protocol which interrupts the coherent signal integration with periodic ancilla-assisted measurements of the same form as in Sec.~\ref{sec:weak_meas} (Fig.~\ref{fig}(c)). These measurements extract signal information throughout the sensor interrogation time and the repeated conditional displacements advantageously perturb the QHO dynamics. We note that the repeated conditional displacements, with amplitude $A$, inject energy into the QHO, which is known to be a resource for improving sensitivity~\cite{Kotler11, Rossi23}. However, if the same $N$ conditional displacements are applied without measurement and reset, the information cannot be efficiently extracted by the ancilla qubit (App.~\ref{app:single_ancilla}), making the periodic measurement and reset itself a resource.

The protocol again starts by initializing the oscillator in the vacuum $\ket{0}$ and the ancilla in $\ket{\downarrow}$, followed by a signal displacement and an ancilla-assisted measurement. Depending on the outcome, the QHO is projected into one of two symmetric superpositions of coherent states, with a phase set by the measurement outcome as seen in  Eq.~\eqref{eq:conditional}. After measurement, the qubit is reset to $\ket{\downarrow}$ while the QHO state is carried forward into the next round (Fig.~\ref{fig}(c)). In this sequential-measurement protocol, information about $\beta$ is encoded in an $N$-bit string with $2^N$ possible measurement outcomes, generally complicating asymptotic analysis and signal estimation~\cite{Direkci26, Monte22}.

To simplify the analysis, we map the $N$ measurements on a single ancilla qubit to one measurement on each of $N$ ancillae (Fig.~\ref{fig}(d)). Since the qubit is reset to $\ket{\downarrow}$ regardless of the outcome, each round applies the same operation to the oscillator, independent of earlier outcomes, so deferring all readout to the end gives the same probabilities, provided the $N$ ancillae are measured in a product basis (App.~\ref{app:one_to_N}). Since the quantum Fisher information of the unmeasured $N$-ancilla state also allows measurements entangled across the ancillae, it upper-bounds the information extractable by the single-ancilla protocol.

With the $N$-ancillae model, we derive the post-protocol state. Since each round of the sequential protocol applies a signal displacement followed by a controlled displacement, these displacements are combined using the Baker--Campbell--Hausdorff (BCH) commutation relation $\hat D(\mu)\hat D(\nu) = e^{\,i\,\operatorname{Im}(\mu\nu^{*})}\hat D(\mu+\nu)$. After $N$ rounds the joint ancilla--oscillator state can be separated as 
\begin{equation}
\label{eq:seq_state_factored}
\ket{\psi_N(\beta)} = \big(\hat U_{\mathrm{BCH}}(\beta)\otimes\mathbbm 1\big)\,\ket{\widetilde{\psi}_N(\beta)},
\end{equation}
where 
\begin{equation}
\label{eq:seq_branch_state}
\ket{\widetilde{\psi}_{N}(\beta)} = \frac{1}{2^{N/2}} \sum_{\mathbf a\in\{ +, - \}^{N}} \ket{\mathbf a} \otimes  \ket{N\beta+\textstyle i\sum_{j=1}^{N}a_{j}A},
\end{equation}
and
\begin{equation}
\label{eq:seq_bch_rotation}
\hat U_{\mathrm{BCH}}(\beta) = \exp\!\left[iA\beta\sum_{j=1}^N (2j-N)\hat\sigma_x^{(j)}\right].
\end{equation}
In Eq.~\ref{eq:seq_branch_state} the QHO coherent state label depends on the branch only through $\sum_{j=1}^{N} a_{j} A$, where $a_{j}$ is the $\hat \sigma ^{(j)}_{x}$ eigenvalue labeling branch $j$. It is therefore invariant under permutations of $\mathbf a$. $\hat U_{\mathrm{BCH}}(\beta)$, however, encodes the order in which outcomes are observed through the weights $(2j-N)$. In Eqs.~\ref{eq:seq_branch_state} and ~\ref{eq:seq_bch_rotation} we have assumed single-parameter estimation, with $\alpha_{j} = iA \, \forall j$ and $\beta \in \mathbb{R}$. 

The ancilla register in Eq.~\ref{eq:seq_branch_state} spans an exponentially large space, with $2^N$ possible records~\cite{Yang23,OConnor24}. To simplify the Fisher information analysis we notice that, due to its invariance under permutations, the state in Eq.~\eqref{eq:seq_branch_state} can be compactly written as $\ket{\widetilde{\psi}_{N}(\beta)} = \sum_{k=0}^{N}\sqrt{\gamma_k}\,\ket{\phi_k}\otimes\ket{N\beta+i(N-2k) A}$ where
\begin{equation}
\label{eq:seq_dicke_basis}
\ket{\phi_{k}} = \binom{N}{k}^{-1/2}\sum_{\mathbf a:\,w(\mathbf a)=k}\ket{\mathbf a},
\end{equation}
is the Dicke basis, $\gamma_k = \tfrac{1}{2^N}\binom{N}{k}$, and  $w(\mathbf a)$ is the number of qubits in $\ket{-}$, such that $\sum_j a_j = N-2w(\mathbf{a})$. The states, $\ket{\widetilde{\psi}_{N}(\beta)}$, thus occupy an $(N+1)$-dimensional symmetric subspace rather than the full $2^N$-dimensional ancilla Hilbert space.

Tracing the oscillator out of $\ket{ \widetilde{\psi}_{N}(\beta) }$ gives
\begin{equation}
\label{eq:seq_reduced_state}
\widetilde{\rho}_{N}(\beta) =  \sum_{k,k'=0}^{N}\sqrt{\gamma_k\gamma_{k'}}\,e^{-2(k-k')^2 A^2}e^{-2i(k-k') N A\beta  }\ket{\phi_k}\bra{\phi_{k'}},
\end{equation}
and the ancilla state that sets the final bit-string probabilities follows by restoring the phase
\begin{equation}
    \label{eq:rotation_on_invariant}
    \rho_N(\beta) = \hat U_{\mathrm{BCH}}(\beta)\,\widetilde{\rho}_N(\beta)\,\hat U_{\mathrm{BCH}}^\dagger(\beta).
\end{equation}
Due to the application of $\hat U_{\text{BCH}}$, $\rho_{N}(\beta)$ is no longer permutationally invariant. While it might appear that this order-dependent phase could be a computational bottleneck when analyzing the protocol, in Sec.~\ref{sec:qfi_single}, we show that its contribution to the Fisher information admits a closed form.

\section{Quantum Fisher Information}
\label{sec:qfi}
\subsection{Single-parameter estimation}
\label{sec:qfi_single}
Here we compare the quantum Fisher information (QFI) of the sequential- and single-measurement protocols for single-parameter estimation with $\alpha = iA$ and $\beta \in \mathbb{R}$. We compute the QFI of the $N$-ancillae register, which implicitly allows for entangling measurements and therefore gives an upper bound on the information accessible to the physical single-ancilla protocol. Full derivations of all results in this section are given in App.~\ref{app:qfi}.

We first consider the QFI of the full QHO--ancilla state which gives an upper bound on what any measurement, on any part of the system, can extract. For the single-measurement protocol,
\begin{equation}
    \label{eq:single_full_qfi}
    \mathcal{F}^{Q}_{\mathrm{single,full}} = 4N^{2},
\end{equation}
independent of $A$, since all of the information comes from the coherently accumulated displacement $N\beta$. For the sequential protocol,
\begin{equation}
    \label{eq:seq_full_qfi}
    \mathcal{F}^{Q}_{\mathrm{seq,full}}
    = 4N^{2} + \frac{8}{3}A^{2}N(N-1)(2N-1)
    = \mathcal{O}(A^{2}N^{3}).
\end{equation}
The additional term arises because the signal and controlled displacements at each round do not commute, so every round imprints a signal-dependent BCH phase conditioned on the state of its ancilla.

In a setting where the QHO is not accessible, we consider the QFI of the ancillae. For the single-measurement protocol, tracing out the QHO leaves a single-qubit state, admitting a closed form QFI,
\begin{equation}
    \label{eq:single_ancilla_qfi}
    \mathcal{F}^{Q}_{\mathrm{single}} = 16 N^{2} A^{2} e^{-4A^{2}},
\end{equation}
which retains the $\mathcal{O}(N^{2})$ scaling of Eq.~\eqref{eq:single_full_qfi}, with an optimum at $A = 0.5$.

The same analysis of sequential-measurement protocols is generally not possible due to the $N$ correlated ancillae and measurement-conditioned QHO dynamics. However, by leveraging the structure identified in Sec.~\ref{sec:sequential}, we formulate a tractable way to compute the ancillae QFI. We express the final ancillae state with an $N$ qubit rotation  
\begin{equation}
    \label{eq:total_generator}
    \rho_{N}(\beta) = e^{i\beta\hat{H}}\,\widetilde{\rho}_{N}(0)\,e^{-i\beta\hat{H}}, \qquad \hat{H} = 2A\sum_{j=1}^{N} j\,\hat{\sigma}_{x}^{(j)},
\end{equation}
where $\widetilde{\rho}_{N}(0) = \sum_{k,k'=0}^{N}\sqrt{\gamma_k\gamma_{k'}}\,e^{-2(k-k')^2A^2}\ket{\phi_k}\bra{\phi_{k'}}$ and $\hat H$ combines the order-dependent phase $\hat U_{\text{BCH}}$ (Eq.~\ref{eq:seq_bch_rotation}) with the accumulated phase from Eq.~\ref{eq:seq_reduced_state}. 

Writing the final ancillae register state as Eq.~\eqref{eq:total_generator} allows for more efficient computation of the QFI. This efficiency boost relies on the permutation-invariance of $\widetilde{\rho}_{N}(0)$ and the order-encoded phases of $\hat U_{\text{BCH}}$, which separate the QFI into respective contributions
\begin{equation}
    \label{eq:qfi_three_term}
    \mathcal{F}^{Q}_{\mathrm{seq}} = \mathcal{F}_{\mathrm{coh}} + \mathcal{F}_{\mathrm{BCH}}.
\end{equation}

The QFI arising from the permutation-invariant state, encoded in $\mathcal{F}_{\text{coh}}$, requires numeric diagonalization of the reference state $\widetilde{\rho}_{N}(0)$ to find its eigenpairs $\lambda_{a}$, $\ket{\widetilde{e}_{a}}$ and takes the form,
\begin{equation}
    \label{eq:qfi_coherence_main}
    \mathcal{F}_{\mathrm{coh}}
    = 2(N{+}1)^{2}A^{2}
    \sum_{\substack{a,a' \\ \lambda_{a}+\lambda_{a'}>0}}
    \frac{(\lambda_{a}-\lambda_{a'})^{2}}{\lambda_{a}+\lambda_{a'}}\, \big|\bra{\widetilde{e}_{a}}\hat{S}_{x}\ket{\widetilde{e}_{a'}}\big|^{2},
\end{equation}
where $\hat{S}_{x} = \sum_{j}\hat{\sigma}_{x}^{(j)}$. This contribution is similar to $\mathcal{F}^{Q}_{\mathrm{single}}$, as it relies on the same Dicke coherences that are suppressed by the Gaussian envelope of Eq.~\eqref{eq:seq_reduced_state}.

Though $\hat H$ of Eq.~\ref{eq:total_generator} breaks permutation-invariance, we show the QFI arising from order dependence, encoded in $\mathcal{F}_{\text{BCH}}$, can be reduced to a closed form (App.~\ref{app:seq_qfi})
\begin{equation}
    \label{eq:qfi_backaction}
    \mathcal{F}_{\mathrm{BCH}} = \frac{4}{3}\,A^{2}\,N(N^{2}-1).
\end{equation}
Since $\mathcal{F}_{\text{coh}} \geq 0$, Eq.~\ref{eq:qfi_backaction} serves as a lower bound on the sequential QFI. Furthermore, Eq.~\eqref{eq:qfi_backaction} grows monotonically with $A$, and the ancillae QFI maintains the $\mathcal{O}(A^{2}N^{3})$ scaling of the full system (Eq.~\ref{eq:seq_full_qfi}).

\begin{figure}[t] \centering \includegraphics{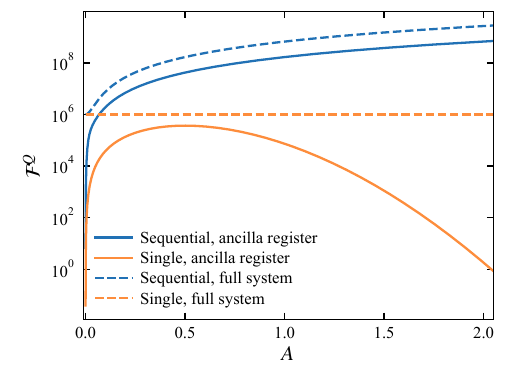} \caption{Quantum Fisher information $\mathcal{F}^{Q}$ as a function of the controlled displacement amplitude $A$ for $N=500$. Blue curves show the sequential-measurement protocol, and orange curves show the single-measurement protocol. Solid and dashed curves correspond to the ancilla register and full system QFI, respectively.} \label{fig:single_qfi} \end{figure}

Fig.~\ref{fig:single_qfi} compares the reduced ancilla QFI with the corresponding full QHO--ancilla QFI for the sequential- and single-measurement protocols as a function of $A$ at $N = 500$. For both protocols, the QFI of the ancillae is less than that of the full system. The sequential-measurement protocol attains higher QFI than the single-measurement protocol due to $\mathcal{F}_{\text{BCH}}$ (Eq.~\ref{eq:qfi_backaction}). The Gaussian suppression of the coherence terms in $\mathcal{F}_{\text{single}}^{Q}$ (Eq.~\ref{eq:single_ancilla_qfi}) is observed with increasing $A$. $\mathcal{F}_{\text{coh}}$ experiences the same suppression, though due to $\mathcal{F}_{\text{BCH}}$, is not visible in the figure. We note that the sequential curve in Fig.~\ref{fig:single_qfi} follows from diagonalizing a single $(501)$-dimensional matrix (Eq.~\ref{eq:total_generator}) and is computed quickly, while a direct treatment of the $2^{500}$-dimensional register would be intractable.

\subsection{Two-parameter estimation}
\label{sec:qfi_two}
We have so far restricted ourselves to single-quadrature estimation after $N$ signal interactions. Here we extend the protocol to estimate both quadratures of the displacement with the same $N$ signal interactions. Full derivations of all results in this section are given in App.~\ref{app:two_param}. 

To estimate two parameters, the scalar QFI generalizes to a $2 \times 2$ QFI matrix with elements $\mathcal{F}_{\mu \nu}^{Q},$ where $ \mu, \nu \in \{\beta_{1}, \beta_{2}\}$ represent $\{ \operatorname{Re}(\beta), \operatorname{Im}(\beta) \}$, respectively. The diagonal elements quantify the information about each quadrature, and the off-diagonal elements capture correlations between the quadrature estimates.

To access both quadratures, we vary the direction of the controlled displacement at each round, $\alpha_{j} = r_{j} A$ with $r_{j} \in \{1, i\}$. To distinguish rounds, we define subsets $\mathcal{I} = \{j : r_{j} = i\}$ and $\mathcal{R} = \{j : r_{j} = 1\}$ with $N_{I} = N_{R} = N/2$. Therefore, information about $\beta_{1}$ ($\beta_{2}$) is recorded during rounds in $\mathcal{I}$ ($\mathcal{R}$). All other elements of the protocol are unchanged. Repeating the steps of Sec.~\ref{sec:sequential}, the register of qubits can be represented as a rotated permutation-invariant state, $\rho^{(2)}_{N}(\beta) = \hat{U}^{(2)}_{\mathrm{BCH}}(\beta)\, \widetilde{\rho}^{(2)}_{N}(\beta)\, \hat{U}^{(2)\dagger}_{\mathrm{BCH}}(\beta)$, where $\widetilde{\rho}^{(2)}_{N}$ is supported on a doubly permutation-invariant subspace spanned by the Dicke states $\ket{\phi_{k_{R},k_{I}}}$, of dimension $(N_{R}+1)(N_{I}+1)$ (App.~\ref{app:two_param}). As in the single-parameter case, the exponentially large trajectory space is thus compressed to one polynomial in size.  

To compute the QFI, we follow a similar procedure to Sec.~\ref{sec:qfi_single} and find the QFI separates similarly: $\mathcal{F}_{\mu\nu}^{Q} = \mathcal{F}^{\mathrm{coh}}_{\mu\nu} + \mathcal{F}^{\mathrm{BCH}}_{\mu\nu}$. We find the QFI matrix to be diagonal, the coherence contribution to be similar to the single-parameter case (Eq.~\ref{eq:qfi_coherence_main}), and the BCH contribution to have the closed form
 \begin{equation}
    \label{eq:two_param_backaction}
    \mathcal{F}^{\mathrm{BCH}}_{\mu\mu} = 4A^{2}\, \mathrm{Var}_{\mathcal{B}_{\mu}}(c),
\end{equation}
where $\mathcal{B}_{\beta_{1}} = \mathcal{I}$ and $\mathcal{B}_{\beta_{2}} = \mathcal{R}$ label the subset of rounds probing each quadrature and $\mathrm{Var}_{\mathcal{B}}(c) = \sum_{j\in\mathcal{B}} (c_{j} - \bar{c}_{\mathcal{B}})^{2}$ is the squared spread of $c_{j} = N - 2j$ around $\bar{c}_{\mathcal{B}} = N_{\mathcal{B}}^{-1}\sum_{j\in\mathcal{B}} c_{j}$ (App.~\ref{app:blocked}). Here $c_{j}$ can be interpreted as the weighted round index of each qubit, encoding the order in which the controlled displacement was applied. In the single-parameter protocol, every round probes the same quadrature, so $\mathcal{B} = \{ 1, ..., N \}$, fixing $\text{Var}_{\mathcal{B}}(c) = \frac{1}{3}N(N^{2} - 1)$, which gives Eq.~\ref{eq:qfi_backaction}. In the two-parameter protocol, splitting rounds between $\mathcal{R}$ and $\mathcal{I}$ can tune $\text{Var}_{\mathcal{B}}(c)$. 

The schedule of $\{\alpha_{j}\}$ is thus a design parameter of two-parameter estimation. We consider two different schedules of applied controlled displacements: blocked, in which the first $N/2$ rounds couple along $\alpha = A$ and the last $N/2$ along $\alpha = iA$, and alternating, $r_{j} = (1,i,1,i,...)$ (App.~\ref{app:blocked}). In the blocked schedule, each subset occupies $N/2$ consecutive rounds, giving a similar BCH contribution per quadrature as in the single-parameter protocol: $\text{Var}_{\mathcal{B}}(c) = \frac{1}{3} N_{\mathcal{B}} (N^{2}_{\mathcal{B}} - 1)$. At large $A$, where $\mathcal{F}^{\text{BCH}}_{\mu \mu} \gg \mathcal{F}^{\text{coh}}_{\mu \mu}$, the protocol reduces to two independent half-length single-parameter runs, giving $\operatorname{Tr}\mathcal{F}^{Q}_{\mathrm{blk}} = \tfrac{1}{3}\,A^{2}\,N(N^{2}-4)$. In the alternating schedule, each subset is sampled every other round. Since its weights are spread across the full interrogation time, the variance term is quadrupled: $\text{Var}_{\mathcal{B}}(c) = \frac{4}{3} N_{\mathcal{B}}(N^{2}_{\mathcal{B}} - 1)$, so $\operatorname{Tr}\mathcal{F}^{Q}_{\mathrm{alt}} = \tfrac{4}{3}\,A^{2}\,N(N^{2}-4)$. Therefore, by tuning the BCH phase through the coupling schedule $\{ \alpha_{j} \}$, we can maximize the QFI gain of the two-parameter protocol.

\begin{figure}[t] \centering \includegraphics{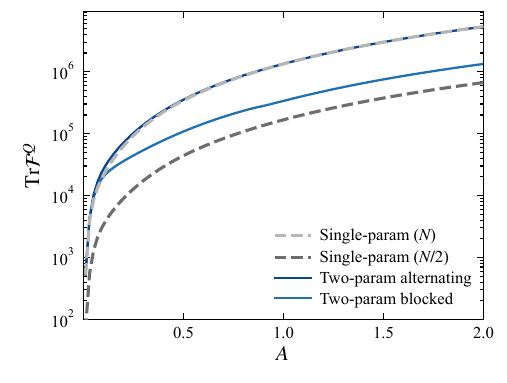} \caption{Total quantum Fisher information $\operatorname{Tr}\mathcal{F}^{Q}$ of the two-parameter sequential protocol as a function of the controlled displacement amplitude $A$ for $N = 100$ signal interactions. Dark and medium blue curves show the alternating and blocked coupling orderings, respectively. Light and dark gray dashed curves show the single-parameter sequential QFI at the same $N$ and at $N/2$ signal interactions, respectively.}\label{fig:two_qfi} \end{figure}

Fig.~\ref{fig:two_qfi} compares the total information $\operatorname{Tr} \mathcal{F}^{Q}$ of the two orderings at $N=100$, alongside two single-parameter references: the sequential QFI at $N$ and $N/2$ signal interactions. The former matches the two-parameter protocol in total signal interrogations while the latter matches the two-parameter protocol in per-quadrature probes. At large $A$, the blocked ordering is similar to two $N/2$ runs, and accordingly is twice the single-parameter $N/2$ curve. The alternating ordering instead outperforms the $N$ single-parameter curve at low $A$ and is equal at large $A$.

In general, there is no  guarantee that both parameters can be estimated simultaneously. The QFI matrix bounds the covariance matrix $\Sigma$ of any unbiased estimator through the multi-parameter quantum Cram\'er--Rao bound (QCRB), $\Sigma \geq (\mathcal{F}^{Q})^{-1}$, so the total variance obeys $\sum_{\mu}\mathrm{Var}(\hat{\beta}_{\mu}) \geq \operatorname{Tr}\!\big[(\mathcal{F}^{Q})^{-1}\big]$. However, the bound may not be attainable if the optimal measurements for the individual parameters are incompatible~\cite{Ragy16, Fiderer21}. Joint saturability is governed by the compatibility condition
\begin{equation}
    \label{eq:weak_commutation}
    \operatorname{Tr}\!\left(\rho^{(2)}_{N}(\beta)\,
    \big[\hat{L}_{\beta_{1}},\hat{L}_{\beta_{2}}\big]\right) = 0,
\end{equation} 
where the $\hat{L}_{\mu}$ are the symmetric logarithmic derivatives. In App.~\ref{app:compatibility}, we find that the QCRB is asymptotically jointly saturable.

\section{Classical Fisher Information}

\subsection{Single-parameter estimation}
\label{sec:cfi}
The QFI bounds of Secs.~\ref{sec:qfi_single} and~\ref{sec:qfi_two} permit arbitrary measurements on the ancilla register. In this section we fix the measurement to the simplest single-ancilla protocol; at every round, the qubit is measured in the computational basis and produces a bit $b_{j}$. After $N$ rounds, a record $\mathbf{b} = (b_{1}, \ldots, b_{N})$ is formed. The classical Fisher information (CFI) of the record distribution $p_{\mathbf{b}}(\beta)$ quantifies the information available to any unbiased estimator built from this record and obeys $\mathcal{F}^{C} \leq \mathcal{F}^{Q}$.

The single-measurement protocol only produces a single outcome, distributed as $p_{ b }(\,\beta) = \tfrac{1}{2}[1 + (-1)^{ b } e^{-2A^{2}} \cos(4NA\beta)]$, which has the closed form CFI, (App.~\ref{app:cfi_single})
\begin{equation}
    \label{eq:cfi_single_main}
    \mathcal{F}^{C}_{\mathrm{single}}
    = \frac{16 N^{2} A^{2}\, e^{-4A^{2}} \sin^{2}\!\big( 4NA\beta \big)}
           {1 - e^{-4A^{2}} \cos^{2}\!\big( 4NA\beta \big)}.
\end{equation}
The CFI of this protocol saturates the QFI bound at $4NA\beta = \pi/2 + m\pi$ and goes to zero at $4NA\beta =  m \pi$ for integer $m$. Estimation of $\beta$ is ambiguous with $m$ which is a well-characterized phase-wrapping problem~\cite{Kimmel15, Higg07}.

The CFI of the sequential protocol does not have a closed form since the measurement record grows as $2^{N}$ and outcomes are correlated across rounds. However, the structure identified in Sec.~\ref{sec:sequential} provides a framework for analysis in polynomial time. Although $\hat{U}_{\mathrm{BCH}}$ breaks the permutation invariance of the register state, Eq.~\eqref{eq:total_generator} shows that the state is generated from the permutation-invariant reference state by a sum of single-qubit rotations, $\hat{H} = 2A\sum_{j} j\,\hat{\sigma}_{x}^{(j)}$. Since each round of the protocol is measured in the computational basis, $\ket{\mathbf{b}} = \bigotimes_{j=1}^{N}\ket{b_{j}}$, the signal dependence of the register state can be incorporated into the measurement, $p_{\mathbf{b}}(\beta) = \bra{\mathbf{b}(\beta)}\, \widetilde{\rho}_{N}(0)\, \ket{\mathbf{b}(\beta)}$, where $\ket{\mathbf{b}(\beta)} = e^{-i\beta\hat{H}}\ket{\mathbf{b}} = \bigotimes_{j} e^{-2iAj\beta\,\hat{\sigma}_{x}^{(j)}}\ket{b_{j}}$. Computing $p_{\mathbf{b}}(\beta)$ then reduces to computing $\braket{\phi_{k} | \mathbf{b}(\beta)}$ for all measurement outcome trajectories. 

Using this technique, we show the probability of a measurement record $\mathbf{b}$ reduces to
\begin{equation}
    \label{eq:cfi_quadratic_main}
    p_{\mathbf{b}}(\beta)
    = 2^{-N}\, \widetilde{c}^{\,\dagger} R\, \widetilde{c},
\end{equation}
where $R_{kk'} = \sqrt{\gamma_{k}\gamma_{k'}}\, e^{-2A^{2}(k-k')^{2}}$ and the coefficients $\widetilde{c}_{k}$ can be computed through recursion at $\mathcal{O}(N^{2})$ per record (App.~\ref{app:cfi_seq}). Therefore, exact likelihoods of bit-string outcomes are obtained without simulating the conditioned QHO--ancilla state, which would otherwise require tracking the exponentially large ancillae space. Exact calculation of the CFI, which sums over all $2^{N}$ records, then costs $\mathcal{O}(2^{N} N^{2})$. At especially large $N$, we can instead estimate it using $M$ records sampled through the exact conditionals $p(b_{s}\,|\,\mathbf{b}_{<s})$, costing $\mathcal{O}(M N^{3})$ (App.~\ref{app:cfi_sampling}).

\begin{figure}[t]
    \centering
    \includegraphics[width=\columnwidth]{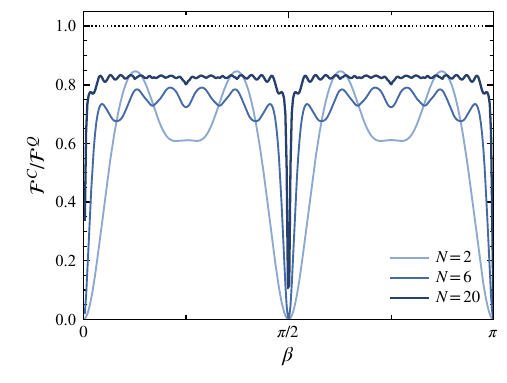}
    \caption{ Ratio of the CFI to the QFI, $\mathcal{F}^{C}/\mathcal{F}^{Q}$, as a function of the signal $\beta$ for $N = \{2, 6, 20\}$ at $A = 0.5$. The dotted line marks saturation of the QFI.}
    \label{fig:cfi_ratio}
\end{figure}

Fig.~\ref{fig:cfi_ratio} plots $\mathcal{F}^{C}/\mathcal{F}^{Q}$ for the sequential protocol at $A = 0.5$ and $N \in \{2, 6, 20\}$. While the CFI does not saturate the QFI bound, the ratio approaches one over a broad range of signals even without adaptive control. The structure of the CFI is captured by the single-round marginals of the measurement record. Tracing out all other rounds (App.~\ref{app:cfi_single_round}), the marginal probability at the $j^{th}$ round is
\begin{equation}
    \label{eq:single_round_marginal_main}
    p_{b_{j}}(\beta)
    = \frac{1}{2}\left[ 1 + (-1)^{b_{j}}\, e^{-2A^{2}} \cos\!\big( 4 j A \beta \big) \right],
\end{equation}
which is similar in structure to the single-measurement probability, with an additional round-dependent modulation since ancilla $j$ interrogates the signal that has accumulated for $j$ rounds. From Eq.~\ref{eq:single_round_marginal_main}, we can approximate the sequential CFI by treating the rounds as independent, $p_{\mathbf{b}}(\beta) \approx \prod_{j=1}^{N} p_{b_{j}}(\beta)$, so that the CFI becomes similar to Eq.~\ref{eq:cfi_single_main} but is additive over the marginals,
\begin{equation}
    \label{eq:cfi_indep}
    \mathcal{F}^{C}_{\mathrm{seq,\,marg}}
    = \sum_{j=1}^{N}
    \frac{16\, j^{2} A^{2}\, e^{-4A^{2}} \sin^{2} 4 j A \beta }
         {1 - e^{-4A^{2}} \cos^{2} 4 j A \beta }.
\end{equation}
Since this approximation omits inter-round correlations, it is neither an upper nor a lower bound on the sequential CFI and is only used to draw conclusions about its structure. In particular, Eq.~\ref{eq:cfi_indep} captures the two periodicities apparent in Fig.~\ref{fig:cfi_ratio}. All rounds of the protocol produce no information when $4 j A \beta = m \pi$ for every $j$, which occurs with the condition $\beta_{\mathrm{slow}} = m\pi/4A$ and is independent of $N$. In between these dips is a round-dependent oscillation with the condition $\beta_{\mathrm{fast}} = m \pi/4NA$, implying that some rounds are informative when others are not. This is in contrast to the single-measurement protocol, where all information is lost at $\beta = m \pi/4NA$.

\begin{figure*}[t!]
    \centering
    \includegraphics[
        width=\textwidth,
        height=0.82\textheight,
        keepaspectratio
    ]{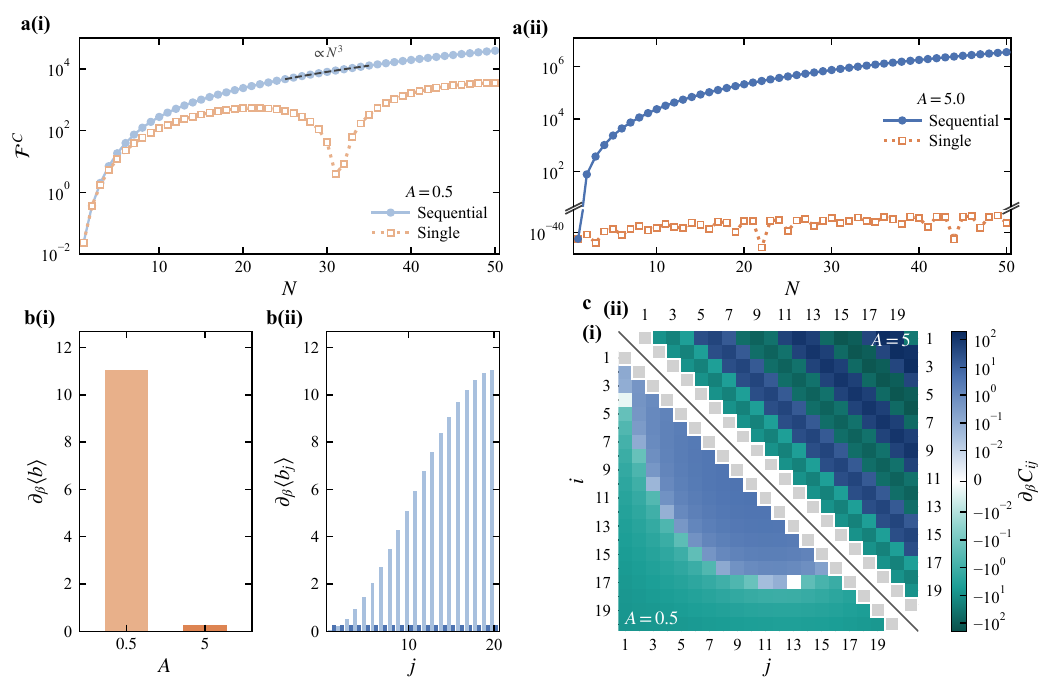}
    \caption{\textbf{a} CFI of the sequential- (blue) and single- (orange) measurement protocols as a function of $N$ at fixed signal $\beta = 0.05$, for (i) $A = 0.5$ and (ii) $A = 5.0$. Dashed gray line is an $N^{3}$ scaling reference for the sequential-measurement protocol. \textbf{b} Signal sensitivity of the outcome means, $\partial_{\beta}\langle b_{j} \rangle$ (Eq.~\eqref{eq:mean_outcome_main}), for (i) the single measurement and (ii) each round of the sequential protocol, with lighter (darker) bars corresponding to $A = 0.5$ ($A = 5.0$). At $A = 5.0$ the sensitivities are suppressed by $e^{-2A^{2}}$ for both protocols. \textbf{c} Signal sensitivity of pairwise correlations, $\partial_{\beta} C_{ij}$, for (i) $A = 0.5$ (ii) $A = 5.0$ with $N = 20$ and $\beta = 0.05$.}
    \label{fig:cfi_comparison}
\end{figure*}

To understand how signal information is encoded in the measurement outcomes, Fig.~\ref{fig:cfi_comparison}(a) compares the two protocols as a function of $N$ at a fixed signal for $A = 0.5$ and $A = 5$. For both couplings, the sequential- and single- measurement protocols grow with $N$ at $\mathcal{O}(N^{3})$ and $\mathcal{O}(N^{2})$, respectively, retaining the QFI scaling (Sec.~\ref{sec:qfi_single}). The $\mathcal{O}(A^{2})$ scaling of the sequential-measurement protocol is also maintained (App.~\ref{app:cfi_scaling}).  

The moments of the measurement record show where the signal information is held. In Fig.~\ref{fig:cfi_comparison}(b,ii) we plot the signal sensitivity of the per-round outcome means for the sequential protocol
\begin{equation}
    \label{eq:mean_outcome_main}
    \partial_{\beta} \langle b_{j} \rangle
    = 2 j A\, e^{-2A^{2}} \sin\!\big( 4 j A \beta \big).
\end{equation}
The single-measurement protocol sets $j = N$ (Fig.~\ref{fig:cfi_comparison}(b,i)). For small $A$ (light blue), the sensitivity grows with the round index, tracing the accumulating signal phase round by round, with the final round matching the single-measurement sensitivity. At large $A$ (dark blue), the prefactor $e^{-2A^{2}}$ suppresses the effect of $\beta$ on the mean, and each round becomes individually uninformative like the single-measurement case. 

However, the signal information still exists in the correlations of the measurement record. In Fig.~\ref{fig:cfi_comparison}(c) we plot the signal sensitivity of pairwise correlations between rounds $i$ and $j$,
\begin{equation}
    \label{eq:correlation_score}
    \begin{split}
    \partial_{\beta} C_{ij} = {}& -2A(i-j) \Big( 1 - e^{-4A^{2}} \Big) \sin\!\big[ 4A\beta\, (i-j) \big] \\ & -2A(i+j) \Big( e^{-8A^{2}} - e^{-4A^{2}} \Big) \sin\!\big[ 4A\beta\, (i+j) \big],
    \end{split}
\end{equation}
where $C_{ij} = \langle z_{i} z_{j} \rangle - \langle z_{i} \rangle \langle z_{j} \rangle$ and $z_{j} = (-1)^{b_{j}}$ (App.~\ref{app:cfi_correlations}). The first term is generated by the accumulating BCH phase and importantly grows monotonically with the coupling strength. In both terms the sensitivity grows linearly with the round indices, like the weights of the generator (Eq.~\eqref{eq:total_generator}). 

At low $A$ the correlations are weak (Fig.~\ref{fig:cfi_comparison}(c,i)) and nearly all information is held in the per-round means (Fig.~\ref{fig:cfi_comparison}(b,ii)). At large $A$, only the lag term survives, $\partial_{\beta}C_{ij} \to -2A(i-j)\sin[4A\beta(i-j)]$ as seen in Fig.~\ref{fig:cfi_comparison}(c,ii) and the information about the signal is carried entirely by the order of the outcomes. Together, Fig.~\ref{fig:cfi_comparison}(b) and Fig.~\ref{fig:cfi_comparison}(c) show how information about the signal is encoded in the moments of the outcome distribution, where the sensitivities of the average outcomes and of the correlations between outcomes mirror the QFI decomposition into $\mathcal{F}_{\mathrm{coh}}$ and $\mathcal{F}_{\mathrm{BCH}}$ from Sec.~\ref{sec:qfi_single}, respectively.

\subsection{Two-parameter estimation}

Finally, we extend the CFI analysis to simultaneous estimation of both quadratures, applying the same fixed computational-basis measurement to the two-parameter protocol of Sec.~\ref{sec:qfi_two}, so that the record $\mathbf{b} = (b_{1}, \ldots, b_{N})$ is distributed according to $p_{\mathbf{b}}(\beta_{1}, \beta_{2})$.

As in the single-parameter case, the BCH unitary can be absorbed into the measurement, and since $\widetilde{\rho}^{(2)}_{N}$ is supported on the doubly permutation-invariant subspace, every record probability can be computed in the $(N_{R}+1)(N_{I}+1)$-dimensional Dicke basis, in the form
\begin{equation}
    \label{eq:cfi_two_quadratic_main}
    p_{\mathbf{b}}(\beta)
    = 2^{-N}\, \widetilde{c}^{\,\dagger} R^{(2)}\, \widetilde{c},
\end{equation}
where $R^{(2)}$ is the signal-dependent reference state and $\widetilde{c}_{k_{R}, k_{I}}$ are the normalized projections of the rotated measurement vector $\hat{U}^{(2)\dagger}_{\mathrm{BCH}}(\beta)\ket{\mathbf{b}}$ onto the Dicke basis, computed by recursion at $\mathcal{O}(N^{3})$ per record (App.~\ref{app:cfi_two}).

The information about $(\beta_{1}, \beta_{2})$ is quantified by the CFI matrix, $\mathcal{F}^{C}_{\mu\nu} = \sum_{\mathbf{b}} p_{\mathbf{b}}^{-1} \big( \partial_{\mu} p_{\mathbf{b}} \big) \big( \partial_{\nu} p_{\mathbf{b}} \big)$. For any measurement $\mathcal{F}^{C} \preceq \mathcal{F}^{Q}$, so the classical Cram\'er--Rao bound implies $\sum_{\mu} \mathrm{Var}(\hat{\beta}_{\mu}) \geq \operatorname{Tr}\!\big[ (\mathcal{F}^{C})^{-1} \big] \geq \operatorname{Tr}\!\big[ (\mathcal{F}^{Q})^{-1} \big]$ for any unbiased estimator built from the record. As the two-parameter figure of merit we take the ratio of the scalar bounds,
\begin{equation}
    \label{eq:crb_ratio}
    r(\beta_{1}, \beta_{2}) = \frac{\operatorname{Tr}\!\big[ (\mathcal{F}^{Q})^{-1} \big]} {\operatorname{Tr}\!\big[ (\mathcal{F}^{C})^{-1} \big]}
    \leq 1,
\end{equation}
which reaches one when the record extracts all of the jointly available information, $\mathcal{F}^{C} = \mathcal{F}^{Q}$.

\begin{figure}[t]
    \centering
    \includegraphics[width=\columnwidth]{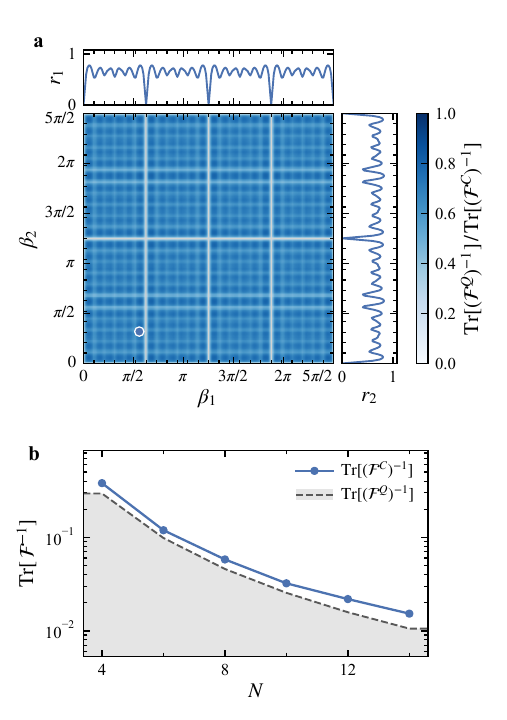}
    \caption{Joint estimation performance of the fixed computational-basis readout with the alternating coupling schedule at $A = 0.2$. \textbf{a}~Ratio of the scalar Cram\'er--Rao bounds, $ \operatorname{Tr}[(\mathcal{F}^{Q})^{-1}]/\operatorname{Tr}[(\mathcal{F}^{C})^{-1}]$ (Eq.~\eqref{eq:crb_ratio}), over one slow-fringe period $\pi/(2A)$ of the phase space at $N = 12$. Gray marks signals where the record CFI matrix is singular. Side panels show the per-quadrature ratios, $r(\beta_{1})$ and $r(\beta_{2})$. The dot marks the signal $(\beta_{1}, \beta_{2}) = (1.83, 0.96)$ used in panel~b. \textbf{b}~$\operatorname{Tr}[\mathcal{F}^{-1}]$ versus $N$ at the marked signal in \textbf{a} for the CFI (solid) and the QFI (dashed). The shaded region below the QFI curve is inaccessible to any measurement.}
    \label{fig:two_cfi}
\end{figure}

Fig.~\ref{fig:two_cfi}(a) shows $\operatorname{Tr}[(\mathcal{F}^{Q})^{-1}]/\operatorname{Tr}[(\mathcal{F}^{C})^{-1}]$ for the two-parameter protocol at $A = 0.2$ and $N = 12$ over one period $\pi/(2A)$ of the signal plane. The ratio is able to reach $r > 0.8$ for a range of signal values meaning the fixed computational-basis measurement recovers most of the available information. The plot also shows an interference pattern like the single-parameter case (Eq.~\eqref{eq:cfi_indep}). 

As in the single-parameter case, the structure of the single-round marginals is
\begin{equation}
    \label{eq:two_param_marginal}
    p_{b_{j}}( \beta_{1}, \beta_{2} )
    = \frac{1}{2}\Big[ 1 + (-1)^{b_{j}}\, \kappa_{j}\, e^{-2A^{2}}
      \cos\!\big( 4 j A\, \beta_{\mu_{j}} \big) \Big],
\end{equation}
where $\beta_{\mu_{j}}$ is the quadrature probed at round $j$
($\beta_{1}$ for $j \in \mathcal{I}$, $\beta_{2}$ for $j \in
\mathcal{R}$) and $\kappa_{j} = \big[ \cos( 4A^{2} )
\big]^{\bar{n}_{j}}$, with $\bar{n}_{j}$ the number of opposite-subset
rounds preceding round $j$. Since each round only depends on the
quadrature of its own subset, the first moments of the record split
into two interleaved copies of the single-parameter record
(Eq.~\ref{eq:single_round_marginal_main}). However, the fringe contrast
of each round is reduced by $\kappa_{j}$, where each earlier round of the
opposite subset displaces the oscillator orthogonally to the branch
separation of round $j$.  At $A = 0.2$ the loss is negligible,
while at $4A^{2} = \pi/2$ the means become non-informative and signal
information is instead carried in the correlations, as in
Fig.~\ref{fig:cfi_comparison}.

Eq.~\eqref{eq:cfi_indep} can be generalized to the two-parameter case, $\mathcal{F}^{C}_{\mathrm{marg}} = \mathrm{diag}\big( f_{1}(\beta_{1}),\, f_{2}(\beta_{2}) \big)$ with
\begin{equation}
    \label{eq:cfi_two_marg_f}
    f_{\mu} = \sum_{j \in \mathcal{B}_{\mu}} \frac{16\, j^{2} A^{2}\, \kappa_{j}^{2}\, e^{-4A^{2}} \sin^{2} 4 j A \beta_{\mu} } {1 - \kappa_{j}^{2}\, e^{-4A^{2}} \cos^{2} 4 j A \beta_{\mu} },
\end{equation}
where $\mathcal{B}_{\beta_{1}} = \mathcal{I}$, and $\mathcal{B}_{\beta_{2}} = \mathcal{R}$ as in Sec.~\ref{sec:qfi_two}. Since each round's mean is dependent on a single quadrature, the bound separates, $\operatorname{Tr}\!\big[(\mathcal{F}^{C}_{\mathrm{marg}})^{-1}\big] = f_{1}^{-1} + f_{2}^{-1}$. 

Like the single-parameter case, the locations of interference dips can be inferred from Eq.~\ref{eq:cfi_two_marg_f}. The fast modulation follows $\beta_{\mathrm{fast}} = \pi/(4NA)$. The slow modulation becomes a property of the coupling schedule. In particular, $f_{\mu} = 0$ requires $\sin 4 j A \beta_{\mu} = 0$ for all $j \in \mathcal{B}_{\mu}$, which occurs at integer multiples of $\beta_{\mathrm{slow}}^{(\mu)}$ with 
\begin{equation}
    \label{eq:dip_lattice}
    \beta_{\mathrm{slow}}^{(\mu)} = \frac{\pi}{4\, g_{\mu} A},
    \qquad
    g_{\mu} = \gcd\{\, j : j \in \mathcal{B}_{\mu} \,\}.
\end{equation}
The single-parameter protocol is the special case in which one subset holds every round, $\mathcal{B} = \{1, \ldots, N\}$ with $g = 1$, recovering $\beta_{\mathrm{slow}} = \pi/(4A)$. For the alternating schedule, $\mathcal{R} = \{1, 3, \ldots\}$ holds the odd rounds and $\mathcal{I} = \{2, 4, \ldots\}$ the even rounds, so $g_{\mathcal{R}} = 1$ while $g_{\mathcal{I}} = 2$, making the $\beta_{1}$ dips twice as dense as the $\beta_{2}$ dips (App.~\ref{app:cfi_two_marg}), which is the asymmetry observed in Fig.~\ref{fig:two_cfi}(a).

Fig.~\ref{fig:two_cfi}(b) plots the scalar CRB as a function of $N$ at the marked signal. The CFI bound tracks the QFI bound and decreases with the same $\mathcal{O}(N^{3})$ scaling, so the fixed computational-basis measurement maintains the two-parameter resource scaling of Sec.~\ref{sec:qfi_two} without adaptive control. The $\mathcal{O}(A^2)$ is also maintained (see App.~\ref{app:cfi_two_scaling}).

\section{Decoherence}
\label{sec:decoherence}

Finally, we examine how decoherence of the QHO affects the two protocols. We assume that the controlled displacements and qubit readout are done quickly with respect to the signal integration time and with high fidelity. Though experimentally tricky, these mid-circuit operations have been used to create non-classical QHO states~\cite{Valahu25, Fluh19, Camp20} with high-fidelity. We also neglect qubit coherence as the requirements are less stringent due to the repeated measurements. The QHO evolves under the Lindblad master equation. We separately consider coupling at rate $\Gamma$ to a bath of mean occupation $\bar n_{\mathrm{th}}$ and dephasing at rate $\Gamma_{\phi}$, with the signal displacement generated by a resonant drive of duration $\tau$ per round. More details are given in App.~\ref{app:decoherence}. Since decoherence breaks the permutation invariance outlined in Sec.~\ref{sec:sequential}, we compute the CFI by numerically integrating the QHO evolution.

\begin{figure}[t]
    \centering
    \includegraphics[width=\columnwidth]{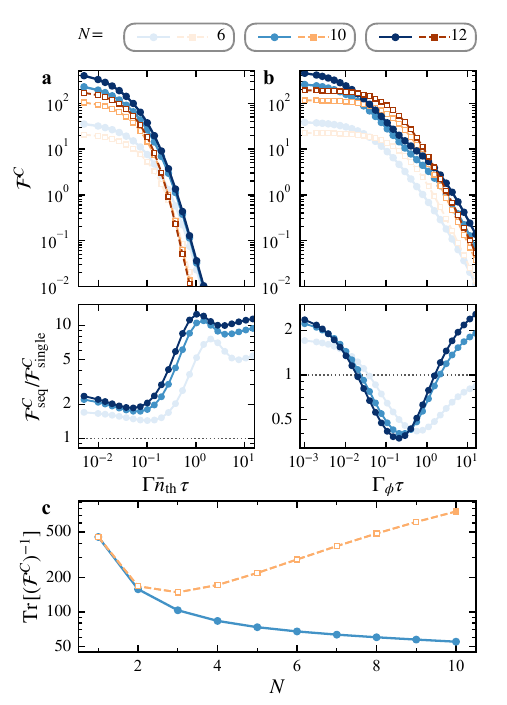}
    \caption{CFI of the sequential- (solid, blue) and single- (dashed, orange) measurement protocols under \textbf{a} heating and loss at $\bar n_{\text{th}} = 5$ and \textbf{b} dephasing at $\Gamma = 0$, for $N = \{6, 10, 12\}$ (light to dark). The leftmost point of each sweep is the noiseless value. Bottom panels show the ratio $\mathcal{F}^{C}_{\mathrm{seq}}/\mathcal{F}^{C}_{\mathrm{single}}$, with the dotted line marking equal performance. \textbf{c} Cram\'er--Rao bound $\operatorname{Tr}[(\mathcal{F}^{C})^{-1}]$ versus $N$ at $\Gamma\tau = 0.15$, $\bar n_{\text{th}} = 5$, and $\Gamma_{\phi}\tau = 0.5$. The filled, orange marker indicates the only measurement. All panels use $A = 0.5$ and $\beta = 0.05$.}
    \label{fig:noise}
\end{figure}

Fig.~\ref{fig:noise}(a) shows the CFI of the two protocols when the QHO is coupled to a thermal bath. Increasing $\Gamma \bar n_{\mathrm{th}}\tau$ decoheres the superposition of coherent states, suppressing the fringe contrast below its noiseless value $e^{-2A^{2}}$ in Eq.~\eqref{eq:single_round_marginal_main}. Additionally, loss damps the signal accumulated across rounds, causing the effective amplitude to saturate at $2\beta/(\Gamma\tau)$ after roughly $2/(\Gamma\tau)$ rounds (App.~\ref{app:amp_decay}), so later rounds contribute less to the sensitivity than in the noiseless case.

With increasing noise the ratio of the two protocols grows and saturates at $N$ (Fig.~\ref{fig:noise}(a), bottom). For $\Gamma \bar n_{\mathrm{th}}\tau \gg 1$ the QHO relaxes to the same displaced thermal state within each round, suppressing the cross-round correlations that carried signal information in the noiseless case (Fig.~\ref{fig:cfi_comparison}(c)). In this limit the record consists of $N$ nearly independent and identically distributed outcomes, with the final round of the sequential protocol coinciding with the single measurement, as already seen in the noiseless marginals (Fig.~\ref{fig:cfi_comparison}(b)). The CFI then becomes approximately additive over rounds, $\mathcal{F}^{C}_{\mathrm{seq}} \approx N \mathcal{F}^{C}_{\mathrm{single}}$, consistent with the observed saturation, and the remaining advantage of the sequential protocol is the number of measurements performed during the interrogation time.

Fig.~\ref{fig:noise}(b) shows the CFI under dephasing. Dephasing applies a random phase-space rotation whose effect grows with the oscillator occupation at the time it acts, and App.~\ref{app:dephased_contrast} estimates the resulting contrast suppression of round $j$ as $\kappa_{j}^{\phi} = \exp[-8A^{2}\Gamma_{\phi}\tau \sum_{k<j}\langle\hat{n}\rangle_{k}]$. In this estimate the two protocols differ only through the occupation entering the exponent. The single-measurement QHO carries only the signal, $\langle\hat{n}\rangle_{k} \approx (k\beta)^{2}$, while each conditional displacement of the sequential protocol adds $A^{2}$, giving $\kappa_{j}^{\phi} \approx \exp[-4A^{4}\Gamma_{\phi}\tau\,j(j-1)]$ for $A^{2} \gg N\beta^{2}$. 

Their ratio is non-monotonic (Fig.~\ref{fig:noise}(b), bottom). We attribute the initial dip to weak dephasing acting first on the sequential protocol, preferentially erasing the latest rounds of the record, which are also the most informative ($\propto j^{2}$, Eq.~\eqref{eq:cfi_indep}), while the single measurement is barely affected at these parameters. This picture is consistent with the observed deepening of the dip with $N$. At stronger dephasing the ratio recovers. Rounds within the coherence window $j \approx (4A^{4}\Gamma_{\phi}\tau)^{-1/2}$ retain nearly full contrast, so the sequential CFI should saturate at that of a $j$-round protocol (App.~\ref{app:coherence_window}), while the single measurement, with no protected early rounds, is expected to continue falling as $\exp\!\big[-\tfrac{8}{3}A^{2}\Gamma_{\phi}\tau\,\beta^{2}N^{3}\big]$, consistent with the ratio climbing back above one.

Figure~\ref{fig:noise}(c) shows the CRB of the two protocols as a function of $N$ with both noise channels present. The two coincide at $N = 1$ and both improve as the signal begins to accumulate, but beyond $N = 3$ the single-measurement bound worsens. The sequential bound instead continues to decrease monotonically, flattening at large $N$, and by $N = 10$, the two differ by roughly an order of magnitude. We note the location of the turnaround depends on the chosen parameters. App.~\ref{app:noise_maps} maps both protocols over the $(N, \Gamma \bar n_{\mathrm{th}}\tau)$ and $(N, \Gamma_{\phi}\tau)$ planes and identifies the regions of optimal operation.

\section{Conclusion}
In this work we have presented a protocol for ancilla-assisted displacement sensing of a QHO that uses sequential measurements as a metrological resource. The protocol borrows the ancilla-assisted structure of quantum phase estimation algorithms and benefits from coherently accumulating signal phase, but replaces the single terminal measurement with interleaved measurement and reset of a single ancilla. These repeated measurements are themselves a source of information, where the non-commuting signal and conditional displacements imprint a signal-dependent geometric phase at every round, yielding a QFI that scales as $\mathcal{O}(A^{2}N^{3})$, exceeding the $\mathcal{O}(N^{2})$ limit of a terminal measurement.

We have also presented a general framework for analyzing where information resides in the measurement records of sequential strategies. By mapping the repeatedly measured ancilla onto a register of $N$-ancillae with deferred measurements, the exponentially large record space is compressed onto a permutation-invariant reference state of polynomial dimension, with signal-dependence mapped to a set of single-qubit rotations. This reduction allows us to separate the QFI into a coherence contribution and a closed-form BCH contribution, isolating the influence of measurement outcome order on estimation performance. 

We also identified a similar information structure in the CFI using the per-round outcome means and inter-round correlations. The framework also presents the coupling schedule $\{\alpha_{j}\}$ and the per-round measurement basis as explicit design parameters that could be optimized. Critically, the presented framework applies to any sequential ancilla-assisted metrology protocol in which a probe is periodically interrogated through a resettable ancilla and can therefore guide the design and analysis of other mid-sensing measurement based protocols.

The presented protocol also has practical advantages. It naturally generalizes to simultaneous estimation of both quadratures of a displacement within a single interrogation time, without complex state preparation or adaptive control, and the resulting QFI bound is asymptotically jointly saturable at strong coupling. Moreover, a simple coupling schedule that samples each quadrature symmetrically across the interrogation time matches the total information of the single-parameter protocol. In both the single- and two-parameter settings, the sequential measurements mitigate phase-wrapping interference dips, extending the sensor's dynamic range, and the periodic information extraction makes the protocol more resilient to QHO decoherence than a terminal measurement.    

In the presented protocol, the qubit measurements are an irreversible action that, over time, allow the QHO to traverse a unique, signal-dependent trajectory, effectively serving as a non-linear resource for sensing~\cite{Senanian24} and allowing the sensor to sample a wider range of operating points during its interrogation time. Future work could leverage this non-linearity for time-dependent or stochastic signals, incorporate variational optimization of the coupling schedule and measurement bases~\cite{Prabhu26}, or add adaptive classical and quantum control, as is conventionally done in single-measurement metrology~\cite{Valahu25}. This work joins a growing interest in leveraging analog quantum system evolution, where a system's inherent dynamics perform processing, sensing, or simulation tasks~\cite{Dong25, Shaw25, Kashyap25, Ryan26}. Beyond displacement sensing, such signal-dependent QHO trajectories could be adapted for non-classical state generation~\cite{Simoni25, Matsos24}, and the principle of monitoring a quantum system through repeated ancilla-assisted measurements generalizes to other architectures, e.g., qubit--qubit~\cite{Monte22} and QHO--QHO couplings. Combined with non-classical probe states~\cite{Labarca26, Bond26} and adaptive control~\cite{Allen25, Neeve25}, mid-sensing measurement-based protocols offer new opportunities for quantum-enhanced parameter estimation.

\section{Acknowledgments}
This work was funded through the Boeing University Partnership Program under the direction of the Disruptive Computing, Networking \& Sensing Group.  We thank Dr. Ben Koltenbah (Boeing Engineering \& Technology Innovation), Christian Pluchar, Ting Rei Tan, Teerawat Chalermpusitarak, and Liam Bond for helpful conversation and comments on the manuscript. K.M. is supported through the Advancing Quantum-Enabled Technologies (AQET) traineeship program at the University of Washington through NSF award DGE-2021540.

%\clearpage

\appendix
\onecolumngrid

%\vspace{5mm}

\startcontents[appx]
%{\noindent\large\bfseries End Matter\par\vskip4pt}
%\printcontents[appx]{}{1}{}
%\vskip8pt

\section{Single-Measurement Protocol}
\label{app:single_meas}
To build intuition for the derivation of the sequential-measurement protocol, here we derive the single-measurement protocol following similar steps as App.~\ref{app:seq_meas}.  Consider $N$ applications of a signal displacement $\hat{D}(\beta) = e^{\beta \hat{a}^{\dagger} - \beta^{*} \hat{a}}$ followed by a single controlled displacement $\hat{D}(\hat{\sigma}_{x}\alpha)$. With the qubit and oscillator prepared in their respective ground states,
\begin{align}
\ket{\psi} &= \hat{D}(\hat{\sigma}_{x}\alpha) \hat D (N \beta) \ket{\downarrow} \otimes \ket{0} .
\end{align}
The controlled displacement yields a qubit-dependent phase,
\begin{equation}
\hat{D}(\hat{\sigma}_{x}\alpha)\,\hat{D}(N\beta) = e^{\,iN\,\operatorname{Im}(\alpha\beta^{*})\,\hat{\sigma}_{x}}\, \hat{D}(N\beta + \hat{\sigma}_{x}\alpha) \equiv \hat{U}_{\mathrm{BCH}}(\beta)\,\hat{D}(N\beta + \hat{\sigma}_{x}\alpha),
\end{equation}
so that
\begin{align}
\ket{\psi} &= \hat{U}_{\mathrm{BCH}}(\beta)\, \hat{D}(N\beta + \hat{\sigma}_{x}\alpha)\, \ket{\downarrow}\otimes\ket{0} \\ &= \hat{U}_{\mathrm{BCH}}(\beta)\left( \ket{\downarrow} \otimes \ket{N\beta + \hat{\sigma}_{x}\alpha} \right),
\end{align}
where $\hat{U}_{\mathrm{BCH}}(\beta)$ is a single-qubit rotation, and $\ket{N\beta + \hat{\sigma}_{x}\alpha}$ denotes the qubit-conditioned coherent state. Since $\hat{U}_{\mathrm{BCH}}(\beta)$ acts only on the qubit, it commutes with the oscillator trace, and the state used to compute the Fisher information factors as $\rho(\beta) = \hat{U}_{\mathrm{BCH}}(\beta) \,\widetilde{\rho}(\beta)\,\hat{U}_{\mathrm{BCH}}^{\dagger}(\beta)$ with
\begin{equation}
\widetilde{\rho}(\beta) = \mathrm{Tr}_{\mathrm{osc}}\!\left( \hat{D}(N\beta + \hat{\sigma}_{x}\alpha) \left( \ket{\downarrow}\bra{\downarrow} \otimes \ket{0}\bra{0} \right) \hat{D}^{\dagger}(N\beta + \hat{\sigma}_{x}\alpha) \right).
\end{equation}
We evaluate $\widetilde{\rho}(\beta)$ in the $\hat{\sigma}_{x}$ eigenbasis $\{\ket{+},\ket{-}\}$, $\hat{\sigma}_{x}\ket{\pm} = \pm\ket{\pm}$. Using $\ket{\downarrow} = \tfrac{1}{\sqrt{2}}\left(\ket{+}+\ket{-}\right)$, the controlled displacement maps each branch onto a distinct coherent amplitude,
\begin{equation}
\hat{D}(N\beta + \hat{\sigma}_{x}\alpha)\,\ket{\downarrow}\otimes\ket{0} = \frac{1}{\sqrt{2}}\left( \ket{+}\otimes\ket{N\beta + \alpha} + \ket{-}\otimes\ket{N\beta - \alpha} \right).
\end{equation}
Tracing out the oscillator and writing the matrix in the $\{\ket{+},\ket{-}\}$ basis gives
\begin{equation}
\widetilde{\rho}(\beta) = \frac{1}{2} \begin{pmatrix} 1 & \braket{N\beta - \alpha | N\beta + \alpha} \\[4pt] \braket{N\beta + \alpha | N\beta - \alpha} & 1 \end{pmatrix}.
\end{equation}
Setting $\alpha = iA$ with $\{A,\beta\}\in\mathbb{R}$, the coherent-state overlap is
\begin{equation}
\braket{N\beta - \alpha | N\beta + \alpha} = e^{-2A^{2}}\,e^{\,2iNA\beta},
\end{equation}
so that
\begin{equation}
\widetilde{\rho}(\beta) = \frac{1}{2}
\begin{pmatrix}
1 & e^{-2A^{2}}\,e^{\,2iNA\beta} \\[4pt]
e^{-2A^{2}}\,e^{-2iNA\beta} & 1
\end{pmatrix}.
\end{equation}
Finally, $\hat{U}_{\mathrm{BCH}}(\beta) = e^{\,iNA\beta\,\hat{\sigma}_{x}}$ is diagonal in this basis, so the rotation leaves the populations unchanged and doubles the phase in the coherence,
\begin{equation}
\label{eq:rho_single}
\rho(\beta) = \hat{U}_{\mathrm{BCH}}(\beta)\,\widetilde{\rho}(\beta)\,\hat{U}_{\mathrm{BCH}}^{\dagger}(\beta) = \frac{1}{2}
\begin{pmatrix}
    1 & e^{-2A^{2}}\,e^{\,4iNA\beta} \\[4pt]
    e^{-2A^{2}}\,e^{-4iNA\beta} & 1
\end{pmatrix},
\end{equation}
which is the state used to evaluate the quantum and classical Fisher information.

\section{Sequential-Measurement Protocol}
\label{app:seq_meas}
\subsection{One qubit to $N$ qubits}
\label{app:one_to_N}

To ease evaluation of the metrological performance of the sequential-measurement protocol, we map $N$ repeated measurements performed on one ancilla to a single measurement performed on each of $N$. Here we show the equivalence and implications of this construction. Each round of the sequential protocol applies a combined unitary, $\hat{U}(\beta, \alpha) \equiv \hat{D}(\hat{\sigma}_{x}\alpha)\,\hat{D}(\beta)$, followed by a measurement of the ancilla qubit and subsequent reset into $\ket{\downarrow}$. Measuring in the $\{ \ket{b} \}$ basis applies the Kraus operator $\hat{K}_{b}^{(j)} = \bra{b} \hat{U}(\beta, \alpha) \ket{\downarrow}$ to the QHO. Since the reset is independent of the outcome $b$, $\hat{K}_{b}^{(j)}$ is independent of the earlier measurement outcomes. Then after $N$ rounds of operation we have a bit-string of outcomes $\mathbf{b} = (b_{1}, \ldots, b_{N})$ which occurs with probability
\begin{equation}
    p( \mathbf{b} ) = \big\| \hat{K}_{b_{N}}^{(N)} \cdots \hat{K}_{b_{1}}^{(1)} \ket{0} \big\|^{2}.
\end{equation}
An equivalent way to obtain the same probability is to assign each round its own ancilla and defer all measurements. The joint ancilla--oscillator system after $N$ interactions is $\ket{\Psi} = \left( \prod_{j=1}^{N} \hat{U}_{j}(\beta,\alpha) \right) \left( \bigotimes_{j=1}^{N} \ket{\downarrow}_{j} \right) \otimes \ket{0}$, where $\hat{U}_{j}(\beta,\alpha) \equiv \hat{D}_{j}(\hat{\sigma}_{x}^{(j)}\alpha)\,\hat{D}(\beta)$ acts only on the $j^{th}$ ancilla and the oscillator. Now measuring the register in $\{ \ket{b} \}$ corresponds to
\begin{equation}
    \left( \bigotimes_{j=1}^{N} \bra{b_{j}}_{j} \right) \ket{\Psi} = \hat{K}_{b_{N}}^{(N)} \cdots \hat{K}_{b_{1}}^{(1)} \ket{0},
\end{equation}
which gives the same $p(\mathbf{b})$ as in the single-ancilla case with reset. Importantly, this equivalence only holds for product measurements across the ancillae.

\subsection{Protocol operation}
Consider $N$ qubits initially in $\ket{\downarrow}$ and a single QHO in the
ground state $\ket{0}$. Each round applies a signal displacement
$\hat{D}(\beta) = e^{\beta \hat{a}^{\dagger} - \beta^{*}\hat{a}}$ to the
oscillator, followed by a controlled displacement
$\hat{D}_{j}(\hat{\sigma}_{x}^{(j)}\alpha)
= e^{\alpha\hat{\sigma}_{x}^{(j)}\hat{a}^{\dagger}
   - \alpha^{*}\hat{\sigma}_{x}^{(j)}\hat{a}}$
conditioned on qubit $j$, leading to
\begin{align}
\ket{\psi}
= \left( \prod_{j=1}^{N}
  \hat{D}_{j}(\hat{\sigma}_{x}^{(j)}\alpha)\,\hat{D}(\beta) \right)
  \ket{\downarrow}^{\otimes N} \otimes \ket{0}.
\end{align}
Take $\alpha = iA$ with $\{A,\beta\}\in\mathbb{R}$. Within each round the two displacements combine through the commutation relation $\hat{D}(\mu)\hat{D}(\nu) = e^{\,i\,\operatorname{Im}(\mu\nu^{*})}\hat{D}(\mu+\nu)$ which gives
\begin{equation}
\hat{D}_{j}(\hat{\sigma}_{x}^{(j)}\alpha)\,\hat{D}(\beta)
= e^{\,iA\beta\hat{\sigma}_{x}^{(j)}}\,
  \hat{D}(\beta + iA\hat{\sigma}_{x}^{(j)}),
\end{equation}
so each per-round controlled kick is absorbed into a displacement of amplitude $\beta + iA\hat{\sigma}_{x}^{(j)}$ together with a single-qubit phase $e^{\,iA\beta  \hat{\sigma}_{x}^{(j)}}$. We then collapse the $N$ displacements using
  \begin{equation}
  \hat{D}(\xi_{N})\cdots\hat{D}(\xi_{2})\hat{D}(\xi_{1})
  = \exp\!\left( -i\!\!\sum_{1\le j<k\le N}\!\!
    \operatorname{Im}(\xi_{j}\xi_{k}^{*}) \right)
    \hat{D}\!\left( \sum_{j=1}^{N}\xi_{j} \right).
  \end{equation}
With $\xi_{j} = \beta + iA\hat{\sigma}_{x}^{(j)}$, the total amplitude is $\sum_{j}\xi_{j} = N\beta + iA\hat{S}_{x}$, where $\hat{S}_{x} = \sum_{j=1}^{N}\hat{\sigma}_{x}^{(j)}$, and the pairwise phases evaluate to $\exp\!\big[iA \beta \sum_{j}(2j-N-1)\hat{\sigma}_{x}^{(j)}\big]$. Combined with the single-round phases $e^{\,iA\beta\hat{\sigma}_{x}^{(j)}}$ this gives
\begin{equation}
\label{eq:bch_phase_unitary}
\prod_{j=1}^{N}\hat{D}_{j}(\hat{\sigma}_{x}^{(j)}\alpha)\,\hat{D}(\beta)
= \hat{U}_{\mathrm{BCH}}(\beta)\,\hat{D}(N\beta + \hat{S}_{x}\alpha),
\qquad
\hat{U}_{\mathrm{BCH}}(\beta) = \exp\!\left( iA\beta
  \sum_{j=1}^{N}(2j-N)\hat{\sigma}_{x}^{(j)} \right).
\end{equation}
Since $\hat{U}_{\mathrm{BCH}}(\beta)$ is a product of single-qubit rotations acting only on the
qubits, the reduced qubit state factorizes as
$\rho_{N}(\beta) = \hat{U}_{\mathrm{BCH}}(\beta)\,\widetilde{\rho}_{N}(\beta)\,\hat{U}_{\mathrm{BCH}}^{\dagger}(\beta)$ with
\begin{equation}
\widetilde{\rho}_{N}(\beta) = \mathrm{Tr}_{\mathrm{osc}}\!\left[
  \hat{D}(N\beta + \hat{S}_{x}\alpha)
  \big( (\ket{\downarrow}\bra{\downarrow})^{\otimes N} \otimes \ket{0}\bra{0} \big)
  \hat{D}^{\dagger}(N\beta + \hat{S}_{x}\alpha)
  \right].
\end{equation}
In the $\hat{\sigma}_{x}$ eigenbasis $\{\ket{\mathbf a}\}$, $\mathbf a\in\{ +,- \}^{N}$, the collective operator $\hat{S}_{x}$ is diagonal with eigenvalue $s(\mathbf a)=\sum_{j}a_{j}$, so the controlled displacement maps each branch onto its own coherent amplitude. Since $\ket{\downarrow}=(\ket{+}+\ket{-})/\sqrt{2}$, the state before the BCH rotation is
\begin{equation}
\ket{\widetilde{\psi}_{N}(\beta)} = \hat{D}(N\beta + \hat{S}_{x}\alpha)\, \ket{\downarrow}^{\otimes N} \otimes \ket{0} = \frac{1}{2^{N/2}}
  \sum_{\mathbf a\in\{ +,- \}^{N}} \ket{\mathbf a} \otimes \ket{N\beta + s(\mathbf a)\alpha},
\end{equation}
a qubit-conditioned superposition of coherent states, matching Eq.~\eqref{eq:seq_branch_state} of the main text. Writing $k=w(\mathbf a)$ for the number of $\ket{-}$ branches in the record, so that $s(\mathbf a)=N-2k$, and tracing out the oscillator yields the reduced state in the permutationally invariant
(Dicke) basis,
\begin{equation}
\widetilde{\rho}_{N}(\beta)
= \sum_{k,k'=0}^{N}
  \sqrt{\gamma_{k}\gamma_{k'}}\;
  e^{-2A^{2}(k-k')^{2}}\,
  e^{-2iNA\beta(k-k')}\,
  \ket{\phi_{k}}\bra{\phi_{k'}},
\end{equation}
with $\gamma_{k}=\binom{N}{k}/2^{N}$ and the Dicke states
$\ket{\phi_{k}}=\binom{N}{k}^{-1/2}\sum_{\mathbf a:\,w(\mathbf a)=k}\ket{\mathbf a}$ (Eq.~\eqref{eq:seq_dicke_basis}). The Gaussian envelope $e^{-2A^{2}(k-k')^{2}}$ is the overlap of coherent states separated by $2A(k-k')$, and the phase $e^{-2iNA\beta(k-k')}$ encodes the signal $\beta$.

\section{Quantum Fisher Information}
\label{app:qfi}
For single-parameter estimation, the QFI is
\begin{equation}
    \label{eq:qfi_def}
    \mathcal{F}^{Q} = 2 \sum_{\substack{a,a' \\ \lambda_{a} + \lambda_{a'} > 0 }} \frac{ |\bra{e_{a}} \partial_{\beta} \rho(\beta) \ket{e_{a'}} |^{2} }{ \lambda_{a} + \lambda_{a'} }.
\end{equation}
This formula can be applied to both pure states and mixed states, provided the eigendecomposition of the sensing state, $\rho(\beta) = \sum_{a} \lambda_{a} \ket{e_{a}} \bra{e_{a}}$, can be computed. In this work, we are mainly concerned with the QFI of the ancilla system.

\subsection{Single measurement}
\label{app:single_meas_qfi_app}

For the single-measurement protocol the reduced qubit state, Eq.~\eqref{eq:rho_single},
accounts for the BCH rotation $\hat{U}_{\mathrm{BCH}}(\beta)$, so the quantum Fisher information
follows directly from its eigendecomposition. With $d \equiv e^{-2A^2}$ for the
coherent-state overlap magnitude and $\phi \equiv 4NA\beta$ for the accumulated phase,
Eq.~\eqref{eq:rho_single} becomes
\begin{equation}
  \rho(\beta) = \frac{1}{2}
  \begin{pmatrix}
    1 & d\,e^{i\phi} \\
    d\,e^{-i\phi} & 1
  \end{pmatrix}
\end{equation}
in the $\{\ket{+},\ket{-}\}$ basis, with $\beta$-independent eigenvalues $\lambda_\pm = \frac{1}{2}\,(1 \pm d)$ and eigenvectors $\ket{e_\pm} = \frac{1}{\sqrt{2}} \left( \ket{+} \pm e^{-i\phi}\ket{-} \right)$. The diagonal matrix elements $\bra{e_a}\partial_\beta\rho\ket{e_a}$ vanish due to the $\beta$-independence of the eigenvalues. Then differentiating gives
\begin{equation}
  \partial_\beta \rho(\beta) = \frac{i d\,\phi'}{2}
  \left( e^{i\phi}\ket{+}\!\bra{-} - e^{-i\phi}\ket{-}\!\bra{+} \right),
  \qquad \phi' = 4NA,
\end{equation}
so that the off-diagonal matrix element evaluates to
\begin{equation}
  \bra{e_+}\partial_\beta\rho(\beta)\ket{e_-} = -\frac{i d\,\phi'}{2} = -2iNA\,d .
\end{equation}
Substituting into Eq.~\eqref{eq:qfi_def} gives
\begin{equation}
  \mathcal{F}^{Q}_{\mathrm{single}}
  = 2\!\!\sum_{\substack{a,a' \\ \lambda_a+\lambda_{a'}>0}}\!\!
    \frac{\big|\bra{e_a}\partial_\beta\rho(\beta)\ket{e_{a'}}\big|^2}{\lambda_a+\lambda_{a'}}
  = 2\cdot\frac{2\,(2NA\,d)^2}{1}
  = 16\,N^2 A^2\, e^{-4A^2}.
\end{equation}
The maximum QFI is achieved when $A = 1/2$ and $\mathcal{F}^{Q}_{\mathrm{single}} = 4N^2/e$. 

\subsection{Splitting of the Hilbert space}
\label{app:splitting}
The QFI of the qubit register is organized by the symmetry structure identified in Sec.~\ref{sec:sequential}. The register state is generated from the reference state by $\hat{H} = 2A\sum_{j} j\,\hat{\sigma}_{x}^{(j)}$ (Eq.~\eqref{eq:total_generator}), which separates into a collective part and the BCH part,
\begin{equation}
    \hat{H} = \hat{G} + \hat{H}_{\phi},
    \qquad
    \hat{G} = NA\,\hat{S}_{x},
    \qquad
    \hat{H}_{\phi} = A\sum_{j=1}^{N} (2j-N)\,\hat{\sigma}_{x}^{(j)},
\end{equation}
corresponding to the accumulated signal weight $AN$ and the order-dependent weight $A(2j-N)$ of each ancilla. While the reference state $\widetilde{\rho}_{N}(0)$ is supported on the $(N{+}1)$-dimensional permutation-invariant Dicke subspace $\mathcal{S} = \mathrm{range}\,\hat{\Pi}$, where $\hat{\Pi} \equiv \sum_{k=0}^{N} \ket{\phi_{k}}\bra{\phi_{k}}$ projects onto the Dicke states of Eq.~\eqref{eq:seq_dicke_basis}, the generator is not permutation invariant. Splitting the ancilla Hilbert space as $\mathcal{H} = \mathcal{S} \oplus \mathcal{S}^{\perp}$, the generator decomposes into blocks,
\begin{equation}
    \hat{H} =
    \begin{pmatrix}
        \hat{\Pi}\hat{H}\hat{\Pi}
        & \hat{\Pi}\hat{H}(\mathbbm{1}-\hat{\Pi}) \\
        (\mathbbm{1}-\hat{\Pi})\hat{H}\hat{\Pi}
        & (\mathbbm{1}-\hat{\Pi})\hat{H}(\mathbbm{1}-\hat{\Pi})
    \end{pmatrix},
\end{equation}
and each block plays a distinct role in computing the QFI. The lower-right block acts entirely on the kernel of $\widetilde{\rho}_{N}(0)$ and carries zero weight, since $\lambda_{a} = \lambda_{a'} = 0$ for such pairs. The upper-left block generates rotations within $\mathcal{S}$. Using permutation invariance, $\hat{\Pi}\hat{\sigma}_{x}^{(j)}\hat{\Pi} = \tfrac{1}{N}\hat{\Pi}\hat{S}_{x}\hat{\Pi}$, together with $\sum_{j}(2j-N) = N$, it reduces to a collective rotation,
\begin{equation}
    \hat{\Pi}\hat{H}\hat{\Pi} = (N+1)A\,\hat{\Pi}\hat{S}_{x}\hat{\Pi},
\end{equation}
which generates the coherence contribution $\mathcal{F}_{\mathrm{coh}}$. The off-diagonal blocks rotate the state out of $\mathcal{S}$. Since $\hat{G} \propto \hat{S}_{x}$ preserves $\mathcal{S}$, only the BCH generator enters these blocks, $\hat{\Pi}\hat{H}(\mathbbm{1}-\hat{\Pi}) = \hat{\Pi}\hat{H}_{\phi}(\mathbbm{1}-\hat{\Pi})$, and their contribution gives $\mathcal{F}_{\mathrm{BCH}}$ of Eq.~\eqref{eq:qfi_backaction}.

\subsection{Sequential measurements}
\label{app:seq_qfi}
For the sequential protocol, the register state is generated from the $\beta$-independent reference state by the total generator of Eq.~\eqref{eq:total_generator},
\begin{equation}
\rho_{N}(\beta) = e^{i\beta\hat{H}}\,\widetilde{\rho}_{N}(0)\,e^{-i\beta\hat{H}}, \qquad \hat{H} = \hat{G} + \hat{H}_{\phi},
\end{equation}
with the collective and BCH parts of App.~\ref{app:splitting}. We work in the eigenbasis of the reference state,
\begin{equation}
\widetilde{\rho}_{N}(0) = \sum_{a} \lambda_{a} \ket{\widetilde{e}_{a}}\bra{\widetilde{e}_{a}}, \qquad \sum_{a} \ket{\widetilde{e}_{a}}\bra{\widetilde{e}_{a}} = \mathbbm{1},
\end{equation}
forming a basis for the $2^{N}$-dimensional ancilla Hilbert space. Since $\widetilde{\rho}_{N}(0)$ is supported on the $(N{+}1)$-dimensional Dicke subspace, at least $2^{N}-(N{+}1)$ of the $\lambda_{a}$ are zero. The eigenvectors of $\rho_{N}(\beta)$ are the rotated basis $\ket{e_{a}(\beta)} = e^{i\beta\hat{H}}\ket{\widetilde{e}_{a}}$. Differentiating, $\partial_{\beta}\rho_{N} = i[\hat{H},\rho_{N}]$, the numerator for the QFI becomes
\begin{equation}
\bra{e_{a}(\beta)}\,\partial_{\beta}\rho_{N}(\beta)\,\ket{e_{a'}(\beta)} = i(\lambda_{a'}-\lambda_{a})\, \bra{\widetilde{e}_{a}}\hat{H}\ket{\widetilde{e}_{a'}},
\end{equation} 
plugging this in to Eq.~\ref{eq:qfi_def},
\begin{equation}
\label{eq:seq_qfi_generator}
\mathcal{F}^{Q}_{\mathrm{seq}} = 2 \sum_{\substack{a,a' \\ \lambda_{a}+\lambda_{a'}>0}} \frac{(\lambda_{a}-\lambda_{a'})^{2}}{\lambda_{a}+\lambda_{a'}}\, \big|\bra{\widetilde{e}_{a}}\hat{H}\ket{\widetilde{e}_{a'}}\big|^{2}.
\end{equation}
The block structure of App.~\ref{app:splitting} splits this sum into two contributions. Pairs with both eigenvectors in the support see only the diagonal block, $\hat{\Pi}\hat{H}\hat{\Pi} = (N{+}1)A\, \hat{\Pi}\hat{S}_{x}\hat{\Pi}$, giving the coherence contribution
\begin{equation}
\label{eq:qfi_coherence}
\mathcal{F}_{\mathrm{coh}} = 2(N{+}1)^{2}A^{2} \sum_{\substack{a,a' \leq N+1 \\ \lambda_{a}+\lambda_{a'}>0}} \frac{(\lambda_{a}-\lambda_{a'})^{2}}{\lambda_{a}+\lambda_{a'}}\, \big|\bra{\widetilde{e}_{a}}\hat{S}_{x}\ket{\widetilde{e}_{a'}}\big|^{2},
\end{equation}
which is evaluated by numerically diagonalizing the $(N{+}1)\times(N{+}1)$ Dicke block of $\widetilde{\rho}_{N}(0)$. Pairs that couple the support to the kernel have $\lambda_{a'}=0$, so the weight reduces to $(\lambda_{a}-0)^{2}/(\lambda_{a}+0) = \lambda_{a}$, and only $\hat{H}_{\phi}$ enters the off-diagonal blocks. Counting these pairings gives
\begin{align}
\mathcal{F}_{\mathrm{BCH}} &= 4 \sum_{a=1}^{N+1} \lambda_{a} \bra{\widetilde{e}_{a}} \hat{H}_{\phi} \big( \mathbbm{1} - \hat{\Pi} \big) \hat{H}_{\phi} \ket{\widetilde{e}_{a}} \\ &= 4\left[\operatorname{Tr}\!\big( \widetilde{\rho}_{N}(0)\, \hat{H}_{\phi}^{2} \big) - \operatorname{Tr}\!\big( \widetilde{\rho}_{N}(0)\, \hat{H}_{\phi}\, \hat{\Pi}\, \hat{H}_{\phi} \big)\right],
\end{align}
where $\hat{\Pi} = \sum_{a=1}^{N+1} \ket{\widetilde{e}_{a}}\bra{\widetilde{e}_{a}}$ projects onto the support, coinciding with the Dicke projector of App.~\ref{app:splitting} since $\widetilde{\rho}_{N}(0)$ is full rank on $\mathcal{S}$ for $A > 0$.

The second trace simplifies using permutation invariance. Since $\widetilde{\rho}_{N}(0)$ is permutation invariant, the projected single-qubit operator satisfies $\hat{\Pi}\hat{\sigma}_{x}^{(j)}\hat{\Pi} = \tfrac{1}{N}\hat{\Pi}\hat{S}_{x}\hat{\Pi}$ with $\hat{S}_{x} = \sum_{j=1}^{N}\hat{\sigma}_{x}^{(j)}$, which, together with $[\hat{S}_{x}, \hat{\Pi}] = 0$, gives
\begin{align}
\operatorname{Tr}\!\big( \widetilde{\rho}_{N}(0)\, \hat{H}_{\phi}\, \hat{\Pi}\, \hat{H}_{\phi} \big) &= A^{2} \sum_{j,k=1}^{N} (2j-N)(2k-N)\, \operatorname{Tr}\!\big( \widetilde{\rho}_{N}(0)\, \hat{\sigma}_{x}^{(j)} \hat{\Pi} \hat{\sigma}_{x}^{(k)} \big) \\ &= A^{2} \sum_{j,k=1}^{N} (2j-N)(2k-N)\, \frac{1}{N^{2}}\, \operatorname{Tr}\!\big( \widetilde{\rho}_{N}(0)\, \hat{S}_{x}^{2} \big) \\ &= A^{2}\, \operatorname{Tr}\!\big( \widetilde{\rho}_{N}(0)\, \hat{S}_{x}^{2} \big),
\end{align}
where we used $\sum_{j=1}^{N}(2j-N) = N$. For the first trace,
\begin{align}
\operatorname{Tr}\!\big( \widetilde{\rho}_{N}(0)\, \hat{H}_{\phi}^{2} \big) &= A^{2} \sum_{j,k=1}^{N} (2j-N)(2k-N)\, \operatorname{Tr}\!\big( \widetilde{\rho}_{N}(0)\, \hat{\sigma}_{x}^{(j)} \hat{\sigma}_{x}^{(k)} \big) \\ &= A^{2} \Bigg[ \sum_{j=1}^{N} (2j-N)^{2}\, \operatorname{Tr}\!\big( \widetilde{\rho}_{N}(0)\, (\hat{\sigma}_{x}^{(j)})^{2} \big) + \sum_{j \neq k} (2j-N)(2k-N)\, \operatorname{Tr}\!\big( \widetilde{\rho}_{N}(0)\, \hat{\sigma}_{x}^{(j)} \hat{\sigma}_{x}^{(k)} \big) \Bigg] \\ &= A^{2} \left[ \frac{N(N^{2}+2)}{3} + \left( N^{2} - \frac{N(N^{2}+2)}{3} \right) \operatorname{Tr}\!\big( \widetilde{\rho}_{N}(0)\, \hat{\sigma}_{x}^{(j)} \hat{\sigma}_{x}^{(k)} \big) \right],
\end{align}
with $j \neq k$ in the final trace. Combining the two traces and using $\hat{S}_{x}^{2} = \sum_{j=1}^{N} (\hat{\sigma}_{x}^{(j)})^{2} + \sum_{j \neq k} \hat{\sigma}_{x}^{(j)} \hat{\sigma}_{x}^{(k)}$ gives
\begin{equation}
\operatorname{Tr}\!\big( \widetilde{\rho}_{N}(0)\, \hat{H}_{\phi}^{2} \big) - \operatorname{Tr}\!\big( \widetilde{\rho}_{N}(0)\, \hat{H}_{\phi}\, \hat{\Pi}\, \hat{H}_{\phi} \big) = \frac{A^{2}(N+1)}{3} \left[ N^{2} - \operatorname{Tr}\!\big( \widetilde{\rho}_{N}(0)\, \hat{S}_{x}^{2} \big) \right].
\end{equation}
Since $\hat{S}_{x}$ is diagonal in the Dicke basis, $\hat{S}_{x}\ket{\phi_{k}} = (N-2k)\ket{\phi_{k}}$, and the diagonal of $\widetilde{\rho}_{N}(0)$ is binomial, $\operatorname{Tr}\!\big(\widetilde{\rho}_{N}(0)\,\hat{S}_{x}^{2}\big) = \sum_{k}\gamma_{k}(N-2k)^{2} = N$, giving the closed form
\begin{equation}
\label{eq:qfi_backaction_app}
\mathcal{F}_{\mathrm{BCH}} = \frac{4A^{2}(N+1)}{3}\big(N^{2}-N\big) = \frac{4}{3}\,A^{2}\,N\big(N^{2}-1\big),
\end{equation} 
which together with Eq.~\eqref{eq:qfi_coherence} yields $\mathcal{F}^{Q}_{\mathrm{seq}} = \mathcal{F}_{\mathrm{coh}} + \mathcal{F}_{\mathrm{BCH}}$ of Eq.~\eqref{eq:qfi_three_term}.

\subsection{Full system QFI}
\label{app:full_qfi}
For the presented single-parameter protocols, we can compute the QFI of the QHO--ancilla system using the QFI of a pure state \cite{Braunstein94}
\begin{equation}
    \label{eq:full_qfi}
    \mathcal{F}^{Q} = 4 \left( \bra{\partial_{\beta} \psi } \partial_{\beta} \psi \rangle - \left| \bra{\psi} \partial_{\beta} \psi \rangle \right|^{2} \right).
\end{equation}
For the single-measurement protocol in App.~\ref{app:single_meas} with $\alpha = iA$ and $\beta \in \mathbb{R}$, the joint state before readout is
\begin{equation}
    \ket{\psi(\beta)}
    = \hat{U}_{\mathrm{BCH}}(\beta)\, \hat{D}(N\beta + \hat{\sigma}_{x}\alpha)\, \ket{\downarrow}\otimes\ket{0}
    = \frac{1}{\sqrt{2}} \sum_{a \in \{ +,- \} }
      e^{\,i N A \beta\, a}\, \ket{a} \otimes \ket{N\beta + i a A}.
\end{equation}
Differentiating using the identity $\partial_{\beta}\ket{N\beta + iaA} = \left( N\hat{a}^{\dagger} - N^{2}\beta \right)\ket{N\beta + iaA}$, and including the accumulated phase gives
\begin{equation}
    \ket{\partial_{\beta}\psi}
    = \frac{1}{\sqrt{2}} \sum_{a \in \{+,- \} }
      e^{\,i N A \beta\, a}\,
      \ket{a} \otimes
      \left[ i N A a + N \hat{a}^{\dagger} - N^{2}\beta \right]
      \ket{N\beta + i a A}.
\end{equation}
The overlaps of Eq.~\eqref{eq:full_qfi} are
\begin{equation}
    \braket{\psi | \partial_{\beta}\psi}
    = \frac{1}{2} \sum_{a \in \{ + , -  \} }
      \left[ i N A a + N\left( N\beta - i a A \right) - N^{2}\beta \right]
    = 0,
\end{equation}
\begin{align}
    \braket{\partial_{\beta}\psi | \partial_{\beta}\psi}
    &= \frac{1}{2} \sum_{a \in \{ + , - \} }
       \bra{N\beta + iaA}
       \left[ -i N A a + N \hat{a} - N^{2}\beta \right]
       \left[ i N A a + N \hat{a}^{\dagger} - N^{2}\beta \right]
       \ket{N\beta + iaA} \\
    &= \frac{1}{2} \sum_{a \in \{ +,- \} }
       \left\{ \left| i N A a + N\left( N\beta - iaA \right) - N^{2}\beta \right|^{2}
       + N^{2} \right\}
    = N^{2},
\end{align}
so that
\begin{equation}
    \mathcal{F}^{Q}_{\mathrm{single}, \mathrm{full}}
    = 4 \left( \braket{\partial_{\beta}\psi | \partial_{\beta}\psi}
    - \left| \braket{\psi | \partial_{\beta}\psi} \right|^{2} \right)
    = 4 N^{2}.
\end{equation}
For the sequential protocol with $\alpha = iA$ and $\beta \in \mathbb{R}$, the joint state before measurement follows from Eq.~\eqref{eq:bch_phase_unitary},
\begin{equation}
    \ket{\psi(\beta)}
    = \hat{U}_{\mathrm{BCH}}(\beta)\, \hat{D}(N\beta + \hat{S}_{x}\alpha)\,
      \ket{\downarrow}^{\otimes N} \otimes \ket{0}
    = \frac{1}{2^{N/2}} \sum_{\mathbf{a} \in \{ +, -  \}^{N}}
      e^{\,i A \beta \sum_{j}(2j-N)a_{j}}\,
      \ket{\mathbf{a}} \otimes \ket{\xi_{\mathbf{a}}},
    \qquad
    \xi_{\mathbf{a}} = N\beta + iA\sum_{j=1}^{N} a_{j}.
\end{equation}
Again using the identity $\partial_{\beta}\ket{\xi_{\mathbf{a}}} = \left( N\hat{a}^{\dagger} - N^{2}\beta \right)\ket{\xi_{\mathbf{a}}}$, and including the accumulated phase gives
\begin{equation}
    \ket{\partial_{\beta}\psi}
    = \frac{1}{2^{N/2}} \sum_{\mathbf{a} \in \{ +,- \}^{N}}
      e^{\,i A \beta \sum_{j}(2j-N)a_{j}}\,
      \ket{\mathbf{a}} \otimes
      \left[ iA\sum_{j=1}^{N}(2j-N)a_{j} + N \hat{a}^{\dagger} - N^{2}\beta \right]
      \ket{\xi_{\mathbf{a}}}.
\end{equation}
The overlaps of Eq.~\eqref{eq:full_qfi} are
\begin{align}
    \braket{\psi | \partial_{\beta}\psi}
    &= 2^{-N} \sum_{\mathbf{a} \in \{ +, - \}^{N}}
      \left[ iA\sum_{j=1}^{N}(2j-N)a_{j}
      + N\left( N\beta - iA\sum_{j=1}^{N} a_{j} \right) - N^{2}\beta \right] \\
    &= 2^{-N} \sum_{\mathbf{a} \in \{ +, - \}^{N}}
      2iA \sum_{j=1}^{N} (j-N)\, a_{j}
    = 0,
\end{align}
\begin{align}
    \braket{\partial_{\beta}\psi | \partial_{\beta}\psi}
    &= 2^{-N} \sum_{\mathbf{a} \in \{+,-\}^{N}}
       \bra{\xi_{\mathbf{a}}}
       \left[ -iA\sum_{j=1}^{N}(2j-N)a_{j} + N \hat{a} - N^{2}\beta \right]
       \left[ iA\sum_{j=1}^{N}(2j-N)a_{j} + N \hat{a}^{\dagger} - N^{2}\beta \right]
       \ket{\xi_{\mathbf{a}}} \\
    &= 2^{-N} \sum_{\mathbf{a} \in \{+,-\}^{N}}
       \left\{ \left| 2iA \sum_{j=1}^{N} (j-N)\, a_{j} \right|^{2}
       + N^{2} \right\} \\
    &= 4A^{2} \sum_{j=1}^{N} (j-N)^{2} + N^{2}
    = 4A^{2}\, \frac{N(N-1)(2N-1)}{6} + N^{2},
\end{align}
using $2^{-N}\sum_{\mathbf{a}} a_{j} a_{k} = \delta_{jk}$, so that
\begin{equation}
    \mathcal{F}^{Q}_{\mathrm{seq,full}}
    = 4 \left( \braket{\partial_{\beta}\psi | \partial_{\beta}\psi}
    - \left| \braket{\psi | \partial_{\beta}\psi} \right|^{2} \right)
    = 4 N^{2} + \frac{8}{3}\, A^{2}\, N (N-1)(2N-1).
\end{equation}

\section{Classical Fisher Information}
\label{app:cfi}
To obtain the CFI, we pick the computational basis for measurement of the sequential- and single-measurement protocols with $\alpha = i A$ and $\beta \in \mathbb{R}$. For an outcome record $\mathbf{b}$ occurring with probability $p_{\mathbf{b}}(\beta)$, the CFI is 
\begin{equation}
    \label{eq:cfi_def}
    \mathcal{F}^{C} = \sum_{\mathbf{b}} \frac{\big[\partial_{\beta}\, p_{\mathbf{b}}(\beta)\big]^{2}}{p_{\mathbf{b}}(\beta)} = \mathbb{E}_{\mathbf{b} \sim p}\!\left[ \big( \partial_{\beta} \ln p_{\mathbf{b}} \big)^{2} \right],
\end{equation}
which satisfies $\mathcal{F}^{C} \leq \mathcal{F}^{Q}$ for any chosen measurement. From now on, we denote the computational basis in terms of the $\hat \sigma_{x}$ eigenbasis as $\ket{b} = \frac{1}{\sqrt{2}} \left( \ket{+} + (-1)^{b} \ket{-} \right) $ with $b \in \{ 0,1 \}$, where $b = 0$ $(1)$ corresponds to the outcome $\downarrow$ $(\uparrow)$ of the main text.

\subsection{Single-measurement protocol}
\label{app:cfi_single}

For the single-measurement protocol, the post-protocol qubit state is given by Eq.~\eqref{eq:rho_single}. Measuring in the computational basis gives the outcome distribution
\begin{equation}
    \label{eq:cfi_single_prob}
    p_{b}(\beta)
    = \bra{b}\, \rho(\beta)\, \ket{b}
    = \frac{1}{2}\left[ 1 + (-1)^{b}\, e^{-2A^{2}} \cos\!\big( 4NA\beta \big) \right],
\end{equation}
where the diagonal entries of Eq.~\eqref{eq:rho_single} each contribute $1/2$ and the off-diagonals carry signal-dependent information. Differentiating, $\partial_{\beta}\, p_{b}(\beta) = -(-1)^{b}\, 2NA\, e^{-2A^{2}} \sin(4NA\beta)$, and substituting into Eq.~\eqref{eq:cfi_def} for the two outcomes yields
\begin{equation}
    \label{eq:cfi_single}
    \mathcal{F}^{C}_{\mathrm{single}}
    = \frac{16 N^{2} A^{2}\, e^{-4A^{2}} \sin^{2}\!\big( 4NA\beta \big)}
           {1 - e^{-4A^{2}} \cos^{2}\!\big( 4NA\beta \big)}.
\end{equation}
Unlike the QFI, which is independent of the signal, the CFI oscillates with $\beta$. At the optimal working points $\beta = (2m+1) \pi / (8NA)$ for integer $m$, Eq.~\eqref{eq:cfi_single} saturates the qubit QFI, so the computational basis is optimal. Conversely, when $\beta = m \pi/ (4NA)$, the CFI experiences interference dips, which decrease in width as $N$ grows.

\subsection{Exact outcome probabilities}
\label{app:cfi_seq}
For the sequential protocol, the register state is generated from a $\beta$-independent reference state by the total generator of Eq.~\eqref{eq:total_generator}, $\rho_{N}(\beta) = e^{i\beta\hat{H}}\, \widetilde{\rho}_{N}(0)\, e^{-i\beta\hat{H}}$ with $\hat{H} = 2A\sum_{j=1}^{N} j\, \hat{\sigma}_{x}^{(j)}$. It is therefore convenient to rotate the measurement rather than the state. Then, the probability of the record $\mathbf{b} = (b_{1}, \ldots, b_{N})$ is $p_{\mathbf{b}}(\beta) = \bra{\mathbf{b}(\beta)}\, \widetilde{\rho}_{N}(0)\, \ket{\mathbf{b}(\beta)}$, where $\ket{\mathbf{b}(\beta)} = e^{-i\beta\hat{H}} \ket{\mathbf{b}} = \bigotimes_{j=1}^{N} e^{-i\vartheta_{j}\hat{\sigma}_{x}^{(j)}} \ket{b_{j}}$ with $\vartheta_{j} = 2A\, j\, \beta$. In this case, all of the $\beta$-dependence is carried by a product of single-qubit rotations, while the reference state $\widetilde{\rho}_{N}(0)$ is fixed and supported on the permutation-invariant subspace $\mathcal{S}$, with real Dicke-basis matrix elements
\begin{equation}
    \label{eq:cfi_ref_state}
    R_{kk'} = \bra{\phi_{k}}\, \widetilde{\rho}_{N}(0)\, \ket{\phi_{k'}}
    = \sqrt{\gamma_{k}\gamma_{k'}}\; e^{-2A^{2}(k-k')^{2}},
    \qquad
    \gamma_{k} = \binom{N}{k} 2^{-N}.
\end{equation}
Since the state is supported only on $\mathcal{S}$, only the projection of the measurement vector onto the Dicke basis enters the probability,
\begin{equation}
    \label{eq:cfi_dicke_prob}
    p_{\mathbf{b}}(\beta) = \sum_{k,k'=0}^{N} \braket{\mathbf{b}(\beta) | \phi_{k}}\, R_{kk'}\, \braket{\phi_{k'} | \mathbf{b}(\beta)},
\end{equation}
which reduces the evaluation of each of the $2^{N}$ outcome probabilities to the $N+1$ overlaps $\braket{\phi_{k} | \mathbf{b}(\beta)}$. Each single-qubit factor of $\ket{\mathbf{b}(\beta)}$ expands in the $\hat{\sigma}_{x}$ eigenbasis as
\begin{equation}
    \label{eq:cfi_amplitudes}
    e^{-i\vartheta_{j}\hat{\sigma}_{x}} \ket{b_{j}}
    = \zeta_{j}^{(+)} \ket{+} + \zeta_{j}^{(-)} \ket{-},
    \qquad
    \zeta_{j}^{(+)} = \frac{e^{-i\vartheta_{j}}}{\sqrt{2}},
    \qquad
    \zeta_{j}^{(-)} = \frac{(-1)^{b_{j}}\, e^{i\vartheta_{j}}}{\sqrt{2}} .
\end{equation}
Writing the Dicke state as the normalized sum over branch strings with $k$ minus branches (Eq.~\eqref{eq:seq_dicke_basis}), the overlap becomes
\begin{equation}
    \label{eq:cfi_coeff}
    \braket{\phi_{k} | \mathbf{b}(\beta)}
    = \binom{N}{k}^{-1/2}
    \sum_{\mathbf{a} :\, w(\mathbf{a}) = k}\; \prod_{j=1}^{N} \zeta_{j}^{(a_{j})}
    = \binom{N}{k}^{-1/2}
    \left( \prod_{j=1}^{N} \zeta_{j}^{(+)} \right)
    e_{k}(y_{1}, \ldots, y_{N}),
    \qquad
    y_{j} \equiv \frac{\zeta_{j}^{(-)}}{\zeta_{j}^{(+)}} = (-1)^{b_{j}}\, e^{4i j A \beta},
\end{equation}
where $e_{k}(y_{1}, \ldots, y_{N}) = \sum_{|\mathcal{J}| = k} \prod_{j \in \mathcal{J}} y_{j}$ is the elementary symmetric polynomial: the sum over branch strings with $w(\mathbf{a}) = k$ selects every set $\mathcal{J}$ of $k$ positions carrying $y_{j}$. The prefactor $\prod_{j}\zeta_{j}^{(+)} = 2^{-N/2} e^{-i\sum_{j}\vartheta_{j}}$ is common to every overlap; its phase cancels in Eq.~\eqref{eq:cfi_dicke_prob}, and its magnitude contributes the overall factor $2^{-N}$. With $\widetilde{c}_{k} \equiv \binom{N}{k}^{-1/2} e_{k}$, the probability is
\begin{equation}
    \label{eq:cfi_quadratic}
    p_{\mathbf{b}}(\beta) = 2^{-N}\, \widetilde{c}^{\,\dagger} R\, \widetilde{c} .
\end{equation}
The elementary symmetric polynomials are generated by the recursion
\begin{equation}
    \label{eq:esp_recursion}
    e_{k}^{(s)} = e_{k}^{(s-1)} + y_{s}\, e_{k-1}^{(s-1)},
    \qquad
    e_{0}^{(s)} = 1, \qquad e_{k}^{(s)} = 0 \;\; \text{for } k > s,
\end{equation}
where $e_{k}^{(s)} = e_{k}(y_{1}, \ldots, y_{s})$, at cost $\mathcal{O}(N^{2})$ per record. Since $\widetilde{\rho}_{N}(0)$ is $\beta$-independent, the
derivative needed for the CFI acts only on the coefficients,
\begin{equation}
    \label{eq:cfi_derivative}
    \partial_{\beta}\, p_{\mathbf{b}}
    = 2^{-N} \cdot 2\, \operatorname{Re}\!\left[ \big( \partial_{\beta} \widetilde{c} \big)^{\dagger} R\, \widetilde{c} \right],
\end{equation}
and $\partial_{\beta} e_{k}$ follows from $\partial_{\beta} y_{j} = 4 i j A\, y_{j}$ together with the companion recursion obtained by differentiating Eq.~\eqref{eq:esp_recursion},
\begin{equation}
    \label{eq:esp_derivative_recursion}
    \dot{e}_{k}^{(s)} = \dot{e}_{k}^{(s-1)} + \dot{y}_{s}\, e_{k-1}^{(s-1)} + y_{s}\, \dot{e}_{k-1}^{(s-1)},
\end{equation}
at the same $\mathcal{O}(N^{2})$ cost. Equations~\eqref{eq:cfi_quadratic}--\eqref{eq:esp_derivative_recursion} give the exact probability and score of any individual record in polynomial time. The remaining obstacle is the sum over all $2^{N}$ records in Eq.~\eqref{eq:cfi_def}.

\subsection{Monte Carlo estimation of the CFI}
\label{app:cfi_sampling}
To avoid the exponential sum over all measurement records, we estimate the CFI by sampling records from the exact distribution,
\begin{equation}
    \label{eq:cfi_mc}
    \mathcal{F}^{C}
    \approx \frac{1}{M} \sum_{m=1}^{M}
    \left( \partial_{\beta} \ln p_{\mathbf{b}^{(m)}} \right)^{2},
\end{equation}
where $\mathbf{b}^{(m)} \sim p_{\mathbf{b}}$. Records are drawn in the same order the physical protocol produces them, by factorizing the distribution into conditionals,
\begin{equation}
    \label{eq:cfi_conditionals}
    p_{\mathbf{b}} = \prod_{s=1}^{N} p( b_{s} \,|\, b_{1} \cdots b_{s-1} ),
    \qquad
    p( b_{s} \,|\, b_{<s} ) = \frac{ p_{b_{1} \cdots b_{s}} }{ p_{b_{1} \cdots b_{s-1}} },
\end{equation}
so that only the marginal probabilities $p_{b_{1} \cdots b_{s}}$ of the first $s$ outcomes are required. Since each rotation $e^{-i\vartheta_{j}\hat{\sigma}_{x}^{(j)}}$ acts only on qubit $j$, tracing over the unmeasured qubits removes their rotations, and the marginal takes the
same form as Eq.~\eqref{eq:cfi_dicke_prob}, evaluated on the reduced reference state,
\begin{equation}
    \label{eq:cfi_marginal_def}
    p_{b_{1} \cdots b_{s}}
    = \bra{ \mathbf{b}_{\leq s}(\beta) }\, \widetilde{\rho}_{s}(0)\, \ket{ \mathbf{b}_{\leq s}(\beta) },
    \qquad
    \widetilde{\rho}_{s}(0) \equiv \operatorname{Tr}_{s+1, \ldots, N}\!\left[ \widetilde{\rho}_{N}(0) \right],
    \qquad
    \ket{ \mathbf{b}_{\leq s}(\beta) } = \bigotimes_{j=1}^{s} e^{-i\vartheta_{j}\hat{\sigma}_{x}^{(j)}} \ket{b_{j}} .
\end{equation}
The reduced states $\widetilde{\rho}_{s}(0)$ are permutationally invariant, like $\widetilde{\rho}_{N}(0)$ and are therefore supported on the $s$-qubit Dicke basis $\{ \ket{\phi_{l}^{(s)}} \}_{l=0}^{s}$. They are obtained by iteratively tracing out one qubit at a time using the identity
\begin{equation}
    \label{eq:dicke_branching}
    \ket{\phi_{k}^{(s)}}
    = \sqrt{\frac{s-k}{s}}\; \ket{\phi_{k}^{(s-1)}} \otimes \ket{+}_{s}
    + \sqrt{\frac{k}{s}}\; \ket{\phi_{k-1}^{(s-1)}} \otimes \ket{-}_{s},
\end{equation}
which yields the recursion for the Dicke-basis matrix elements $(R_{s})_{ll'} = \bra{\phi_{l}^{(s)}} \widetilde{\rho}_{s}(0) \ket{\phi_{l'}^{(s)}}$,
\begin{equation}
    \label{eq:marginal_recursion}
    (R_{s-1})_{l,l'}
    = \frac{1}{s} \left[
    \sqrt{(s-l)(s-l')}\; (R_{s})_{l,l'}
    + \sqrt{(l+1)(l'+1)}\; (R_{s})_{l+1,\,l'+1}
    \right],
    \qquad
    l, l' = 0, \ldots, s-1,
\end{equation}
initialized at $(R_{N})_{kk'} = R_{kk'}$ [Eq.~\eqref{eq:cfi_ref_state}]. Producing the set $\{ R_{s} \}_{s=1}^{N}$ costs $\mathcal{O}(N^{3})$. Since it is independent of both $\beta$ and the sampled records, it can be computed once before sampling. The record-dependent coefficients take the form
\begin{equation}
    \label{eq:marginal_coeff}
    \widetilde{c}^{\,(s)}_{l}
    = \binom{s}{l}^{-1/2} e_{l}(y_{1}, \ldots, y_{s}),
    \qquad
    p_{b_{1} \cdots b_{s}} = 2^{-s}\, \big( \widetilde{c}^{\,(s)} \big)^{\dagger} R_{s}\, \widetilde{c}^{\,(s)},
\end{equation}
where each round updates the coefficients of the previous round through Eq.~\eqref{eq:esp_recursion}. To build a full record, we start at round $s$ and compute the marginal $p_{b_{1}\cdots b_{s}}$ for both possible outcomes $b_{s} \in \{0,1\}$. Then, the outcome is drawn from the conditional of Eq.~\eqref{eq:cfi_conditionals} and the coefficient recursion is moved to the next round. For a single record the coefficient updates cost $\mathcal{O}( N^{2})$, while computing Eq.~\eqref{eq:marginal_coeff}, at $\mathcal{O}(s^{2})$ per round, accumulate to $\mathcal{O}(N^{3})$. To approximate the CFI using this technique then amounts to drawing $M$ records costing $\mathcal{O}(MN^{3})$. Both the exact and approximate forms are polynomial-speed-ups from standard propagation of the density matrix which is enabled by treating the signal-dependent dynamics as qubit rotations.

\subsection{Single-round marginal probabilities}
\label{app:cfi_single_round}
The conditional structure of Eq.~\eqref{eq:cfi_conditionals} can be summarized by the marginal distribution of a single round, $p_{b_{j}}(\beta) = \sum_{\{b_{i}\}_{i \neq j}} p_{\mathbf{b}}$, which has a closed form. Since $\widetilde{\rho}_{N}(0)$ is permutation invariant, the reduced state of any single qubit is the same for every round. The round index enters only through the rotation angle $\vartheta_{j}$ of the measurement vector $\ket{\mathbf{b}(\beta)}$. Tracing all but one qubit using Eq.~\eqref{eq:dicke_branching}, the off-diagonal contributions collapse onto neighboring Dicke components ($k' = k \pm 1$),
\begin{equation}
    \operatorname{Tr}_{2, \ldots, N}\!\left( \ket{\phi_{k}^{(N)}} \bra{\phi_{k'}^{(N)}} \right)
    = \frac{\sqrt{(N-k)(N-k')}}{N}\, \delta_{k,k'}\, \ket{+}\!\bra{+}
    + \frac{\sqrt{(N-k)\,k'}}{N}\, \delta_{k,\,k'-1}\, \ket{+}\!\bra{-}
    + \mathrm{h.c.}
    + \frac{\sqrt{k k'}}{N}\, \delta_{k,k'}\, \ket{-}\!\bra{-} .
\end{equation}
Weighting each term by the matrix elements $R_{kk'}$ of Eq.~\eqref{eq:cfi_ref_state}, the diagonal sums evaluate to $\sum_{k} \gamma_{k} (N-k)/N = \sum_{k} \gamma_{k}\, k/N = 1/2$. For the off-diagonal sum, using $\sqrt{\gamma_{k}\gamma_{k+1}}\sqrt{(N-k)(k+1)}/N = \gamma_{k}(N-k)/N$, the result is also $1/2$. This gives the single-qubit reference state
\begin{equation}
    \label{eq:single_qubit_reduced}
    \rho^{(1)}
    = \frac{1}{2}
    \begin{pmatrix}
        1 & e^{-2A^{2}} \\[1mm]
        e^{-2A^{2}} & 1
    \end{pmatrix}
\end{equation}
in the $\{\ket{+}, \ket{-}\}$ basis. A single qubit therefore carries signal information only in its coherence between neighboring Dicke components. Applying the round-$j$ rotation, $e^{i\vartheta_{j}\hat{\sigma}_{x}}\, \rho^{(1)}\, e^{-i\vartheta_{j}\hat{\sigma}_{x}}$, multiplies the coherence by $e^{2i\vartheta_{j}} = e^{4ijA\beta}$. Measuring in the computational basis gives
\begin{equation}
    \label{eq:single_round_marginal}
    p_{b_{j}}(\beta)
    = \frac{1}{2} \left[ 1 + (-1)^{b_{j}}\, e^{-2A^{2}} \cos\!\big( 4 j A \beta \big) \right].
\end{equation}
Each round realizes the single-measurement distribution of Eq.~\eqref{eq:cfi_single_prob} with the effective interaction number $N \to j$, since ancilla $j$ interrogates a signal that has accumulated for $j$ rounds (Eq.~\eqref{eq:total_generator}). The marginals are related to the conditionals of Eq.~\eqref{eq:cfi_conditionals} through $p_{b_{s}}(\beta) = \sum_{b_{<s}} p_{b_{1}\cdots b_{s-1}}\, p(b_{s} \,|\, b_{<s})$.
If the rounds were to be independent, summing the CFI of each single-round marginal [Eq.~\eqref{eq:single_round_marginal}] gives an approximation to the sequential CFI,
\begin{equation}
    \label{eq:cfi_marginal}
    \mathcal{F}^{C}_{\mathrm{marg}} = \sum_{j=1}^{N} \frac{16\, j^{2} A^{2}\, e^{-4A^{2}} \sin^{2}4 j A \beta} {1 - e^{-4A^{2}} \cos^{2}4 j A \beta},
\end{equation}
which is a sum of single-measurement contributions (Eq.~\eqref{eq:cfi_single}) with an additional round-dependent modulation. Equation~\eqref{eq:cfi_marginal} has two periodicities. One is a slow oscillation at $\beta_{\mathrm{slow}} = \pi/(4A)$, which is independent of $N$. The other is a fast oscillation at $\beta_{\mathrm{fast}} = \pi/(4NA)$. The round-dependent CFI has these periodicities but also has inter-round correlations captured by the sampling procedure of App.~\ref{app:cfi_sampling}.

\subsection{CFI scaling with $A$}
\label{app:cfi_scaling}
\begin{figure*}
    \centering
    \includegraphics[width=0.5\columnwidth]{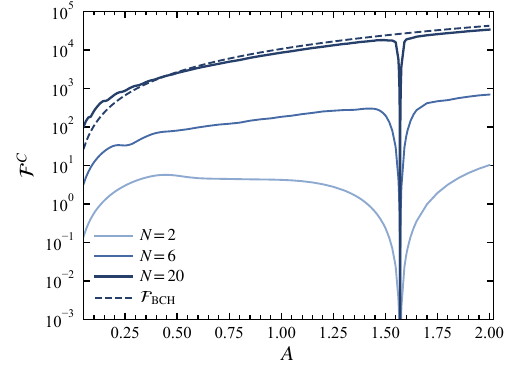}
    \caption{ CFI as a function of the controlled displacement amplitude $A$ at fixed $\beta=0.05$ for $N \in \{2,6,20\}$. The dashed dark blue line plots the $\mathcal{F}_{\text{BCH}}$ contribution.}
    \label{fig:cfi_scaling}
\end{figure*}

The dependence of the extracted information on the coupling strength is shown in Fig.~\ref{fig:cfi_scaling}, where the CFI is evaluated as a function of $A$ at a fixed signal $\beta = 0.05$. For all $N$, the CFI maintains the $A^{2}$ scaling of the QFI, modulated by the round-dependent fringes of Eq.~\eqref{eq:cfi_marginal}: round $j$ is uninformative at $A = m\pi/(4 j \beta)$, so the fastest modulation, set by the final round, has period $\pi/(4N\beta)$ in $A$ ($\pi/4$ for the $N=20$ curve). $\mathcal{F}_{\mathrm{BCH}} = \tfrac{4}{3}A^{2}N(N^{2}-1)$ (Eq.~\eqref{eq:qfi_backaction}) is plotted for reference.

\subsection{Outcome means and correlations}
\label{app:cfi_correlations}
The single-round marginal of Eq.~\eqref{eq:single_round_marginal} can also be interpreted as being the first moment of the record. To be more concise, set $z_{j} \equiv (-1)^{b_{j}}$ giving
\begin{equation}
    \label{eq:mean_outcome}
    \langle z_{j} \rangle
    = e^{-2A^{2}} \cos\!\big( 4 j A \beta \big),
    \qquad
    \langle b_{j} \rangle = \frac{1}{2}\Big[ 1 - \langle z_{j} \rangle \Big],
\end{equation}
which oscillates with the round index $j$. Round-dependency can be observed through the joint marginal of a pair of rounds $i \neq j$, obtained by tracing out the other $N-2$ qubits. Then, generalizing the partial trace of App.~\ref{app:cfi_single_round} to keep an arbitrary subset of $s$ qubits gives a closed form for the reduced reference state,
\begin{equation}
    \label{eq:sigma_closed}
    \big( R_{s} \big)_{qq'}
    = \sqrt{ \gamma^{(s)}_{q}\, \gamma^{(s)}_{q'} }\;
    e^{-2A^{2}(q-q')^{2}},
    \qquad
    \gamma^{(s)}_{q} = \binom{s}{q} 2^{-s} .
\end{equation}
On any subset of $s$ qubits, the reference state of the $N$-round protocol reduces to that of an $s$-round protocol. For this case, the marginal of $s = 2$ is obtained in  $\{ \ket{++}, \ket{+-}, \ket{-+}, \ket{--} \}$ using $\ket{\phi_{1}^{(2)}} = \big( \ket{+-} + \ket{-+} \big)/\sqrt{2}$,
\begin{equation}
    \label{eq:pair_state}
    \widetilde{\rho}_{2}(0)
    = \frac{1}{4}
    \begin{pmatrix}
        1 & d & d & d^{4} \\
        d & 1 & 1 & d \\
        d & 1 & 1 & d \\
        d^{4} & d & d & 1
    \end{pmatrix},
    \qquad
    d \equiv e^{-2A^{2}} .
\end{equation}
The round indices enter only through the rotations $e^{-i\vartheta_{j}\hat{\sigma}_{x}}$ with $\vartheta_{j} = 2A\, j\, \beta$, defined in App.~\ref{app:cfi_seq}. The second moment $\langle z_{i} z_{j} \rangle$ detects the coherences of $\widetilde{\rho}_{2}(0)$ in which both qubits change branch and can be expressed for $i \neq j$ as
\begin{equation}
    \label{eq:second_moment}
    \langle z_{i} z_{j} \rangle
    = \frac{1}{2} \cos\!\big[ 4A\beta\, (i-j) \big]
    + \frac{1}{2}\, e^{-8A^{2}} \cos\!\big[ 4A\beta\, (i+j) \big] .
\end{equation}
Subtracting the product of first order moments (Eq.~\eqref{eq:mean_outcome}) gives the inter-round correlations,
\begin{equation}
    \label{eq:round_correlations}
    C_{ij}
    = \langle z_{i} z_{j} \rangle - \langle z_{i} \rangle \langle z_{j} \rangle
    = \frac{1}{2} \Big( 1 - e^{-4A^{2}} \Big) \cos\!\big[ 4A\beta\, (i-j) \big]
    + \frac{1}{2} \Big( e^{-8A^{2}} - e^{-4A^{2}} \Big) \cos\!\big[ 4A\beta\, (i+j) \big] .
\end{equation}
The first term depends on the lag, $i - j$, and grows monotonically with the coupling strength. This is the information coming from the accumulating BCH phase from repeated conditional displacements. The second term depends on the absolute round indices through $i + j$ and follows the coherently accumulated signal phase, which resembles the contribution the single-measurement protocol benefits from. In the large $A$ limit, $\langle z_{j} \rangle \to 0$ but $C_{ij} \to \tfrac{1}{2}\cos[4A\beta(i-j)]$, so signal information is not suppressed by strong coupling. This split resembles the two separate contributions in the QFI (Sec.~\ref{sec:qfi_single}).

\subsection{Arbitrary product-measurement bases}
\label{app:cfi_basis}
The sequential protocol implements an arbitrary product measurement in which qubit $j$ is read out in the orthonormal basis $\{\ket{m_{j}^{(0)}}, \ket{m_{j}^{(1)}}\}$. In this work we mainly considered measurement of the computational basis, but generalizing another basis is possible. First consider,
\begin{equation}
    \label{eq:general_basis}
    \ket{m_{j}^{(0)}} = u_{j}\ket{+} + v_{j}\ket{-},
    \qquad
    \ket{m_{j}^{(1)}} = -v_{j}^{*}\ket{+} + u_{j}^{*}\ket{-},
    \qquad
    |u_{j}|^{2} + |v_{j}|^{2} = 1 .
\end{equation}
which corresponds to inserting a single-qubit rotation before each readout of the sequential protocol. Writing $\big(u_{j}(0), v_{j}(0)\big) = (u_{j}, v_{j})$ and $\big(u_{j}(1), v_{j}(1)\big) = (-v_{j}^{*}, u_{j}^{*})$ for the amplitudes selected by outcome $b_{j}$, the branch amplitudes of Eq.~\eqref{eq:cfi_amplitudes} become
\begin{equation}
    \label{eq:general_amplitudes}
    \zeta_{j}^{(+)} = u_{j}(b_{j})\, e^{-i\vartheta_{j}},
    \qquad
    \zeta_{j}^{(-)} = v_{j}(b_{j})\, e^{i\vartheta_{j}},
    \qquad
    y_{j} = \frac{v_{j}(b_{j})}{u_{j}(b_{j})}\, e^{2i\vartheta_{j}},
\end{equation}
and the record probability of Eq.~\eqref{eq:cfi_quadratic} becomes
\begin{equation}
    \label{eq:general_quadratic}
    p_{\mathbf{b}}(\beta)
    = \Bigg( \prod_{j=1}^{N} |u_{j}(b_{j})|^{2} \Bigg)\,
    \widetilde{c}^{\,\dagger} R\, \widetilde{c},
    \qquad
    \widetilde{c}_{k} = \binom{N}{k}^{-1/2} e_{k}(y_{1}, \ldots, y_{N}) .
\end{equation}
The record-dependent weight carries no $\beta$ dependence, so the score acts only on the coefficients like in Eq.~\eqref{eq:cfi_derivative}, and the recursions of Eqs.~\eqref{eq:esp_recursion} and~\eqref{eq:esp_derivative_recursion} remain the same but with the modified ratios. Since records are generated round by round, the basis parameters $(u_{j}, v_{j})$ may potentially be conditioned on earlier outcomes without changing the cost, so the framework could also be applied to adaptive product strategies.
Consider $u_{j} = 0$ or $v_{j} = 0$, which means qubit $j$ is read out in the $\hat{\sigma}_{x}$ eigenbasis. This is because $\hat H = 2A \sum_{j} j \hat \sigma_{x}^{(j)}$ [Eq.~\eqref{eq:total_generator}] and
\begin{equation}
    e^{-i \vartheta_{j} \hat \sigma_{x}} \ket{+} = e^{-i \vartheta_{j} } \ket{+} , \qquad e^{-i \vartheta_{j} \hat \sigma_{x}} \ket{-} = e^{i \vartheta_{j}} \ket{-}.
\end{equation} 
which is only a global phase to the measurement and cancels in Eq.~\eqref{eq:general_quadratic}, implying that round $j$ carries no signal dependence. Then consider $|u_{j}| = |v_{j}| = 1/\sqrt{2}$ which gives
\begin{equation}
    \ket{m_{j}^{(b)}} = \frac{1}{\sqrt{2}} \left( \ket{+} + (-1)^{b} e^{i \chi_{j}} \ket{-} \right)
\end{equation}
where $\chi_{j} = \arg(v_{j}/u_{j})$. Since $\vartheta_{j} = 2j A \beta$, the polynomial terms become  
\begin{equation}
    \label{eq:equatorial_ratios}
    y_{j} = (-1)^{b_{j}}\, e^{i(4jA\beta + \chi_{j})},
    \qquad
    \prod_{j=1}^{N} |u_{j}(b_{j})|^{2} = 2^{-N},
\end{equation}
and measurement in the computational basis is achieved when $\chi_{j} = 0$. Repeating the partial trace of App.~\ref{app:cfi_single_round} for a general basis, the single-round marginal becomes
\begin{equation}
    \label{eq:general_marginal}
    p_{b_{j}}(\beta)
    = \frac{1}{2}\left[ 1 + (-1)^{b_{j}}\, 2|u_{j}v_{j}|\, e^{-2A^{2}} \cos\!\big( 4jA\beta + \chi_{j} \big) \right].
\end{equation}

\section{Two-Parameter Protocol}
\label{app:two_param}
\subsection{Density matrix}
If the controlled displacement is varied round-to-round, $\hat{D}_{j}(\hat{\sigma}_{x}^{(j)}\alpha_{j})$ with $\alpha_{j} = r_{j} A$ and $r_{j} \in \{ 1, i \}$, the joint state after $N$ rounds is
\begin{equation}
    \ket{\psi} = \left( \prod_{j=1}^{N} \hat{D}_{j} ( \hat{\sigma}_{x}^{(j)} \alpha_{j})\, \hat{D}(\beta)  \right) \ket{\downarrow}^{\otimes N} \otimes \ket{0}.
\end{equation}
Applying the phase relation to combine the displacements within each round, $\hat{D}_{j}( \hat{\sigma}_{x}^{(j)} \alpha_{j} )\, \hat{D}(\beta) = \exp \left( -i A\, \operatorname{Im} (\beta r_{j}^{*})\, \hat{\sigma}_{x}^{(j)} \right) \hat{D}(\beta + \alpha_{j} \hat{\sigma}_{x}^{(j)})$, turns the full product into
\begin{equation}
    \left( \prod_{j=1}^{N} \hat{D}_{j} ( \hat{\sigma}_{x}^{(j)} \alpha_{j})\, \hat{D}(\beta) \right)
    = \left( \prod_{j=1}^{N}  \exp \left( -i A\, \operatorname{Im}(\beta r_{j}^{*})\, \hat{\sigma}_{x}^{(j)} \right) \right) \prod_{j=1}^{N} \hat{D}(\beta + A r_{j} \hat{\sigma}_{x}^{(j)}).
\end{equation}
For the remaining product, define $\xi_{j} = \beta + r_{j} A \hat{\sigma}_{x}^{(j)}$. This makes
\begin{equation}
    \prod_{j =1}^{N} \hat{D}(\xi_{j}) = \exp \left[ -i \sum_{1 \leq j < k \leq N } \operatorname{Im} (\xi_{j} \xi_{k}^{*}) \right] \hat{D} \left( \sum_{j =1}^{N} \xi_{j} \right),
\end{equation}
where $\sum_{j = 1}^{N} \xi_{j} = N \beta + A \hat{S}_{x}^{\,r}$ with the collective operator
\begin{equation}
    \hat{S}_{x}^{\,r} \equiv \sum_{j=1}^{N} r_{j}\, \hat{\sigma}_{x}^{(j)} = \hat{S}_{x}^{\mathcal{R}} + i\, \hat{S}_{x}^{\mathcal{I}},
\end{equation}
where $\mathcal{R} = \{ j : r_{j} = 1 \}$ and $\mathcal{I} = \{ j : r_{j} = i \}$ as in the main text, with $N_{R} = |\mathcal{R}|$ and $N_{I} = |\mathcal{I}|$. The final product becomes
\begin{equation}
    \prod_{j = 1}^{N} \hat{D}(\xi_{j}) = \exp \left[ i \sum_{j<k} \left( A\, \operatorname{Im} (\beta r_{j}^{*})\, \hat{\sigma}_{x}^{(j)} - A\, \operatorname{Im} ( \beta r_{k}^{*} )\, \hat{\sigma}_{x}^{(k)} - A^{2}\, \operatorname{Im} (r_{j} r_{k}^{*})\, \hat{\sigma}_{x}^{(j)} \hat{\sigma}_{x}^{(k)} \right)   \right] \hat{D}(N \beta + A \hat{S}_{x}^{\,r}).
\end{equation}
Adding back the single-round contribution,
\begin{equation}
    \prod_{j=1}^{N} \hat{D}_{j} (\hat{\sigma}_{x}^{(j)} \alpha_{j})\, \hat{D}(\beta) = \hat{U}^{(2)}_{\mathrm{BCH}}(\beta)\, \hat{D}(N \beta + A \hat{S}_{x}^{\,r}),
\end{equation}
with
\begin{equation}
    \label{eq:two_param_bch}
    \hat{U}^{(2)}_{\mathrm{BCH}}(\beta) = \exp \left[ i \sum_{j =1}^{N} (N - 2j)\, A\, \operatorname{Im} (\beta r_{j}^{*})\, \hat{\sigma}_{x}^{(j)} - i A^{2} \sum_{j < k} \operatorname{Im} (r_{j} r_{k}^{*})\, \hat{\sigma}_{x}^{(j)} \hat{\sigma}_{x}^{(k)} \right].
\end{equation}
The final state of the qubit register is
\begin{equation}
    \rho^{(2)}_{N}(\beta) = \hat{U}^{(2)}_{\mathrm{BCH}}(\beta)\, \widetilde{\rho}^{(2)}_{N} (\beta)\, \hat{U}^{(2)\dagger}_{\mathrm{BCH}}(\beta),
\end{equation}
where
\begin{equation}
    \widetilde{\rho}^{(2)}_{N}(\beta) = \mathrm{Tr}_{\mathrm{osc}} \left[ \hat{D}(N \beta + A \hat{S}_{x}^{\,r}) \left( ( \ket{\downarrow} \bra{\downarrow} )^{\otimes N} \otimes \ket{0} \bra{0} \right) \hat{D}^{\dagger} (N \beta + A \hat{S}_{x}^{\,r}) \right],
\end{equation}
which results in a state supported on the doubly permutation-invariant subspace of dimension $(N_{R}+1)(N_{I}+1)$, in analogy with the single-parameter case. To see this, note that the two-parameter Dicke states
\begin{equation}
    \ket{\phi_{k_{R}, k_{I}}} = \binom{N_{R}}{k_{R}}^{-1/2} \binom{N_{I}}{k_{I}}^{-1/2}
    \sum_{\substack{\mathbf{a} :\, w_{\mathcal{R}}(\mathbf{a}) = k_{R} \\ \phantom{\mathbf{a} :\,} w_{\mathcal{I}}(\mathbf{a}) = k_{I}}} \ket{\mathbf{a}},
\end{equation}
where $w_{\mathcal{B}}(\mathbf{a})$ counts the number of $\ket{-}$ branches among the rounds $j \in \mathcal{B}$, are eigenstates of the collective operator,
\begin{equation}
    \hat{S}_{x}^{\,r} \ket{\phi_{k_{R}, k_{I}}} = (M_{R} + i M_{I}) \ket{\phi_{k_{R}, k_{I}}},
    \qquad
    M_{R} = N_{R} - 2k_{R}, \quad M_{I} = N_{I} - 2k_{I}.
\end{equation}
Tracing out the oscillator, the reduced qubit state is
\begin{equation}
    \widetilde{\rho}^{(2)}_{N} (\beta) = \sum_{k_{R}, k_{I}} \sum_{k_{R}', k_{I}'} \sqrt{\gamma_{k_{R}, k_{I}}\, \gamma_{k_{R}', k_{I}'} }\;
    e^{-\frac{1}{2} A^{2} \left(\Delta M_{R}^{2} + \Delta M_{I}^{2}\right)}\, e^{i \Phi_{\mathrm{coh}} }\,
    \ket{\phi_{k_{R}, k_{I}}} \bra{\phi_{k_{R}', k_{I}'} },
\end{equation}
with
\begin{equation}
    \gamma_{k_{R}, k_{I}} = 2^{-N} \binom{ N_{R} }{ k_{R}} \binom{ N_{I} }{ k_{I}},
    \qquad
    \Delta M_{R} = M_{R} - M_{R}', \quad \Delta M_{I} = M_{I} - M_{I}',
\end{equation}
where $M_{R}' = N_{R} - 2k_{R}'$ and $M_{I}' = N_{I} - 2k_{I}'$. The signal-dependent phase is
\begin{equation}
    \label{eq:two_param_phase}
    \Phi_{\mathrm{coh}} = N A \left( \beta_{1}\, \Delta M_{I} - \beta_{2}\, \Delta M_{R} \right)  + A^{2} \left( M_{R}' M_{I} - M_{I}' M_{R} \right).
\end{equation}

\subsection{Quantum Fisher information matrix}
\label{app:two_param_qfi}
Each element of the QFI matrix is
\begin{equation}
    \mathcal{F}^{Q}_{\mu \nu}
    = 2 \sum_{\substack{a,a' \\ \lambda_{a} + \lambda_{a'} > 0}}
    \frac{ \operatorname{Re} \left[ (Q_{\mu})_{aa'} (Q_{\nu})_{aa'}^{*} \right] }{\lambda_{a} + \lambda_{a'}},
\end{equation}
where
\begin{equation}
    (Q_{\mu})_{aa'} = \bra{\widetilde{e}_{a}} \partial_{\mu} \widetilde{\rho}^{(2)}_{N}(\beta) \ket{ \widetilde{e}_{a'}} + i(\lambda_{a'} - \lambda_{a}) \bra{\widetilde{e}_{a}} \hat{H}_{\beta_{\mu}} \ket{\widetilde{e}_{a'}},
    \qquad
    \mu, \nu \in \{ \beta_{1}, \beta_{2} \},
\end{equation}
which follows from $\partial_{\mu}\rho^{(2)}_{N} = \hat{U}^{(2)}_{\mathrm{BCH}} \big( \partial_{\mu} \widetilde{\rho}^{(2)}_{N} + i \big[ \hat{H}_{\beta_{\mu}}, \widetilde{\rho}^{(2)}_{N} \big] \big) \hat{U}^{(2)\dagger}_{\mathrm{BCH}}$, in analogy with App.~\ref{app:seq_qfi}. The estimated parameter is now the full displacement $\beta = \beta_{1} + i\beta_{2}$ and $\widetilde{\rho}^{(2)}_{N}(\beta) = \sum_{a} \lambda_{a} \ket{\widetilde{e}_{a}}\bra{\widetilde{e}_{a}}$ is the reduced ancilla state of the two-parameter protocol. Recalling the subsets $\mathcal{R}$ and $\mathcal{I}$ of the previous subsection, with $N_{R} + N_{I} = N$, the BCH phase generators of each parameter follow from Eq.~\eqref{eq:two_param_bch},
\begin{equation}
    \hat{H}_{\beta_{1}} = A \sum_{j \in \mathcal{I}} (2j - N)\, \hat{\sigma}_{x}^{(j)},
    \qquad
    \hat{H}_{\beta_{2}} = A \sum_{j \in \mathcal{R}} (N-2j)\, \hat{\sigma}_{x}^{(j)}.
\end{equation}
In analogy with the single-parameter case (App.~\ref{app:splitting}), permutation invariance within each subset gives $\hat{\Pi}^{(2)}\hat{\sigma}_{x}^{(j)}\hat{\Pi}^{(2)} = \tfrac{1}{N_{\mathcal{B}}}\hat{\Pi}^{(2)}\hat{S}_{x}^{\mathcal{B}}\hat{\Pi}^{(2)}$ for $j \in \mathcal{B} \in \{\mathcal{R}, \mathcal{I}\}$, which together with the phase weights
\begin{equation}
    s_{R} \equiv \sum_{j \in \mathcal{R}} (N-2j),
    \qquad
    s_{I} \equiv \sum_{j \in \mathcal{I}} (N-2j),
\end{equation}
yields the projection identities
\begin{equation}
    \hat{\Pi}^{(2)}\, \hat{H}_{\beta_{1}}\, \hat{\Pi}^{(2)}
    = -A\, \frac{s_{I}}{N_{I}}\, \hat{\Pi}^{(2)}\hat{S}_{x}^{\mathcal{I}}\hat{\Pi}^{(2)},
    \qquad
    \hat{\Pi}^{(2)}\, \hat{H}_{\beta_{2}}\, \hat{\Pi}^{(2)}
    = A\, \frac{s_{R}}{N_{R}}\, \hat{\Pi}^{(2)}\hat{S}_{x}^{\mathcal{R}}\hat{\Pi}^{(2)},
\end{equation}
where the sign difference arises from $\sum_{j \in \mathcal{I}}(2j-N) = -s_{I}$. Here,
\begin{equation}
    \hat{\Pi}^{(2)} = \sum_{a=1}^{(N_{R} + 1)(N_{I}+ 1)} \ket{\widetilde{e}_{a}} \bra{\widetilde{e}_{a}}
\end{equation}
is the projector onto the support of $\widetilde{\rho}^{(2)}_{N}(\beta)$, coinciding with the projector onto the doubly permutation-invariant subspace $\mathcal{S}^{(2)}$ spanned by the Dicke states $\ket{\phi_{k_{R}, k_{I}}}$, since $\widetilde{\rho}^{(2)}_{N}$ is full rank there for $A > 0$, in analogy with $\hat{\Pi}$ of the single-parameter protocol. Since the subset operators are diagonal in the two-parameter Dicke basis, $\hat{S}_{x}^{\mathcal{B}}\ket{\phi_{k_{R}, k_{I}}} = M_{\mathcal{B}}\ket{\phi_{k_{R}, k_{I}}}$, they commute with $\hat{\Pi}^{(2)}$, which we use below. For pairs with both eigenvectors in the support, the matrix elements then simplify to
\begin{gather}
    (Q_{\beta_{1}})_{aa'} = \bra{\widetilde{e}_{a}} \partial_{\beta_{1}} \widetilde{\rho}^{(2)}_{N}(\beta) \ket{ \widetilde{e}_{a'}}
    + i(\lambda_{a} - \lambda_{a'})\, A \frac{s_{I}}{N_{I}}\, \bra{\widetilde{e}_{a}} \hat{S}_{x}^{\mathcal{I}} \ket{\widetilde{e}_{a'}}, \\
    (Q_{\beta_{2}})_{aa'} = \bra{\widetilde{e}_{a}} \partial_{\beta_{2}} \widetilde{\rho}^{(2)}_{N}(\beta) \ket{ \widetilde{e}_{a'}}
    - i(\lambda_{a} - \lambda_{a'})\, A \frac{s_{R}}{N_{R}}\, \bra{\widetilde{e}_{a}} \hat{S}_{x}^{\mathcal{R}} \ket{\widetilde{e}_{a'}}.
\end{gather}
Next, consider the orthogonal complement, projected by $\mathbbm{1} - \hat{\Pi}^{(2)}$, which gives the geometric-phase contribution
\begin{align}
    \mathcal{F}^{\mathrm{BCH}}_{\mu \nu}
    &= 4\, \operatorname{Re} \left[   \operatorname{Tr} \left( \widetilde{\rho}^{(2)}_{N}(\beta)\, \hat{H}_{\beta_{\mu}} \left( \mathbbm{1} - \hat{\Pi}^{(2)} \right) \hat{H}_{\beta_{\nu}} \right) \right] \\
    &= 4\, \operatorname{Re} \left[ \operatorname{Tr}\!\left( \widetilde{\rho}^{(2)}_{N}(\beta)\, \hat{H}_{\beta_{\mu}} \hat{H}_{\beta_{\nu}} \right)
    - \operatorname{Tr}\!\left(\widetilde{\rho}^{(2)}_{N}(\beta)\, \hat{H}_{\beta_{\mu}}\, \hat{\Pi}^{(2)}\, \hat{H}_{\beta_{\nu}} \right) \right],
\end{align}
in analogy with $\mathcal{F}_{\mathrm{BCH}}$ of App.~\ref{app:seq_qfi}. The traces simplify similarly, but must be split into diagonal ($\mu = \nu$) and off-diagonal ($\mu \neq \nu$) contributions. For the diagonal contribution of the second trace term above,
\begin{align}
    \operatorname{Tr} \left(  \widetilde{\rho}^{(2)}_{N}(\beta)\, \hat{H}_{\beta_{1}}\, \hat{\Pi}^{(2)}\, \hat{H}_{\beta_{1}} \right)
    &= A^{2} \left(  \frac{s_{I}}{N_{I}} \right)^{2} \operatorname{Tr}\!\left(\widetilde{\rho}^{(2)}_{N}(\beta)\, (\hat{S}_{x}^{\mathcal{I}})^{2}\right).
\end{align}
For the other diagonal contribution, replace $\mathcal{I}$ with $\mathcal{R}$. For the off-diagonal contribution,
\begin{align}
    \operatorname{Tr} \left(  \widetilde{\rho}^{(2)}_{N}(\beta)\, \hat{H}_{\beta_{1}}\, \hat{\Pi}^{(2)}\, \hat{H}_{\beta_{2}} \right)
    &= - A^{2}\, \frac{s_{R} s_{I} }{N_{R} N_{I} }\,
    \operatorname{Tr}\!\left(\widetilde{\rho}^{(2)}_{N} (\beta)\, \hat{S}_{x}^{\mathcal{R}} \hat{S}_{x}^{\mathcal{I}}\right).
\end{align}
Now for the first trace term's diagonal contributions. Recall $c_{j} = N-2j$ from the main text, so that $s_{R} = \sum_{j \in \mathcal{R}} c_{j}$, and define $q_{R} \equiv \sum_{j \in \mathcal{R}} c_{j}^{2}$ (and similarly $q_{I}$). Then
\begin{align}
    \operatorname{Tr}\!\left( \widetilde{\rho}^{(2)}_{N}(\beta)\, \hat{H}_{\beta_{1}} \hat{H}_{\beta_{1}} \right)
    &= A^{2} \sum_{j,k \in \mathcal{I}} c_{j} c_{k}\,
    \operatorname{Tr} \left( \widetilde{\rho}^{(2)}_{N}(\beta)\, \hat{\sigma}_{x}^{(j)} \hat{\sigma}_{x}^{(k)}  \right) \\
    &= A^{2} \left[  q_{I} + \sum_{\substack{j,k \in \mathcal{I} \\ j \neq k}} c_{j} c_{k}\,
    \operatorname{Tr}\!\left(\widetilde{\rho}^{(2)}_{N}(\beta)\, \hat{\sigma}_{x}^{(j)} \hat{\sigma}_{x}^{(k)} \right)  \right] \\
    &= A^{2} \left[ q_{I} + \left( s_{I}^{2} - q_{I} \right)
    \operatorname{Tr}\!\left(\widetilde{\rho}^{(2)}_{N}(\beta)\, \hat{\sigma}_{x}^{(j)} \hat{\sigma}_{x}^{(k)} \right) \right],
\end{align}
where $j \neq k$ for $j,k \in \mathcal{I}$ in the final trace. The other diagonal term is obtained by replacing $\mathcal{I}$ with $\mathcal{R}$. Now for the off-diagonal contribution of the first trace:
\begin{equation}
    \operatorname{Tr} \left( \widetilde{\rho}^{(2)}_{N}(\beta)\, \hat{H}_{\beta_{1}} \hat{H}_{\beta_{2}} \right)
    = - A^{2}\, \frac{s_{R} s_{I} }{N_{R} N_{I} }\,
    \operatorname{Tr}\!\left(\widetilde{\rho}^{(2)}_{N} (\beta)\, \hat{S}_{x}^{\mathcal{R}} \hat{S}_{x}^{\mathcal{I}}\right).
\end{equation}
This is identical to the off-diagonal contribution of the second trace, which means the off-diagonal terms of $\mathcal{F}^{\mathrm{BCH}}$ go to zero. Combining both diagonal contributions gives
\begin{align}
    \mathcal{F}^{\mathrm{BCH}} = 4 A^{2}
    \begin{pmatrix}
    \dfrac{q_{I} - s^{2}_{I} / N_{I}}{N_{I}(N_{I} - 1)}
    \left( N^{2}_{I} - \operatorname{Tr}\!\left(\widetilde{\rho}^{(2)}_{N}(\beta)\, (\hat{S}_{x}^{\mathcal{I}})^{2}\right) \right) & 0 \\[3mm]
    0 & \dfrac{q_{R} - s^{2}_{R} / N_{R}}{N_{R}(N_{R} - 1)}
    \left( N^{2}_{R} - \operatorname{Tr}\!\left(\widetilde{\rho}^{(2)}_{N}(\beta)\, (\hat{S}_{x}^{\mathcal{R}})^{2}\right) \right)
    \end{pmatrix},
\end{align}
where we used the relation $\operatorname{Tr} \left( \widetilde{\rho}^{(2)}_{N}(\beta)\, ( \hat{S}_{x}^{\mathcal{B}} )^{2} \right) = N_{\mathcal{B}} + N_{\mathcal{B}} (N_{\mathcal{B}} - 1)\, \operatorname{Tr} \left( \widetilde{\rho}^{(2)}_{N}(\beta)\, \hat{\sigma}_{x}^{(j)} \hat{\sigma}_{x}^{(k)} \right)$, for $j \neq k \in \mathcal{B}$ with $\mathcal{B} \in \{ \mathcal{R}, \mathcal{I} \}$, to eliminate the two-point correlator. Finally, since the diagonal weights of $\widetilde{\rho}^{(2)}_{N}$ are binomial, $\operatorname{Tr}\!\left( \widetilde{\rho}^{(2)}_{N}(\beta)\, (\hat{S}_{x}^{\mathcal{B}})^{2} \right) = \sum_{k_{R}, k_{I}} \gamma_{k_{R}, k_{I}}\, M_{\mathcal{B}}^{2} = N_{\mathcal{B}}$, so the brackets reduce to $N_{\mathcal{B}}(N_{\mathcal{B}}-1)$ and the diagonal entries collapse to
\begin{equation}
    \mathcal{F}^{\mathrm{BCH}}_{\mu\mu}
    = 4A^{2} \left( q_{\mathcal{B}_{\mu}} - \frac{s_{\mathcal{B}_{\mu}}^{2}}{N_{\mathcal{B}_{\mu}}} \right)
    = 4A^{2}\, \mathrm{Var}_{\mathcal{B}_{\mu}}(c),
\end{equation}
which is Eq.~\eqref{eq:two_param_backaction} of the main text, with $\mathcal{B}_{\beta_{1}} = \mathcal{I}$ and $\mathcal{B}_{\beta_{2}} = \mathcal{R}$. To summarize, the final QFI matrix is diagonal with elements
\begin{equation}
    \mathcal{F}^{Q}_{\mu \nu}
    = 2 \sum_{\substack{a, a' \leq (N_{R} + 1)(N_{I} + 1) \\ \lambda_{a} + \lambda_{a'} > 0 }}
    \frac{\operatorname{Re} \left[ (Q_{\mu})_{aa'} (Q_{\nu})_{aa'}^{*} \right]}{\lambda_{a} + \lambda_{a'}}
    + \mathcal{F}^{\mathrm{BCH}}_{\mu \nu},
\end{equation}
where the first term is the coherence contribution $\mathcal{F}^{\mathrm{coh}}_{\mu\nu}$ of the main text, evaluated by numerically diagonalizing the $(N_{R}+1)(N_{I}+1)$-dimensional Dicke block of $\widetilde{\rho}^{(2)}_{N}(\beta)$, in analogy with Eq.~\eqref{eq:qfi_coherence}.

\subsection{Blocked coupling schedule}
\label{app:blocked}
Here we compare the alternating schedule of the main text, $\mathcal{R} = \{1,3,\ldots,N-1\}$ and $\mathcal{I} = \{2,4,\ldots,N\}$, with the blocked schedule, $\mathcal{R} = \{1,\ldots,N/2\}$ and $\mathcal{I} = \{N/2+1,\ldots,N\}$, both with $N_{R} = N_{I} = N/2$. Up to a relabeling of the ancillae, the reduced state
$\widetilde{\rho}^{(2)}_{N}$ is identical for every schedule with the same subset sizes. All ordering dependence is expressed in $\hat{U}^{(2)}_{\mathrm{BCH}}$ and therefore changes the QFI only through the first moment $s_{\mathcal{B}}$, entering $\mathcal{F}^{\mathrm{coh}}_{\mu\mu}$, and the spread $\mathrm{Var}_{\mathcal{B}}(c)$, entering $\mathcal{F}^{\mathrm{BCH}}_{\mu\mu}$, of the time weights $c_{j} = N-2j$ (Eq.~\eqref{eq:two_param_backaction}). Since $\mathcal{F}^{\mathrm{BCH}}_{\mu\mu} = 4A^{2}\, \mathrm{Var}_{\mathcal{B}_{\mu}}(c)$, we can isolate the effect of each schedule. Within each subset the $c_{j}$ form an arithmetic progression with spacing $h$, for which $\mathrm{Var}_{\mathcal{B}}(c) = h^{2}N_{\mathcal{B}}(N_{\mathcal{B}}^{2}-1)/12$; consecutive rounds give $h = 2$ for the blocked schedule and every-other rounds give $h = 4$ for the alternating schedule, so
\begin{equation}
    \label{eq:blocked_variances}
    \mathrm{Var}^{\mathrm{blk}}_{\mathcal{B}}(c)
    = \tfrac{1}{3} N_{\mathcal{B}}\big(N_{\mathcal{B}}^{2}-1\big),
    \qquad
    \mathrm{Var}^{\mathrm{alt}}_{\mathcal{B}}(c)
    = \tfrac{4}{3} N_{\mathcal{B}}\big(N_{\mathcal{B}}^{2}-1\big).
\end{equation}
At large $A$, where $\mathcal{F}^{\mathrm{coh}}_{\mu\mu}$ is negligible, summing the two diagonal entries with $N_{\mathcal{B}} = N/2$ gives $\operatorname{Tr}\mathcal{F}^{Q}_{\mathrm{blk}} \approx \tfrac{1}{3}A^{2}N(N^{2}-4)$ and $\operatorname{Tr}\mathcal{F}^{Q}_{\mathrm{alt}} \approx \tfrac{4}{3}A^{2}N(N^{2}-4)$, the values quoted in Sec.~\ref{sec:qfi_two}.

\subsection{Compatibility condition}
\label{app:compatibility}
Here we evaluate the commutation condition (Eq.~\eqref{eq:weak_commutation}), which, when satisfied, guarantees that the multi-parameter QCRB is asymptotically jointly saturable~\cite{Ragy16}. The SLDs are defined by $\partial_{\mu}\rho^{(2)}_{N}(\beta) = \tfrac{1}{2}\big[ \hat{L}_{\mu}\rho^{(2)}_{N} + \rho^{(2)}_{N}\hat{L}_{\mu} \big]$ with $\mu\in\{\beta_{1},\beta_{2}\}$. The two-parameter BCH unitary [Eq.~\eqref{eq:two_param_bch}] factors into signal-dependent and signal-independent parts,
\begin{equation}
    \hat{U}^{(2)}_{\mathrm{BCH}}(\beta)
    = \hat{W}\, e^{\,i\left(\beta_{1}\hat{H}_{\beta_{1}} + \beta_{2}\hat{H}_{\beta_{2}}\right)}, \qquad \hat{W} = \exp\!\Big[ -iA^{2} \sum_{j<k} \operatorname{Im}(r_{j}r_{k}^{*})\, \hat{\sigma}_{x}^{(j)}\hat{\sigma}_{x}^{(k)} \Big].
\end{equation}
We define the total generators
\begin{equation}
    \hat{K}_{\beta_{1}} \equiv NA\hat{S}_{x}^{\mathcal{I}} + \hat{H}_{\beta_{1}},
    \qquad
    \hat{K}_{\beta_{2}} \equiv -NA\hat{S}_{x}^{\mathcal{R}} + \hat{H}_{\beta_{2}},
\end{equation}
where the collective parts generate the signal-dependent overlap phases of $\widetilde{\rho}^{(2)}_{N}$ [Eq.~\eqref{eq:two_param_phase}] and the $\hat{H}_{\beta_{\mu}}$ generate the BCH rotations. $\hat{W}$, $\hat{K}_{\beta_{1}}$, and $\hat{K}_{\beta_{2}}$ are diagonal in the $\hat{\sigma}_{x}$ product basis and therefore commute. Isolating the signal-dependence of the register state gives
\begin{equation}
    \label{eq:compat_family}
    \rho^{(2)}_{N}(\beta)
    = \hat{W}\, e^{\,i\beta\cdot\hat{K}}\,
      \widetilde{\rho}^{(2)}_{N}(0)\,
      e^{-i\beta\cdot\hat{K}}\, \hat{W}^{\dagger},
    \qquad
    \beta\cdot\hat{K} \equiv \beta_{1}\hat{K}_{\beta_{1}} + \beta_{2}\hat{K}_{\beta_{2}}.
\end{equation}
The SLDs can be factored in a similar way: $\hat{L}_{\mu} (\beta) = \hat{V} \widetilde{L}_{\mu} \hat{V}^{\dagger}$ with $\hat{V} = \hat{W} e^{i \beta \cdot \hat{K}}$, where $\widetilde{L}_{\mu}$ is the SLD at $\beta = 0$. Like the QFI then, the commutation condition is independent of $\beta$,
\begin{equation}
    \operatorname{Tr}\!\left( \rho^{(2)}_{N}(\beta)
    \big[\hat{L}_{\beta_{1}},\hat{L}_{\beta_{2}}\big] \right) = \operatorname{Tr}\!\left( \widetilde{\rho}^{(2)}_{N}(0) \big[\widetilde{L}_{\beta_{1}},\widetilde{L}_{\beta_{2}}\big] \right).
\end{equation}
Writing $\widetilde{\rho}^{(2)}_{N}(0) = \sum_{a}\lambda_{a}\ket{\widetilde{e}_{a}}\bra{\widetilde{e}_{a}}$ and using $\partial_{\mu}\rho^{(2)}_{N}\,\big|_{\beta=0} = i\big[\hat{K}_{\beta_{\mu}},\widetilde{\rho}^{(2)}_{N}(0)\big]$, the SLD matrix elements are
\begin{equation}
    \big(\widetilde{L}_{\mu}\big)_{aa'}
    = \frac{2i(\lambda_{a'}-\lambda_{a})}{\lambda_{a}+\lambda_{a'}}\,
      \bra{\widetilde{e}_{a}}\hat{K}_{\beta_{\mu}}\ket{\widetilde{e}_{a'}}
\end{equation}
for $\lambda_{a}+\lambda_{a'} > 0$. Substituting gives
\begin{equation}
    \label{eq:compat_eigen}
    \operatorname{Tr}\!\left( \widetilde{\rho}^{(2)}_{N}(0) \big[\widetilde{L}_{\beta_{1}},\widetilde{L}_{\beta_{2}}\big] \right) = 8i \sum_{\substack{a<a' \\ \lambda_{a}+\lambda_{a'}>0}} \frac{(\lambda_{a}-\lambda_{a'})^{3}}{(\lambda_{a}+\lambda_{a'})^{2}}\, \operatorname{Im}\!\left[ \bra{\widetilde{e}_{a}}\hat{K}_{\beta_{1}}\ket{\widetilde{e}_{a'}} \bra{\widetilde{e}_{a'}}\hat{K}_{\beta_{2}}\ket{\widetilde{e}_{a}} \right].
\end{equation}
This sum can be split with respect to the doubly permutation-invariant subspace $\mathcal{S}^{(2)}$ of App.~\ref{app:two_param_qfi}, in analogy with App.~\ref{app:splitting}. Since the collective parts commute with $\hat{\Pi}^{(2)}$, only the BCH generators couple outside the subspace, $\hat{\Pi}^{(2)}\hat{K}_{\beta_{\mu}}(\mathbbm{1}-\hat{\Pi}^{(2)}) = \hat{\Pi}^{(2)}\hat{H}_{\beta_{\mu}}(\mathbbm{1}-\hat{\Pi}^{(2)})$, giving the contribution
\begin{equation}
    \label{eq:compat_back}
    8i\, \operatorname{Im}\, \operatorname{Tr}\!\left[ \widetilde{\rho}^{(2)}_{N}(0)\,
    \hat{H}_{\beta_{1}}\big(\mathbbm{1}-\hat{\Pi}^{(2)}\big)
    \hat{H}_{\beta_{2}} \right] = 2i\, \operatorname{Im}\, T_{\beta_{1}\beta_{2}},
    \qquad
    T_{\mu\nu} \equiv \operatorname{Tr}\!\left[ \widetilde{\rho}^{(2)}_{N}(0)\, \widetilde{L}_{\mu}\big(\mathbbm{1}-\hat{\Pi}^{(2)}\big) \widetilde{L}_{\nu} \right],
\end{equation}
where the SLD elements coupling the support to the kernel have been substituted, $(\widetilde{L}_{\mu})_{ak} = -2i(\hat{H}_{\beta_{\mu}})_{ak}$, so that $T_{\mu\nu} = 4\operatorname{Tr}\!\big[\widetilde{\rho}^{(2)}_{N}(0)\, \hat{H}_{\beta_{\mu}}\big(\mathbbm{1}-\hat{\Pi}^{(2)}\big)\hat{H}_{\beta_{\nu}}\big]$. The real part of this quantity defines the geometric-phase QFI, $\operatorname{Re}\,T_{\mu\nu} = \mathcal{F}^{\mathrm{BCH}}_{\mu\nu}$ (App.~\ref{app:two_param_qfi}), and its imaginary part enters the commutation condition. Since App.~\ref{app:two_param_qfi} shows the two traces composing $T_{\beta_{1}\beta_{2}}$ to be identical, $T_{\beta_{1}\beta_{2}} = 0$ and the BCH phase does not contribute to the commutation condition.

The coherence contribution comes from the remaining pairs, with both eigenvectors in $\mathcal{S}^{(2)}$. There, the projection identities of Eq.~\eqref{eq:two_param_projection} replace each generator by a collective operator,
\begin{equation}
    \label{eq:two_param_projection}
    \hat{\Pi}^{(2)}\hat{K}_{\beta_{1}}\hat{\Pi}^{(2)} = A\Big(N - \frac{s_{I}}{N_{I}}\Big)\, \hat{\Pi}^{(2)}\hat{S}_{x}^{\mathcal{I}}\hat{\Pi}^{(2)},
    \qquad 
    \hat{\Pi}^{(2)}\hat{K}_{\beta_{2}}\hat{\Pi}^{(2)} = -A\Big(N - \frac{s_{R}}{N_{R}}\Big)\, \hat{\Pi}^{(2)}\hat{S}_{x}^{\mathcal{R}}\hat{\Pi}^{(2)},
\end{equation}
so, since the BCH sector vanishes, the full commutation condition is carried by the support pairs,
\begin{equation}
    \label{eq:compat_dicke}
    \operatorname{Tr}\!\left( \rho^{(2)}_{N}(\beta) \big[\hat{L}_{\beta_{1}},\hat{L}_{\beta_{2}}\big] \right) = -8i\, A^{2} \Big(N-\frac{s_{I}}{N_{I}}\Big)\Big(N-\frac{s_{R}}{N_{R}}\Big) \sum_{\substack{a<a' \in \mathcal{S}^{(2)} \\ \lambda_{a}+\lambda_{a'}>0}} \frac{(\lambda_{a}-\lambda_{a'})^{3}}{(\lambda_{a}+\lambda_{a'})^{2}}\, \operatorname{Im}\!\left[ \big(\hat{S}_{x}^{\mathcal{I}}\big)_{aa'} \big(\hat{S}_{x}^{\mathcal{R}}\big)_{a'a} \right],
\end{equation}
where the matrix elements are evaluated in the eigenbasis of $\widetilde{\rho}^{(2)}_{N}(0)$, requiring diagonalization of its $(N_{R}+1)(N_{I}+1)$-dimensional Dicke block. Writing the collective round weighting as $N - s_{\mathcal{B}}/N_{\mathcal{B}} = 2\bar{j}_{\mathcal{B}}$, with $\bar{j}_{\mathcal{B}} \equiv N_{\mathcal{B}}^{-1}\sum_{j\in\mathcal{B}} j$ the mean round index of the subset, the prefactor becomes $-32iA^{2}\,\bar{j}_{\mathcal{I}}\,\bar{j}_{\mathcal{R}}$, which is $-8iA^{2}N(N+2)$ for the alternating schedule and $-2iA^{2}(N+2)(3N+2)$ for the blocked schedule.

\begin{figure*}
    \centering
    \includegraphics[width=0.5\columnwidth]{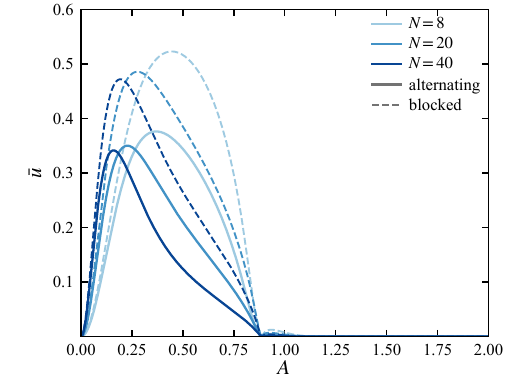}
    \caption{Incompatibility $\bar{u}$ [Eq.~\eqref{eq:norm_commutator}] versus $A$ for $N = \{8, 20, 40\}$ (light to dark), alternating (solid) and blocked (dashed) orderings.}
    \label{fig:sld_coherence}
\end{figure*}

Since the BCH contribution goes to zero, any violation of Eq.~\eqref{eq:weak_commutation} is carried by the coherence contribution, Eq.~\eqref{eq:compat_dicke}, which we evaluate numerically and find to be nonzero (Fig.~\ref{fig:sld_coherence}). The absolute value of the commutator is not meaningful on its own, since it grows with the same factors as the QFI. We therefore normalize it by its maximum value permitted by the QFI matrix~\cite{Carollo19},
\begin{equation}
    \label{eq:norm_commutator}
    \bar{u} = \frac{\left|\operatorname{Tr}\!\left(\rho^{(2)}_{N}
      \big[\hat{L}_{\beta_{1}},\hat{L}_{\beta_{2}}\big]\right)\right|}
      {2\sqrt{\mathcal{F}^{Q}_{\beta_{1}\beta_{1}}\,
      \mathcal{F}^{Q}_{\beta_{2}\beta_{2}}}} \in [0,1].
\end{equation}
The incompatibility is maximal at $A \approx 0.37$, with a value of $\bar{u} \approx 0.38$ for $N = 8$, decreasing to $\bar{u} \approx 0.34$ at $N = 40$. For the blocked ordering, $\bar{u} \approx 0.47$, which is slightly larger than the alternating ordering. As $A \to 0$ the cross phase $\propto A^{2}$ of $\widetilde{\rho}^{(2)}_{N}(0)$ is small, and at large $A$, $\bar{u}$ is suppressed exponentially ($\bar{u} < 10^{-6}$ for $A \gtrsim 1.3$), where the information is instead generated by the BCH phase. The QCRB of Sec.~\ref{sec:qfi_two} may therefore understate the attainable total variance by up to $38\%$ at the worst-case $A$. The ultimate bound on variance is set by the Holevo bound $C_{H}$ and is related to the QCRB by $\operatorname{Tr}[(\mathcal{F}^{Q})^{-1}] \leq C_{H} \leq (1+\bar{u})\operatorname{Tr}[(\mathcal{F}^{Q})^{-1}]$. In the BCH-dominated regime, $\bar u \rightarrow 0$ making the two bounds coincide.

\subsection{Classical Fisher information matrix}
\label{app:cfi_two}

The method of computing the probability introduced in App.~\ref{app:cfi_seq} can be extended to the two-parameter protocol. For a measurement record $\mathbf{b} = (b_{1}, \ldots, b_{N})$ occurring with probability $p_{\mathbf{b}} (\beta) $, the CFI matrix is
\begin{equation}
    \label{eq:cfi_two_def}
    \mathcal{F}^{C}_{\mu\nu}
    = \sum_{\mathbf{b}}
    \frac{ \big[ \partial_{\mu}\, p_{\mathbf{b}}(\beta) \big]
           \big[ \partial_{\nu}\, p_{\mathbf{b}}(\beta) \big] }
         { p_{\mathbf{b}}(\beta) },
    \qquad
    \mu, \nu \in \{ \beta_{1}, \beta_{2} \},
\end{equation}
which satisfies $\mathcal{F}^{C} \preceq \mathcal{F}^{Q}$ for any measurement.

Since the register state factorizes as $\rho^{(2)}_{N}(\beta) = \hat{U}^{(2)}_{\mathrm{BCH}}(\beta)\, \widetilde{\rho}^{(2)}_{N}(\beta)\, \hat{U}^{(2)\dagger}_{\mathrm{BCH}}(\beta)$ (App.~\ref{app:two_param}), we can absorb the BCH unitary into the measurement as in App.~\ref{app:cfi_seq},
\begin{equation}
    \label{eq:cfi_two_rotated}
    p_{\mathbf{b}}(\beta)
    = \bra{ \mathbf{b}(\beta) }\, \widetilde{\rho}^{(2)}_{N}(\beta)\, \ket{ \mathbf{b}(\beta) },
    \qquad
    \ket{ \mathbf{b}(\beta) } = \hat{U}^{(2)\dagger}_{\mathrm{BCH}}(\beta) \ket{ \mathbf{b} }.
\end{equation}
In contrast to the single-parameter case, where $\ket{\mathbf{b}(\beta)}$ absorbed the full generator $\hat{H}$, here only the BCH unitary is absorbed, and the reduced state retains the signal dependence of its Dicke coherences through $\Phi_{\mathrm{coh}}$ [Eq.~\eqref{eq:two_param_phase}]. Both the state and the measurement therefore carry part of the $\beta$ dependence: the state retains the collective phases generated by $NA\hat{S}_{x}^{\mathcal{I}}$ and $-NA\hat{S}_{x}^{\mathcal{R}}$ [App.~\ref{app:compatibility}], while the measurement carries the BCH phases. Since $\widetilde{\rho}^{(2)}_{N}$ is supported on the doubly permutation-invariant subspace $\mathcal{S}^{(2)}$, only the projections of the measurement vector onto the two-parameter Dicke basis enter,
\begin{equation}
    \label{eq:cfi_two_quadratic}
    p_{\mathbf{b}}(\beta)
    = 2^{-N}\, \widetilde{c}^{\,\dagger} R^{(2)}\, \widetilde{c},
    \qquad
    \widetilde{c}_{k_{R}, k_{I}} \equiv 2^{N/2}\, \braket{ \phi_{k_{R}, k_{I}} | \mathbf{b}(\beta) },
\end{equation}
where the matrix elements of the reduced state (App.~\ref{app:two_param}),
\begin{equation}
    \label{eq:cfi_two_refstate}
    R^{(2)}_{(k_{R} k_{I}),(k_{R}' k_{I}')}
    = \sqrt{ \gamma_{k_{R}, k_{I}}\, \gamma_{k_{R}', k_{I}'} }\;
    e^{ -\frac{1}{2} A^{2} \left( \Delta M_{R}^{2} + \Delta M_{I}^{2} \right) }\,
    e^{ i \Phi_{\mathrm{coh}} },
\end{equation}
reduce the evaluation of each of the $2^{N}$ record probabilities to $(N_{R}+1)(N_{I}+1)$ overlaps, and the normalization of $\widetilde{c}$ is chosen to match Eq.~\eqref{eq:cfi_quadratic}.

Since $\hat{U}^{(2)}_{\mathrm{BCH}}$ is diagonal in the $\hat{\sigma}_{x}$ product basis, its action on the measurement is a branch-dependent phase,
\begin{equation}
    \label{eq:cfi_two_bch_phase}
    \hat{U}^{(2)\dagger}_{\mathrm{BCH}}(\beta) \ket{\mathbf{a}}
    = e^{ -i \theta(\mathbf{a}) } \ket{\mathbf{a}},
    \qquad
    \theta(\mathbf{a})
    = A \beta_{2} \sum_{j \in \mathcal{R}} c_{j}\, a_{j}
    - A \beta_{1} \sum_{j \in \mathcal{I}} c_{j}\, a_{j}
    - A^{2} \sum_{j < k} \operatorname{Im}( r_{j} r_{k}^{*} )\, a_{j} a_{k},
\end{equation}
with the time weights $c_{j} = N - 2j$ of the main text. The linear terms are the per-subset BCH phases generated by $\hat{H}_{\beta_{1}}$ and $\hat{H}_{\beta_{2}}$, while the pairwise term is the signal-independent cross phase generated by $\hat{W}$ [App.~\ref{app:compatibility}], arising because orthogonal controlled displacements do not commute. Expanding the computational basis as $\ket{\mathbf{b}} = 2^{-N/2} \sum_{\mathbf{a}} \eta_{\mathbf{b}}(\mathbf{a}) \ket{\mathbf{a}}$ with $\eta_{\mathbf{b}}(\mathbf{a}) = (-1)^{ \sum_{j} b_{j} (1 - a_{j})/2 }$, the coefficients take the form
\begin{equation}
    \label{eq:cfi_two_coeff}
    \widetilde{c}_{k_{R}, k_{I}}
    = \left[ \binom{N_{R}}{k_{R}} \binom{N_{I}}{k_{I}} \right]^{-1/2}
    \sum_{ \mathbf{a} \in \mathcal{Z}_{k_{R}, k_{I}} }
    \eta_{\mathbf{b}}(\mathbf{a})\; e^{ -i \theta(\mathbf{a}) },
\end{equation}
where $\mathcal{Z}_{k_{R}, k_{I}} = \{ \mathbf{a} : w_{\mathcal{R}}(\mathbf{a}) = k_{R},\, w_{\mathcal{I}}(\mathbf{a}) = k_{I} \}$ collects the branch configurations with fixed weights in each subset.

In the single-parameter case of App.~\ref{app:cfi_seq}, the phase accrues as a sum over single rounds, $\theta(\mathbf{a}) = \sum_{j} \theta_{j} (a_{j}) $, where $\theta_{j}$ depends on the branch $a_{j}$ of round $j$ alone, so that $e^{-i \theta(\mathbf{a})} = \prod_{j} e^{-i \theta_{j} (a_{j})}$ factorizes and the branch sums reduce to the elementary symmetric polynomials $e_{k}(y_{1}, \ldots, y_{N}) = \sum_{|\mathcal{J}| = k} \prod_{j \in \mathcal{J}} y_{j}$ of Eq.~\eqref{eq:cfi_coeff}. In the two-parameter case, however, $\hat W$ does not allow for this factorization. Despite this, the cross phase can be rewritten as a sum over rounds, each weighted by the earlier branches of the opposite subset,
\begin{equation}
    \label{eq:cfi_two_running}
    \sum_{j < k} \operatorname{Im}( r_{j} r_{k}^{*} )\, a_{j} a_{k}
    = \sum_{s=1}^{N} a_{s}
    \left[ [s \in \mathcal{R}]\, m_{I}^{(<s)}
         - [s \in \mathcal{I}]\, m_{R}^{(<s)} \right],
    \qquad
    m_{\mathcal{B}}^{(<s)} = \sum_{\substack{j \in \mathcal{B} \\ j < s}} a_{j}
    = N_{\mathcal{B}}^{(s-1)} - 2 k_{\mathcal{B}},
\end{equation}
where $[P]$ is the Iverson bracket, equal to one if $P$ holds and zero otherwise, $N_{\mathcal{B}}^{(s-1)}$ counts the rounds of subset $\mathcal{B}$ among the first $s-1$, and $k_{\mathcal{B}}$ counts the $\ket{-}$ branches of subset $\mathcal{B}$ among them. This shows that the phase acquired at round $s$ depends on the earlier configuration only through the weights $(k_{R}, k_{I})$. We therefore define the partial branch sums over the first $s$ rounds, $S^{(s)}_{k_{R}, k_{I}}$, by $S^{(0)}_{0,0} = 1$ and the recursion
\begin{align}
    \label{eq:cfi_two_recursion}
    s \in \mathcal{R}: \quad
    S^{(s)}_{k_{R}, k_{I}}
    &= \sum_{z = \pm 1} \eta_{s}(z)\,
    e^{ -i z \left( A c_{s} \beta_{2} - A^{2}\, m_{I}^{(<s)} \right) }\,
    S^{(s-1)}_{k_{R} - \delta_{z,-1},\, k_{I}}, \\
    s \in \mathcal{I}: \quad
    S^{(s)}_{k_{R}, k_{I}}
    &= \sum_{z = \pm 1} \eta_{s}(z)\,
    e^{ +i z \left( A c_{s} \beta_{1} - A^{2}\, m_{R}^{(<s)} \right) }\,
    S^{(s-1)}_{k_{R},\, k_{I} - \delta_{z,-1}}, \notag
\end{align}
where $\eta_{s}(+1) = 1$ and $\eta_{s}(-1) = (-1)^{b_{s}}$ carry the computational-basis signs, and $m_{\mathcal{B}}^{(<s)}$ is evaluated at the indices of the state being extended, so that $S^{(N)}_{k_{R}, k_{I}} = \sum_{\mathbf{a} \in \mathcal{Z}_{k_{R}, k_{I}}} \eta_{\mathbf{b}}(\mathbf{a})\, e^{-i\theta(\mathbf{a})}$ and
\begin{equation}
    \label{eq:cfi_two_ctilde}
    \widetilde{c}_{k_{R}, k_{I}} = \left[ \binom{N_{R}}{k_{R}} \binom{N_{I}}{k_{I}} \right]^{-1/2} S^{(N)}_{k_{R}, k_{I}}.
\end{equation}
Equation~\eqref{eq:cfi_two_recursion} generalizes the elementary-symmetric-polynomial recursion of Eq.~\eqref{eq:esp_recursion}, but with a two-branch structure. Now the phase increments depend on the recursion indices through $m_{\mathcal{B}}^{(<s)}$. When one subset is empty the cross phase is zero, the increments become index-independent, and the recursion reduces to Eq.~\eqref{eq:esp_recursion}, provided the collective phases of $R^{(2)}$ are accounted for in the coefficients. Each round updates the full $(N_{R}+1)(N_{I}+1)$ grid, so the coefficients of Eq.~\eqref{eq:cfi_two_ctilde} are computed in $\mathcal{O}(N N_{R} N_{I}) = \mathcal{O}(N^{3})$ per record.

Since Eq.~\eqref{eq:cfi_two_quadratic} carries signal dependence in both the coefficients and the reduced state, the score has two contributions,
\begin{equation}
    \label{eq:cfi_two_score}
    \partial_{\mu}\, p_{\mathbf{b}}
    = 2^{-N} \left\{ 2\, \operatorname{Re} \left[ \big( \partial_{\mu} \widetilde{c} \big)^{\dagger} R^{(2)}\, \widetilde{c} \right]
    + \widetilde{c}^{\,\dagger} \big( \partial_{\mu} R^{(2)} \big)\, \widetilde{c} \right\}.
\end{equation}
The derivative of the reduced state acts on the coherent phase contributions [Eq.~\eqref{eq:cfi_two_refstate}] as
\begin{equation}
    \label{eq:cfi_two_drho}
    \partial_{\beta_{1}} R^{(2)}_{(k_{R} k_{I}),(k_{R}' k_{I}')}
    = i N A\, \Delta M_{I}\; R^{(2)}_{(k_{R} k_{I}),(k_{R}' k_{I}')},
    \qquad
    \partial_{\beta_{2}} R^{(2)}_{(k_{R} k_{I}),(k_{R}' k_{I}')}
    = -i N A\, \Delta M_{R}\; R^{(2)}_{(k_{R} k_{I}),(k_{R}' k_{I}')}.
\end{equation}
The derivative $\partial_{\mu} \widetilde{c}$ follows by differentiating the recursion, as in Eq.~\eqref{eq:esp_derivative_recursion}, now carrying the index-dependent phase along. Suppressing the $(k_{R}, k_{I})$ indices, in the $s \in \mathcal{R}$ case
\begin{equation}
    S^{(s)} = \sum_{z} \eta_{s} (z)\, J_{s}(z)\, S^{(s-1)}, \qquad J_{s}(z) = e^{-i z (A c_{s} \beta_{2} - A^{2} m_{I}^{(<s)} )},
\end{equation}
and differentiating gives
\begin{equation}
    \partial_{\beta_{2}} S^{(s)} = \sum_{z} \eta_{s}(z) \left[ J_{s}(z)\, \partial_{\beta_{2}} S^{(s-1)} + \left( \partial_{\beta_{2}} J_{s}(z)  \right) S^{(s-1)} \right],
\end{equation}
where, since $J_{s}$ is a phase, $\partial_{\beta_{2}} J_{s} = (-i z A c_{s})\, J_{s}$. Similarly, for $s \in \mathcal{I}$ the phase becomes
\begin{equation}
    J_{s}(z) = e^{ +i z ( A c_{s} \beta_{1} - A^{2} m_{R}^{(<s)} ) },
    \qquad
    \partial_{\beta_{1}} J_{s} = ( +i z A c_{s} )\, J_{s} .
\end{equation}
Since the phase of each round contains only the quadrature probed by its own subset, the opposite derivative remains unchanged, $\partial_{\beta_{1}} S^{(s)} = \sum_{z} \eta_{s}(z)\, J_{s}(z)\, \partial_{\beta_{1}} S^{(s-1)}$ for $s \in \mathcal{R}$, and likewise for $\partial_{\beta_{2}} S^{(s)}$ when $s \in \mathcal{I}$. The signal-independent cross phase $A^{2} m_{\bar{\mathcal{B}}}^{(<s)}$, where $\bar{\mathcal{B}}$ denotes the subset opposite to that of round $s$, enters only through $J_{s}(z)$ and never generates a derivative term of its own. The full score is therefore obtained by recursively computing the three arrays, $S^{(s)}$, $\partial_{\beta_{1}} S^{(s)}$, and $\partial_{\beta_{2}} S^{(s)}$, over the $(k_{R}, k_{I})$ grid.

\subsection{Single-round marginals}
\label{app:cfi_two_marg}
The single-round marginal probabilities of the two-parameter protocol carry information about the signal, as in App.~\ref{app:cfi_single_round}. Let $\mathbf{a}, \mathbf{a}' \in \{+,-\}^{N}$ be branch strings in the $\hat{\sigma}_{x}$ product basis of App.~\ref{app:two_param}, with one entry per round. Since the register state can be written as $\rho^{(2)}_{N}(\beta) = \hat{W}\, e^{i\beta\cdot\hat{K}}\, \widetilde{\rho}^{(2)}_{N}(0)\, e^{-i\beta\cdot\hat{K}}\, \hat{W}^{\dagger}$ [Eq.~\eqref{eq:compat_family}], with every factor diagonal in this basis,
\begin{equation}
    \label{eq:two_param_generators}
    \hat{W}\ket{\mathbf{a}} = e^{i\theta_{W}(\mathbf{a})}\ket{\mathbf{a}},
    \quad
    \theta_{W}(\mathbf{a}) = -A^{2} \sum_{j<k} \operatorname{Im}(r_{j} r_{k}^{*})\, a_{j} a_{k},
    \qquad
    e^{i\beta\cdot\hat{K}}\ket{\mathbf{a}} = e^{i\theta_{K}(\mathbf{a})}\ket{\mathbf{a}},
    \quad
    \theta_{K}(\mathbf{a}) = 2A\beta_{1} \sum_{j \in \mathcal{I}} j\, a_{j} - 2A\beta_{2} \sum_{j \in \mathcal{R}} j\, a_{j},
\end{equation}
a general matrix element is
\begin{equation}
    \bra{\mathbf{a}}\, \rho^{(2)}_{N}(\beta)\, \ket{\mathbf{a}'} = e^{i[\theta_{W}(\mathbf{a}) - \theta_{W}(\mathbf{a}')]}\, e^{i[\theta_{K}(\mathbf{a}) - \theta_{K}(\mathbf{a}')]}\, \bra{\mathbf{a}}\, \widetilde{\rho}^{(2)}_{N}(0)\, \ket{\mathbf{a}'},
    \qquad
    \bra{\mathbf{a}}\, \widetilde{\rho}^{(2)}_{N}(0)\, \ket{\mathbf{a}'} = 2^{-N} \braket{ \xi_{\mathbf{a}'} | \xi_{\mathbf{a}} },
\end{equation}
where $\xi_{\mathbf{a}} = A \big( M_{R}(\mathbf{a}) + i M_{I}(\mathbf{a}) \big)$ is the oscillator amplitude of branch $\mathbf{a}$ at $\beta = 0$. In $\theta_{K}$ the collective and BCH weights of $\hat{K}_{\beta_{\mu}}$ have combined into $2Aj$, in analogy with Eq.~\eqref{eq:total_generator}; the BCH phase $\theta(\mathbf{a})$ of Eq.~\eqref{eq:cfi_two_bch_phase} is $\theta_{W} + \theta_{K}$ with the collective part $NA[\beta_{1} M_{I}(\mathbf{a}) - \beta_{2} M_{R}(\mathbf{a})]$ removed. The marginal of round $j$ is set by the reduced state $\varrho_{j}$ of qubit $j$, obtained by tracing over the remaining rounds,
\begin{equation}
    \bra{a_{j}}\, \varrho_{j}\, \ket{a_{j}'} = \sum_{\{a_{k}\}_{k \neq j}} \bra{\mathbf{a}}\, \rho^{(2)}_{N}(\beta)\, \ket{\mathbf{a}'},
\end{equation}
where $\mathbf{a}$ and $\mathbf{a}'$ share the same value $a_{k}$ on every round $k \neq j$, and the sum runs over the $2^{N-1}$ configurations of these spectator rounds. Like in the single-parameter case, the diagonal elements are $\bra{\pm}\varrho_{j}\ket{\pm} = 1/2$. The off-diagonal elements $(a_{j}, a_{j}') = (+1, -1)$ carry signal information.
Take $j \in \mathcal{R}$. The two branches differ by the single displacement $2A r_{j}$, so the overlap magnitude is $e^{-2A^{2}}$ for every spectator configuration, while its phase is the cross term of $\Phi_{\mathrm{coh}}$ evaluated at $\Delta M_{R} = 2$ [Eq.~\eqref{eq:two_param_phase}], giving $\arg \braket{ \xi_{\mathbf{a}'} | \xi_{\mathbf{a}} } = -2A^{2}\, M_{I}(\mathbf{a})$, with $M_{I}(\mathbf{a}) = \sum_{k \in \mathcal{I}} a_{k}$ shared by both branches. The $\hat{W}$ term contributes, using $\operatorname{Im}(1 \cdot i^{*}) = -1$ for an $\mathcal{R}$ round preceding an $\mathcal{I}$ round and $\operatorname{Im}(i \cdot 1^{*}) = +1$ for the reverse order,
\begin{equation}
    \theta_{W}(\mathbf{a}) - \theta_{W}(\mathbf{a}') = -2A^{2} \Bigg[ \sum_{\substack{k \in \mathcal{I} \\ k < j}} a_{k} - \sum_{\substack{k \in \mathcal{I} \\ k > j}} a_{k} \Bigg].
\end{equation}
Adding the two phases, the later opposite-subset rounds cancel and the earlier ones double, leaving the total spectator phase $-4A^{2} \sum_{k \in \mathcal{I},\, k < j} a_{k}$. Physically, round $j$ is read out before the later cross terms are applied, so they cannot affect its outcome. Averaging each earlier opposite-subset spectator over $a_{k} = \pm 1$ gives a factor $\cos(4A^{2})$. With the generator phase $\theta_{K}(\mathbf{a}) - \theta_{K}(\mathbf{a}') = -4A j \beta_{2}$,
\begin{equation}
    \label{eq:two_param_qubit_coherence}
    \bra{+} \varrho_{j} \ket{-}
    = \frac{1}{2}\, e^{-2A^{2}}
      \big[ \cos( 4A^{2} ) \big]^{\bar{n}_{j}}\,
      e^{-4 i A j \beta_{2}},
    \qquad
    \bar{n}_{j} = \big| \{ k \in \mathcal{I} : k < j \} \big|,
\end{equation}
and identically for $j \in \mathcal{I}$ with $\mathcal{I} \leftrightarrow \mathcal{R}$, $\beta_{2} \to \beta_{1}$, and conjugated phases. Computational-basis readout gives Eq.~\eqref{eq:two_param_marginal}, with $\kappa_{j} = [\cos(4A^{2})]^{\bar{n}_{j}}$, which reduces to Eq.~\eqref{eq:single_round_marginal} when one subset is empty. Each marginal then contributes the single-measurement CFI [Eq.~\eqref{eq:cfi_single}] with $N \to j$ and contrast $\kappa_{j}\, e^{-2A^{2}}$, giving Eq.~\eqref{eq:cfi_two_marg_f}. Therefore the general condition for interference dips [Eq.~\eqref{eq:dip_lattice}] comes from $\sin( 4 j A \beta_{\mu} ) = 0$ for every $j \in \mathcal{B}_{\mu}$, and with $4A\beta_{\mu} = \pi t$ this holds if and only if $t \in g_{\mu}^{-1}\mathbb{Z}$, $g_{\mu} = \gcd \mathcal{B}_{\mu}$.

\subsection{Two-parameter CFI scaling}
\label{app:cfi_two_scaling}
\begin{figure*}
    \centering
    \includegraphics[width=\columnwidth]{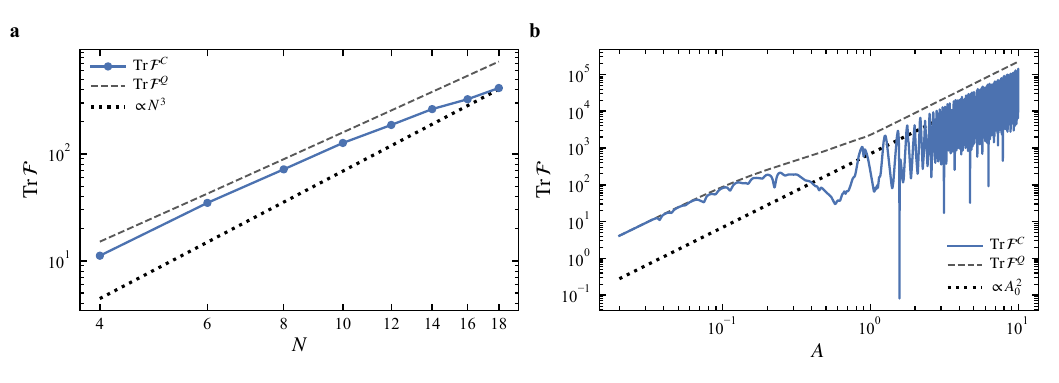}
    \caption{ Scaling comparison of the CFI of the two-parameter protocol with respect to \textbf{a} $N$, at $A=0.2$, and \textbf{b} $A$, at $N=12$. Both plots are computed for $(\beta_{1}, \beta_{2}) = (1.75, 1)$.}
    \label{fig:two_param_cfi_scaling}
\end{figure*}

Figure~\ref{fig:two_param_cfi_scaling} shows the CFI of the two-parameter protocol as a function of $N$ at fixed $A$ and of $A$ at fixed $N$, evaluated at $(\beta_{1}, \beta_{2}) = (1.75, 1)$. In both cases the record CFI retains the $\mathcal{O}(A^{2}N^{3})$ scaling of the QFI (Sec.~\ref{sec:qfi_two}), while following the interference structure of Eq.~\eqref{eq:cfi_two_marg_f}.

\section{$N$ Controlled Displacements on a Single Ancilla}
\label{app:single_ancilla}
We consider a variation of the single-measurement protocol in which $N$ controlled displacements are interleaved with the signal interactions. Such a protocol can be described as 
\begin{align}
    \ket{\psi} = \left( \prod_{j=1}^{N} \hat{D}(\hat{\sigma}_{x}\alpha)\,\hat{D}(\beta) \right) \ket{\downarrow} \otimes \ket{0}.
\end{align}
Taking $\alpha = iA$ and $\beta \in\mathbb{R}$ for single-parameter estimation. A single round gives
\begin{equation}
    \hat{D}(\hat{\sigma}_{x}\alpha)\,\hat{D}(\beta) = e^{\,iA\beta\hat{\sigma}_{x}}\, \hat{D}(\beta + iA\hat{\sigma}_{x}),
\end{equation} 
The full system state is
\begin{equation}
    \label{eq:full_system_single_ancilla}
    \ket{ \psi} = \frac{1}{ \sqrt{2} } \left( e^{i N A \beta} \ket{+} \ket{N \beta + i N A}  + e^{-i N A \beta } \ket{-} \ket{N \beta - iNA } \right),
\end{equation}
and the qubit state is 
\begin{equation}
    \label{eq:ancilla_system_single_ancilla}
    \rho(\beta) = \frac{1}{2} \begin{pmatrix}  1 & e^{-2N^{2} A^{2}} e^{2iN(N+1)A \beta } \\ e^{-2N^{2} A^{2}} e^{-2iN(N+1)A \beta } & 1
    \end{pmatrix}.
\end{equation}

Using Eq.~\ref{eq:full_qfi} we can compute the QFI of the full system by first taking the derivative,
\begin{equation}
    \ket{\partial_{\beta}\psi}
    = \frac{1}{\sqrt{2}} \sum_{a \in \{+,- \}}
      e^{\,iNA\beta\, a}\,
      \ket{a} \otimes
      \left[ i N A a + N \hat{a}^{\dagger} - N^{2}\beta \right]
      \ket{N\beta + iaNA}.
\end{equation}
The overlaps of Eq.~\eqref{eq:full_qfi} are
\begin{equation}
    \braket{\psi | \partial_{\beta}\psi}
    = \frac{1}{2} \sum_{a \in \{ + , - \}}
      \left[ i N A a + N\left( N\beta - i a N A \right) - N^{2}\beta \right]
    = 0,
\end{equation}
\begin{align}
    \braket{\partial_{\beta}\psi | \partial_{\beta}\psi}
    &= \frac{1}{2} \sum_{a \in \{ + , - \}}
       \bra{N\beta + iaNA}
       \left[ -i N A a + N \hat{a} - N^{2}\beta \right]
       \left[ i N A a + N \hat{a}^{\dagger} - N^{2}\beta \right]
       \ket{N\beta + iaNA} \\
    &= \frac{1}{2} \sum_{a \in \{ +,- \}}
       \left\{ \left| i N A a + N\left( N\beta - i a N A \right) - N^{2}\beta \right|^{2}
       + N^{2} \right\}
    = N^{2} A^{2} (N-1)^{2} + N^{2},
\end{align}
so that
\begin{equation}
    \label{eq:single_ancilla_full_qfi}
    \mathcal{F}^{Q}_{\mathrm{full}}
    = 4 \left( \braket{\partial_{\beta}\psi | \partial_{\beta}\psi}
    - \left| \braket{\psi | \partial_{\beta}\psi} \right|^{2} \right)
    = 4 N^{2} + 4 A^{2} N^{2} (N-1)^{2}.
\end{equation}
 
 Using Eq.~\ref{eq:qfi_def} we can compute the QFI of the ancillae as
 \begin{align}
    \label{eq:single_ancilla_ancilla_qfi}
    \mathcal{F}_{Q}^{\,\mathrm{ancilla}}
    &= 2\left[
    \frac{\left| -\tfrac{i}{2}\,\big(2N(N{+}1)A\big)\, e^{-2N^{2}A^{2}} \right|^{2}}{\tfrac{1+e^{-2N^{2}A^{2}}}{2}+\tfrac{1-e^{-2N^{2}A^{2}}}{2}}
    + \frac{\left| \tfrac{i}{2}\,\big(2N(N{+}1)A\big)\, e^{-2N^{2}A^{2}} \right|^{2}}{\tfrac{1-e^{-2N^{2}A^{2}}}{2}+\tfrac{1+e^{-2N^{2}A^{2}}}{2}}
    \right] \nonumber\\[4pt]
    &= 2\left[ N^{2}(N{+}1)^{2}A^{2}\,e^{-4N^{2}A^{2}} + N^{2}(N{+}1)^{2}A^{2}\,e^{-4N^{2}A^{2}} \right]
    = 4N^{2}(N{+}1)^{2}A^{2}\,e^{-4N^{2}A^{2}}.
\end{align}
Eq.~\ref{eq:single_ancilla_full_qfi} scales as $\mathcal{O}(A^{2}N^{4})$. The $N^{4}$ is greater than the $N^{3}$ of the presented sequential-measurement protocol (App.~\ref{app:full_qfi}). However, if only the ancilla is available for measurements, Eq.~\ref{eq:single_ancilla_ancilla_qfi} shows the optimal scaling is damped by the $N$-dependent factor $e^{-4N^{2} A^{2} }$. Comparing this to the single-measurement protocol ancilla QFI (App.~\ref{app:single_meas_qfi_app}) we see an additional $N$-dependent damping factor on the QFI, such that both $A$ and $N$ are not leveraged as resources. Meanwhile, we see the ancillae QFI of the sequential protocol uses $A$, $N$ as resources (Eq.~\ref{eq:qfi_backaction_app}). Though the QFI of the sequential-measurement protocol is computed for $N$ ancillae, we note its equivalence to using a single ancilla for product measurements (App.~\ref{app:one_to_N}, Fig.~\ref{fig}).

\section{Noise Model}
\label{app:decoherence}
Since decoherence breaks permutation invariance, we instead simulate the protocol numerically by evolving the oscillator under the Lindblad master equation
\begin{equation}
    \label{eq:master_eq}
    \dot{\rho} = -\frac{i}{\hbar}\left[\hat{H}_{\mathrm{sig}}, \rho\right] + \Gamma( \bar n_{\mathrm{th}}+1)\,\mathcal{D}[\hat{a}]\rho + \Gamma \bar n_{\mathrm{th}}\,\mathcal{D}[\hat{a}^{\dagger}]\rho + \Gamma_{\phi}\,\mathcal{D}[\hat{a}^{\dagger}\hat{a}]\rho,
\end{equation}
where $\mathcal{D}[\hat{L}]\rho = \hat{L}\rho\hat{L}^{\dagger} - \tfrac{1}{2}\{\hat{L}^{\dagger}\hat{L}, \rho\}$. The first two terms describe loss into and heating from a bath with mean occupation $\bar n_{\mathrm{th}} = (e^{\hbar\omega/k_{B}T}-1)^{-1}$ at loss rate $\Gamma$, and the third describes pure dephasing at rate $\Gamma_{\phi}$. The signal displacement is created by a resonant drive that creates a coherent state whose magnitude increases linearly in time. Therefore, we can model the signal Hamiltonian as
\begin{equation}
    \label{eq:signal_hamiltonian}
    \hat{H}_{\mathrm{sig}} = i\hbar\,\varepsilon_{\beta} \left( \hat{a}^{\dagger} - \hat{a} \right), \qquad \varepsilon_{\beta} = \beta/\tau,
\end{equation}
which in the absence of decoherence integrates to the ideal signal displacement, $e^{-\tfrac{i}{\hbar}\hat{H}_{\mathrm{sig}}\tau}$. Here the signal integrates each round for duration $\tau$ at drive strength $\varepsilon_{\beta}$. The noise channels are independent of $\hat H_{\text{sig}}$ and therefore Eq.~\ref{eq:master_eq} becomes
\begin{equation}
    \mathcal{L} [\rho] \equiv \frac{\beta}{\tau}\left[\hat{a}^{\dagger} - \hat{a}, \rho\right] + \Gamma( \bar n_{\mathrm{th}}+1)\,\mathcal{D}[\hat{a}]\rho + \Gamma \bar n_{\mathrm{th}}\,\mathcal{D}[\hat{a}^{\dagger}]\rho + \Gamma_{\phi}\,\mathcal{D}[\hat{a}^{\dagger}\hat{a}]\rho.
\end{equation}
The above is time-independent. To obtain the final state after noise we integrate $\rho(t) = e^{t \mathcal{L}} \rho(0)$ to integer multiples of $\tau$.

\subsection{Signal attenuation under loss}
\label{app:amp_decay}

Taking the mean of Eq.~\eqref{eq:master_eq} with the drive of Eq.~\eqref{eq:signal_hamiltonian}, the mean amplitude obeys
\begin{equation}
    \label{eq:amp_ode}
    \partial_{t}\langle\hat{a}\rangle
    = \varepsilon_{\beta} - \frac{\Gamma}{2}\langle\hat{a}\rangle,
\end{equation}
where the $\bar n_{\mathrm{th}}$ contributions of the loss and heating dissipators cancel. The drive and the damping balance at $2\varepsilon_{\beta}/\Gamma = 2\beta/(\Gamma\tau)$. Integrating Eq.~\eqref{eq:amp_ode} from the vacuum, the amplitude at round $j$ is
\begin{equation}
    \label{eq:amp_rounds}
    \langle\hat{a}\rangle_{j}
    = \frac{2\beta}{\Gamma\tau} \left( 1 - e^{-j\Gamma\tau/2} \right),
\end{equation}
which follows the noiseless accumulation, $\langle\hat{a}\rangle_{j} \approx j\beta$, for $j \ll 2/\Gamma\tau$ and converges to $2\beta/(\Gamma\tau)$ afterwards. The fringe phase of round $j$ (Eq.~\eqref{eq:single_round_marginal}) is then set by $\langle\hat{a}\rangle_{j}$ in place of $j\beta$. For $\Gamma\tau \gg 1$, the amplitude converges within a single round, so every round interrogates the same state regardless of the measurement history.

\subsection{Dephasing of the sequential fringe contrast}
\label{app:dephased_contrast}
Pure dephasing acts on the matrix elements as $\rho_{nm} \to e^{-\Gamma_{\phi}\tau(n-m)^{2}/2}\,\rho_{nm}$, which is equivalent to a rotation by a random angle,
\begin{equation}
    e^{\Gamma_{\phi}\tau\,\mathcal{D}[\hat{a}^{\dagger}\hat{a}]}\rho
    = \big\langle\,
    e^{-i\theta\hat{a}^{\dagger}\hat{a}}\,\rho\,
    e^{\,i\theta\hat{a}^{\dagger}\hat{a}}
    \,\big\rangle_{\theta},
    \qquad
    \theta \sim \mathcal{N}(0, \Gamma_{\phi}\tau),
\end{equation}
since $\langle e^{-i\theta(n-m)}\rangle_{\theta} = e^{-\Gamma_{\phi}\tau(n-m)^{2}/2}$. Round $k$ therefore contributes an independent random angle $\theta_{k}$, and dephased probabilities are noiseless probabilities averaged over $\{\theta_{k}\}$. Using $e^{-i\theta\hat{a}^{\dagger}\hat{a}}\,\hat{D}(\gamma) = \hat{D}(\gamma e^{-i\theta})\,e^{-i\theta\hat{a}^{\dagger}\hat{a}}$, the rotations commute through the circuit and are absorbed by the ground state, rotating each displacement by the angle accumulated after it, $\gamma_{m} \to \gamma_{m}\,e^{-i\Theta_{m}}$ where $ \Theta_{m} = \sum_{k \geq m} \theta_{k}$. Rotations applied after the readout of round $j$ cannot affect its outcome, so only rounds $k < j$ enter its contrast. The fringe of round $j$ oscillates as $\cos(4A\,\operatorname{Re}\,\mu_{j})$, where $\mu_{j}$ is the oscillator amplitude at readout. Then the rotation $\theta_{k}$ roughly displaces the state by $-i\theta_{k}\mu_{k}$ and shifts the fringe phase by $ \delta\varphi_{j} = 4A\,\theta_{k}\,\operatorname{Im}\,\mu_{k}$. Therefore, averaging over $\theta_{k}$ multiplies the contrast by
\begin{equation}
    \big\langle e^{\,i\delta\varphi_{j}} \big\rangle_{\theta_{k}} = \exp\!\left[ -8A^{2}\,\Gamma_{\phi}\tau\, (\operatorname{Im}\,\mu_{k})^{2} \right],
\end{equation}
and the independent rounds multiply. Replacing $(\operatorname{Im}\,\mu_{k})^{2}$ with the mean occupation,
\begin{equation}
    \langle\hat{n}\rangle_{k} = (k\beta)^{2} + kA^{2},
\end{equation}
where the second term is the occupation added by the conditional displacements, the fringe contrast of round $j$ is suppressed by
\begin{equation}
    \label{eq:deph_contrast}
    \kappa_{j}^{\phi} = \exp\!\left[ -8A^{2}\,\Gamma_{\phi}\tau \sum_{k<j} \langle\hat{n}\rangle_{k} \right].
\end{equation}
For $A^{2} \gg N\beta^{2}$, the geometric phase term dominates, $\sum_{k<j}\langle\hat{n}\rangle_{k} \approx A^{2}j(j-1)/2$, giving
\begin{equation}
    \kappa_{j}^{\phi} \approx \exp\!\left[ -4A^{4}\,\Gamma_{\phi}\tau\, j(j-1) \right].
\end{equation}
The same argument applies to the single-measurement protocol, where the QHO carries only the signal occupation before its one measurement, $\langle\hat{n}\rangle_{k} \approx (k\beta)^{2}$. Summing over the $N$ drive intervals preceding the measurement, $\sum_{k<N}\langle\hat{n}\rangle_{k} \approx \beta^{2}N^{3}/3$, gives the contrast suppression $\kappa^{\phi}_{\mathrm{single}} \approx \exp\!\left[-\tfrac{8}{3}A^{2}\Gamma_{\phi}\tau\,\beta^{2}N^{3}\right]$, with exponential growth throughout the entire interrogation time.

\subsection{Dephasing coherence window}
\label{app:coherence_window}

The round-dependent contrast suppression of Eq.~\eqref{eq:deph_contrast} sets the number of informative rounds under dephasing. Inserting the suppressed contrast $\kappa_{j}^{\phi}\,e^{-2A^{2}}$ into the marginal CFI (Eq.~\eqref{eq:cfi_marginal}), similar to the two-parameter case in Eq.~\eqref{eq:cfi_two_marg_f}, gives
\begin{equation}
    \label{eq:cfi_marg_dephased}
    \mathcal{F}^{C}_{\mathrm{seq}, \mathrm{marg}}
    = \sum_{j=1}^{N}
    \frac{16\, j^{2} A^{2}\, \big(\kappa_{j}^{\phi}\big)^{2}\, e^{-4A^{2}} \sin^{2}\varphi_{j}}
         {1 - \big(\kappa_{j}^{\phi}\big)^{2}\, e^{-4A^{2}} \cos^{2}\varphi_{j}},
    \qquad
    \varphi_{j} = 4 j A \beta .
\end{equation}
In the geometric-phase-dominated regime, $\kappa_{j}^{\phi} \approx \exp[-4A^{4}\Gamma_{\phi}\tau\, j(j-1)]$, so the weight of round $j$ decays super-exponentially with the round index. Defining the coherence window $j^{\ast}$ as the round at which the contrast has fallen to $e^{-1}$, $4A^{4}\,\Gamma_{\phi}\tau\; j^{\ast}\big(j^{\ast}-1\big) = 1$ giving $j^{\ast} \approx \big( 4A^{4}\,\Gamma_{\phi}\tau \big)^{-1/2}$ for $j^{\ast} \gg 1$. Rounds $j < j^{\ast}$ contribute to Eq.~\eqref{eq:cfi_marg_dephased} with nearly their noiseless weight, while the contribution of round $j$ is suppressed by $\exp[-2(j/j^{\ast})^{2}]$ beyond it, so the sum converges within a few multiples of $j^{\ast}$. For $N \gtrsim j^{\ast}$ the marginal CFI therefore saturates at approximately its noiseless value evaluated at $N = j^{\ast}$, further rounds add negligible information, and increasing the coupling shrinks the window as $j^{\ast} \propto A^{-2}$. Equation~\eqref{eq:cfi_marg_dephased} treats the rounds as independent where the inter-round correlations (Eq.~\eqref{eq:round_correlations}) are carried by the same oscillator coherences suppressed by the random rotations.

\subsection{Regions of optimal operation}
\label{app:noise_maps}

\begin{figure}[!ht]
    \centering
    \includegraphics[width=0.8\textwidth]{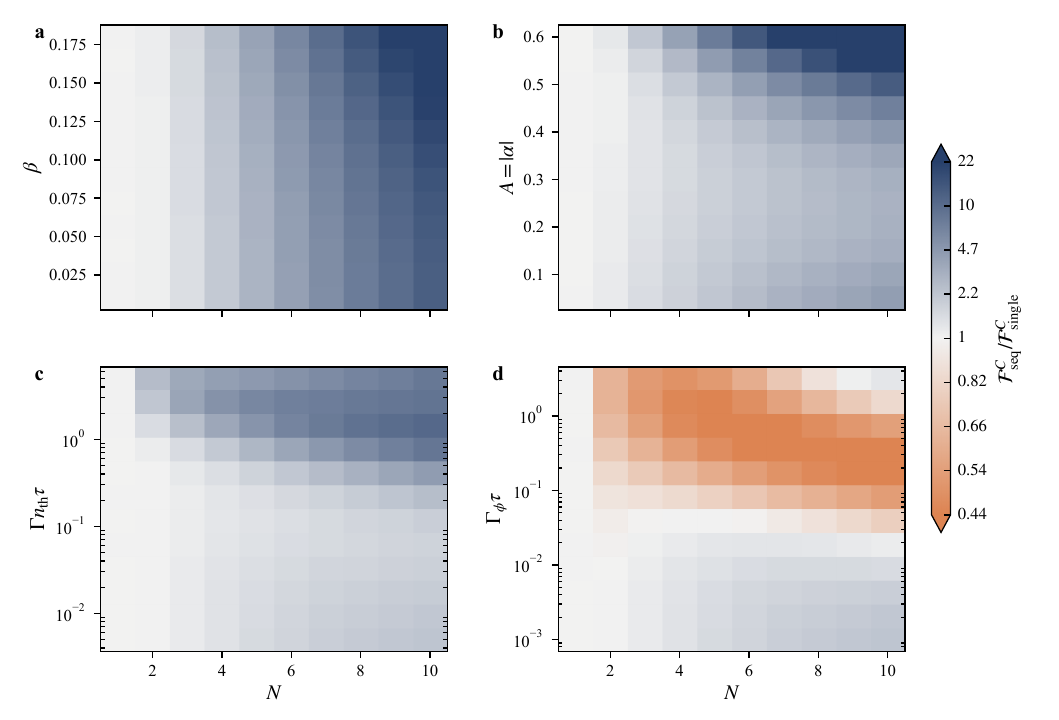}
    \caption{Ratio of the CFI of the two protocols, $\mathcal{F}^{C}_{\mathrm{seq}}/\mathcal{F}^{C}_{\mathrm{single}}$, with computational-basis measurement, as a function of the number of signal interactions $N$. Blue marks $\mathcal{F}^{C}_{\mathrm{seq}} > \mathcal{F}^{C}_{\mathrm{single}}$. The two protocols coincide at $N=1$. \textbf{a} $\beta$ at $A = 0.5$. \textbf{b} $A$ at $\beta = 0.05$. Panels \textbf{a} and \textbf{b} use the noise parameters of Fig.~\ref{fig:noise}(c), $\Gamma\tau = 0.15$, $\bar n_{\mathrm{th}} = 5$, $\Gamma_{\phi}\tau = 0.5$. \textbf{c} Heating and loss at $\Gamma_{\phi} = 0$, $\bar n_{\mathrm{th}} = 5$. \textbf{d} Pure dephasing at $\Gamma = 0$. Panels \textbf{c} and \textbf{d} use $A = 0.5$ and $\beta = 0.05$.}
    \label{fig:appx_regions}
\end{figure}

Figure~\ref{fig:appx_regions} compares the two protocols away from the operating point of Fig.~\ref{fig:noise}(c), varying the signal amplitude, the coupling amplitude, and each noise channel. The sequential protocol maintains its advantage across the amplitudes considered and under heating and loss. The single-measurement protocol is advantageous when dephasing rises above $\Gamma_{\phi}\tau = 10^{-1}$.

%\clearpage
\stopcontents[appx]
\twocolumngrid
\bibliography{resubmission_refs}

\end{document}